\documentclass[aps,prd,twocolumn,letterpaper,showpacs,floatfix]{revtex4}
\usepackage[latin1]{inputenc}

\usepackage[squaren]{SIunits}
\usepackage{amssymb,amsmath}
\usepackage{graphicx}
\usepackage{dcolumn}
\newcolumntype{.}{D{.}{.}{-1}}
\usepackage{hyperref}

\begin{document}

\newcommand{\Eref}[1]{Eq.~(\ref{#1})}
\newcommand{\Fig}[1]{Fig.~\ref{#1}}
\newcommand{\Figure}[1]{Figure~\ref{#1}}
\newcommand{\Sec}[1]{Sec.~\ref{#1}}
\newcommand{\Tab}[1]{Tab.~\ref{#1}}
\newcommand{\Pizza}{\textsc{pizza\ }}
\newcommand{\D}{\partial}
\newcommand{\Rmd}{\rho}
\newcommand{\Rmdp}{\rho_p}
\newcommand{\Sed}{\epsilon}
\newcommand{\Crmd}{D}
\newcommand{\Ced}{E}
\newcommand{\Cmom}{S}
\newcommand{\Vc}{\sqrt{\gamma}}
\newcommand{\Lf}{W}
\newcommand{\Lapse}{\alpha}
\newcommand{\e}[1]{\ensuremath{\cdot 10^{#1}}}
\newcommand{\parsec}{\ensuremath{\mathrm{pc}}}
\newcommand{\Aelec}{A^{E2}}
\newcommand{\Amagn}{A^{B2}}
\newcommand{\Prad}{\ensuremath{F}}
\newcommand{\Paxi}{\ensuremath{{}^2\!f_0}}
\newcommand{\Pcor}{\ensuremath{{}^2\!f_{-2}}}
\newcommand{\Pctr}{\ensuremath{{}^2\!f_{2}}}

\renewcommand{\arraystretch}{1.1}
\addtolength{\tabcolsep}{1pt}

\title{On the saturation amplitude of the f-mode instability}

\author{Wolfgang Kastaun}
\affiliation{International School for Advanced Studies (SISSA) and
National Institute of Nuclear Physics (INFN),
via Bonomea 265, Trieste 34136, Italy}
\affiliation{Theoretical Astrophysics, Eberhard-Karls University of T\"ubingen, T\"ubingen 72076, Germany}
\author{Beatrix Willburger}
\affiliation{Theoretical Astrophysics, Eberhard-Karls University of T\"ubingen, T\"ubingen 72076, Germany}
\author{Kostas D. Kokkotas}
\affiliation{Theoretical Astrophysics, Eberhard-Karls University of T\"ubingen, T\"ubingen 72076, Germany}
\affiliation{Department of Physics, Aristotle University of Thessaloniki, Thessaloniki 54124, Greece}

\date{\today}


\begin{abstract}
We investigate strong nonlinear damping effects which occur
during high amplitude oscillations of neutron stars,
and the gravitational waves they produce.
For this, we use a general relativistic nonlinear hydrodynamics code
in conjunction with a fixed spacetime (Cowling approximation) and
a polytropic equation of state (EOS).
Gravitational waves are estimated using the quadrupole formula.
Our main interest are $l=m=2$ $f$~modes subject to the CFS (Chandrasekhar, Friedman, Schutz)
instability,
but we also investigate axisymmetric and quasiradial modes.
We study various models to determine the influence of rotation rate and EOS.
We find that axisymmetric oscillations at high amplitudes are predominantly damped
by shock formation, while the nonaxisymmetric $f$~modes are
mainly damped by wave breaking and, for rapidly rotating models,
coupling to nonaxisymmetric inertial modes.
From the observed nonlinear damping,
we derive upper limits for the saturation amplitude
of CFS-unstable $f$~modes.
Finally, we estimate that the corresponding gravitational waves
for an oscillation amplitude at the upper limit
should be detectable with the advanced LIGO and VIRGO
interferometers at distances above $10\usk\mega\parsec$.
This strongly depends on the stellar model,
in particular on the mode frequency.
\end{abstract}

\maketitle

\section{Introduction}
\label{sec:introduction}

As a consequence of Einstein's theory of relativity, it is predicted that the violent nonradial
pulsations excited immediately after the birth of a rotating proto-neutron star result
in the emission of significant amounts of gravitational radiation.
In addition, it has been discovered by
Chandrasekhar \cite{1970PhRvL..24..611C} and
Friedman \& Schutz \cite{1978ApJ...221..937F,1978ApJ...222..281F}
that certain pulsation modes of rotating relativistic stars may grow exponentially
due to gravitational radiation backreaction,
especially if the proto-neutron star is rapidly rotating;
this is the so-called  CFS instability.

The exact amount of energy emitted by the stable oscillations of neutron stars
in the form of gravitational waves  is actually unknown and depends uniquely on the initial conditions.
The stable stellar pulsations will emit gravitational waves which are detectable only
if the neutron star is in our galaxy or in the most optimal case in the nearby ones
\cite{2009arXiv0912.0384A}.
Thus, the cases in which these oscillations become unstable are of special interest
for gravitational wave astronomy,
since the corresponding gravitational waves will be detectable
even for sources in the nearby galactic clusters.
It has been suggested that the detection of gravitational waves from pulsating neutron stars
will allow the study of their interior,  see
\cite{1998MNRAS.299.1059A,2001MNRAS.320..307K,2010arXiv1005.5228G}.
It is expected that the identification of specific pulsation frequencies in the observational data will
reveal the properties of matter at densities that currently cannot be probed by any other experiment.

The study of the dynamics of fast rotating neutron stars in a general relativistic framework
was practically impossible until recently.
The main reason why linear theory failed is the form and the size of the relevant perturbation equations.
Thus, it is not surprising that the first results for the oscillations of
fast rotating stars were derived by using evolutions of the nonlinear equations,
see \cite{SAF2004,DSF2006,Kastaun06,Kastaun08}.
Still most of these studies were purely axisymmetric and thus
there was a significant influence of rotation
on the oscillation spectra only for very high rotation rates.
The CFS instability on the other hand is active only for nonaxisymmetric perturbations.
In the last two years there was a significant progress in the study of nonaxisymmetric perturbations,
for the first time in a general relativistic framework,
using both perturbation theory
\cite{2008PhRvD..78f4063G,2009PhRvD..80f4026G,2010PhRvD..81h4019K,2010arXiv1005.5228G}
and nonlinear evolutions of coupled hydrodynamic and Einstein equations \cite{2010PhRvD..81h4055Z}.

For the case of $r$~modes, the CFS instability is active for any rotation rate
\cite{1998ApJ...502..708A,1998ApJ...502..714F,1998PhRvL..80.4843L,1998PhRvD..58h4020O,1999ApJ...510..846A}.
In practice the unstable $r$ modes will be excited even for slowly rotating stars if they are
hot enough ($T > 10^7\usk\kelvin$), so that the instability growth time is shorter
than the shear viscosity damping time.
Initially it was considered as a prime source for gravitational waves,
but later more detailed studies on the effect of the magnetic field on the instability
\cite{2000ApJ...531L.139R},
or the presence of hyperons in the core  \cite{2002PhRvD..65f3006L},
seriously questioned  the potential of the instability.
Detailed studies suggested that the $r$~mode is limited to very small amplitudes due to energy transfer
to a large number of other inertial modes,
in the form of a cascade,
leading to an equilibrium distribution of mode amplitudes \cite{2002PhRvD..65b4001S}.
The small saturation values for the amplitude are supported by recent nonlinear estimations
\cite{2004PhRvD..69h4001S,2005PhRvD..71d4007S} based on the drift,
induced by the $r$~modes,
causing differential rotation.
On the other hand,
hydrodynamical simulations of limited resolution showed that an $r$~mode of  large initial amplitude
does not decay appreciably over several dynamical timescales
\cite{2001PhRvL..86.1148S},
but on a longer timescale a catastrophic decay occurs
\cite{2002PhRvD..66d1303G};
this indicates a transfer of energy to other modes
due to nonlinear mode couplings, and suggests that a hydrodynamical instability may be operating.
A specific resonant 3-mode coupling was identified in \cite{2006MNRAS.370.1295L}
as the cause of the instability and a perturbative analysis of the decay rate
suggests a maximum dimensionless saturation amplitude $\alpha_\text{max} < 10^{-2}-10^{-3}$.
A new computation using second-order perturbation theory finds that the catastrophic decay
seen in the hydrodynamical simulations \cite{2002PhRvD..66d1303G,2006MNRAS.370.1295L}
can indeed be explained by a parametric instability operating in 3-mode couplings
between the $r$~mode and two other inertial modes
\cite{2004PhRvD..70l1501B,2004PhRvD..70l4017B,2005PhRvD..71f4029B,2007PhRvD..76f4019B,2009PhRvD..79j4003B}.

The CFS instability of the $f$~modes
(nonradial modes with no radial nodes, restored by pressure)
can potentially  be
the strongest source for gravitational waves from isolated neutron stars.
Although it was the first to be studied, already 30-years ago,
it was forgotten because at that time it was nearly impossible  to study
rapidly rotating neutron stars in  a general relativistic framework.
Studies using Newtonian theory came to the disappointing conclusion
that (in Newtonian theory) the $l=m=2$ $f$~mode is stable for rotation rates
below the Kepler limit.
The $l=m=4$ $f$~mode does become unstable,
but the growth time is so long that viscosity severely limits the amplitudes,
rendering the process unimportant for gravitational wave astronomy.
Only in the late '90s, Stergioulas \& Friedman  \cite{1998ApJ...492..301S}
showed that for certain EOS the $l=m=2$ $f$~mode can become unstable in full GR (general relativity).
Still, this study was left aside since no proper linear or nonlinear codes
were available to deal with the perturbations of rapidly rotating neutron stars in GR.

If such an instability sets in,
the star will emit copious amounts of gravitational radiation.
First studies suggest that the signal may be detectable from  distances
as far as  $15\usk\mega\parsec$ for the current sensitivity of Virgo \& LIGO,
and from distances greater than $100\usk\mega\parsec$ for the sensitivities of Advanced Virgo \& LIGOs
\cite{1995ApJ...442..259L,2004PhRvD..70h4022S,2004ApJ...617..490O}.
In the latter case the event rate of supernovas resulting in the creation of a proto-neutron star
can be quite high, i.e. more than thousand per year.
However, it should be noted that it is not clear
what will be the initial rotation period of the collapsing core;
it seems to depend strongly on the profile of the angular momentum distribution,
the initial mass and angular momentum of the collapsing star,
and also the strength of the magnetic field, which can
slow down the newly born compact object quite efficiently.
The hope is that still a number of supernova events,
maybe 10\usk\%,
will produce rapidly rotating cores which can be
subject to the CFS instability of the $f$~modes.
In any case, the next generation of gravitational wave detectors,
such as Einstein Telescope (ET) \cite{2010CQGra..27h4007P}
will be ideal for the detection of this type of instability \cite{2009arXiv0912.0384A}.

The two most recent hydrodynamical simulations \cite{2004PhRvD..70h4022S,2004ApJ...617..490O}
(in the Newtonian limit and using an artificially increased
post-Newtonian radiation-reaction potential)
essentially confirm this picture.
In \cite{2004PhRvD..70h4022S} a differentially rotating,
$N=1$ polytropic model with a large $\beta=T/|W| \sim 0.2 \dots 0.26$
was chosen as the initial equilibrium state
($T$ is the rotational kinetic energy and $|W|$ the gravitational binding energy).
The main difference of this simulation compared to the ellipsoidal approximation
\cite{1995ApJ...442..259L} lies in the choice of the EOS.
For $N=1$ Newtonian polytropes it was argued that the secular evolution cannot lead
to a stationary Dedekind-like state.
Instead, the $f$-mode instability will continue to be active until all nonaxisymmetries
are radiated away and an axisymmetric shape is reached.
In another recent simulation \cite{2004ApJ...617..490O},
the initial state was chosen to be a uniformly rotating,
$N=0.5$ polytropic model with $T/|W|\sim 0.18$.
Again, the main conclusions reached in \cite{1995ApJ...442..259L} were confirmed,
however the assumption of uniform initial rotation limits
the available angular momentum that can be radiated away,
leading to a detectable signal only out to about $40\usk\mega\parsec$.
The star appears to be driven towards a Dedekind-like state,
but after about 10 dynamical periods,
the shape is disrupted by growing short-wavelength motions,
which are explained by a shearing type instability
such as the elliptic flow instability \cite{1993ApJ...408..603L}.

The recent progress
\cite{2008PhRvD..78f4063G,2009PhRvD..80f4026G,2010PhRvD..81h4019K,2010arXiv1005.5228G,2010PhRvD..81h4055Z},
demonstrates that the problem can be handled properly,
which should allow to answer the following questions:
First, has the unstable mode any chance to be excited?
Second, what is the instability window of the mode?
Third, what are the exact rotation rates at the onset of the instability for the various EOS?
And finally, what is the maximum amplitude that the mode may reach before
saturated by nonlinear phenomena, e.g. mode coupling, mass loss, or shock formation?
In this work we try to answer the last question.
In particular,
we investigate nonlinear effects active at oscillation  amplitudes
sufficiently large to allow a detection of sources at distances
greater than $1\usk\mega\parsec$.
Since our methods are not restricted to a certain mode,
we map out nonlinear effects occurring for axisymmetric modes as well.
The latter also serves as a numerically less expensive test case
to calibrate our code.

The question about the saturation amplitude of the $f$-mode instability
is a major one that has to be answered since it may completely eliminate
the importance of this instability.
In this article we present
a systematic study of nonlinear damping effects for different amplitudes,
rotation rates, and equations of state.
For this, we use a nonlinear general relativistic evolution code
named \Pizza \cite{Kastaun06,Kastaun08}.
Up to now it has been successfully tested on
small amplitude stable oscillations of rapidly rotating neutron stars and self-gravitating tori.
After a number of improvements the code is well suited to follow even
large amplitude oscillations of rotating stars,
such as those excited during the saturation phase of the $f$-mode instability.
A similar study has been performed for the $r$-mode instability in \cite{2002PhRvD..66d1303G},
where an artificially excited $r$~mode was left to evolve.
The amplitude of the $r$~mode seemed initially to decrease slowly but then it decayed catastrophically.
The decay time was found to be amplitude dependent and the effect attributed to three mode coupling.

The structure of the paper is as follows.
First we describe the numerical methods in \Sec{sec_meth}.
In \Sec{sec_models}, we introduce the stellar models and physical problems we studied.
The numerical results are presented in \Sec{sec_results}.
The main results on the damping and detectability of the CFS-unstable $f$~mode
can be found in \Sec{sec_damp_m2} and \ref{sec_detect},
and a summary is given in \Sec{sec_summary}.

Throughout the paper we use geometrical units $c=G=1$.
Greek indices run from $0\dots 3$, Latin indices from $1\dots 3$.
We denote the lowest order quasi radial mode by \Prad,
the axisymmetric fundamental pressure modes ($f$~modes) by $^l\!f_0$,
and nonaxisymmetric ones by  $^l\!f_m$, where $m>0$ refers to the counter-rotating modes.

\section{Numerical method}
\label{sec_meth}
\subsection{Time evolution}
For our simulations,
we use the \Pizza code described in \cite{Kastaun06}.
It evolves the general relativistic nonlinear hydrodynamic evolution equations
for an ideal fluid without magnetic fields,
while keeping the spacetime metric fixed (Cowling approximation).

The hydrodynamic equations in covariant form are
\begin{align}
  \nabla_\mu T^{\mu\nu} &= 0 , \label{eq_div_tmunu}\\
  \nabla_\mu \left(\Rmd u^\mu \right) &=0 .
\end{align}
The stress energy tensor $T^{\mu\nu}$ of an ideal fluid is given by
\begin{align}
  T^{\mu\nu} = \Rmd h u^\mu u^\nu + P g^{\mu\nu} ,
\end{align}
where $\Rmd$, $h$, and $u^\mu$ are the rest mass density, relativistic specific enthalpy, and
4-velocity of the fluid; $g^{\mu\nu}$ is the metric tensor of signature $(-,+,+,+)$.

The covariant equations can be written as a first order system of hyperbolic evolution equations
in conservation form with source terms:
\begin{align}
  \D_0 q &= -\D_i f^i(q,x^i) + s(q,x^i) , \label{eq_qconslaw} \\
  q &\equiv \left(\Crmd,\Ced,\Cmom_j \right) . 
\end{align}
For our study, it is important to use a conservative formulation,
because it ensures correct shock wave propagation speeds
when evolved by means of a finite volume scheme.

The evolved hydrodynamic variables are
\begin{align}
  \Crmd   &\equiv \Vc\Lf \Rmd ,\\
  \Ced    &\equiv \Vc \left( \Lf^2 \Rmd h - P -\Lf\Rmd \right) ,\\
  \Cmom_i &\equiv \Vc\Lf^2 \Rmd h v_i  , \label{def_Cmom}
\end{align}
where $\gamma$ is the determinant of the 3-metric, $v^i$ the 3-velocity,
and $\Lf$ the corresponding Lorentz factor.
In the Newtonian limit using Cartesian coordinates, $\Crmd$, $\Ced$, and $\Cmom^i$
reduce to mass density, energy density, and momentum density.

The flux terms $f^i=(f^i_{\Crmd},f^i_{\Ced},f^i_{\Cmom_j})$ are given by
\begin{align}
  f^i_{\Crmd}   &= w^i \Crmd , \\
  f^i_{\Ced}    &= w^i \Ced  + \Lapse\Vc v^i P , \\
  f^i_{\Cmom_j} &= w^i \Cmom_j  + \Lapse\Vc P \delta^i_j  ,
\end{align}
where $w^i=u^i/u^0$ is the coordinate velocity of the fluid and $\Lapse$ the lapse function.
The evolution equations need to be completed by an equation of state (EOS) to compute the pressure.
We choose a polytropic EOS defined by
\begin{align}
  P(\Rmd) = \Rmdp\left(\frac{\Rmd}{\Rmdp}\right)^\Gamma \equiv K \Rmd^\Gamma ,
  \label{eq_eos_poly}
\end{align}
where $\Rmdp$ is a constant density scale,
and the constant $\Gamma$ is the polytropic exponent.

When assuming a one-parametric EOS,
the evolution equations (\ref{eq_qconslaw}) are over-determined.
Therefore, we do not evolve the energy density $\Ced$,
but recompute it from the remaining variables and the EOS.
The physical implications will be discussed in the next subsection.

The above formulation is used by many modern relativistic codes.
For a review, we refer to \cite{FontLRR}.
Like most codes,
the \Pizza code is based on a HRSC (high resolution shock capturing) scheme,
which was however optimized for quasistationary simulations.
The difference to standard HRSC schemes is
that source and flux terms are treated more consistently,
making use of a special formulation of the source terms
in \Eref{eq_qconslaw}.
As a consequence, the code is able to preserve a stationary star to high accuracy.
For details, see \cite{Kastaun06}.

One of the weak points of current relativistic hydrodynamic schemes is the treatment
of the stellar surface, where the numerical schemes designed for the interior
would break down without further measures.
The most common remedy is to enforce an artificial atmosphere of low density.
The original \Pizza code described in \cite{Kastaun06} used a different workaround,
which works perfectly for low amplitude oscillations.
Unfortunately, we found that it yields unsatisfactory results for high amplitude oscillations
in conjunction with a stiff EOS.

For the results in this article, we used yet another method of treating the surface.
In short, it is based on a smooth transition of the numerical flux,
between the formally correct expression at and above a certain density,
and a flux which causes no acceleration at zero density.
The local density scale for the beginning of the transition
is computed from the local gravity, using an expression
which corresponds to the average density gradient
of the stationary model near the stellar surface times the grid spacing.
No artificial atmosphere is used. To prevent negative densities, the flux is limited.
The details of the scheme will be described in a forthcoming publication.
We just note that the evolution below the aforementioned density scale is still plain wrong,
like it is the case with other common schemes.
Due to the low densities involved,
the huge local errors at the surface have only limited impact on the global evolution.
Nevertheless, the surface is an important source of numerical errors for
this study, in particular with regard to the numerical damping.

The code has been tested on shock tube problems, linear oscillations
of neutron stars, see \cite{thesis, Kastaun06, Kastaun08},
and recently with self-gravitating relativistic tori.
On the timescales of the simulations presented here,
the code is able to evolve the stationary background model 
without significant changes in density or rotation profile.
In contrast to codes using artificial atmospheres,
our code exactly conserves the total mass.

We use Cartesian grids for three-dimensional simulations,
cylindrical coordinates for two-dimensional ones, and spherical coordinates
for one-dimensional problems.
The rigidly rotating models are evolved in the corotating frame.
To save computation time, we further assume equatorial symmetry.
Finally,
we apply vacuum outer boundary conditions,
such that any material hitting the outer boundary is lost,
and monitor the total mass to detect this case.

\subsection{Treatment of shock waves}
\label{sec_polyshock}
For some of our results, shock formation plays an important role.
The use of a polytropic EOS then becomes one of the main limitations,
as will be explained in the following.

For any ideal fluid system with a two-parametric EOS,
the physically correct solution evolves adiabatically
as long as there are no shock waves present.
For initial data which is isentropic,
the evolution in the absence of shocks is thus the same
as for an one-parametric EOS corresponding to a curve
of constant specific entropy of the two-parametric EOS.

The polytropic EOS used in our simulations are the curves of
constant specific entropy of the ideal gas EOS given by
\begin{align}
  P(\Rmd,\Sed) &= \left(\Gamma-1\right) \Rmd\Sed ,
\end{align}
where $\Sed$ is the specific internal energy and $\Gamma$ matches the polytropic exponent.
Our simulations are therefore equivalent to the case of an ideal gas
EOS with isentropic initial data, until shock waves form.
Hence we can accurately predict the onset of shock formation.

After the shocks form, using the polytropic EOS becomes unphysical.
Formally, the reduced system of evolution equations for the polytropic
case admits solutions with discontinuities.
However those differ from realistic shock waves in two aspects:
there is no entropy production, i.e. shock heating,
and the local conservation of energy is violated.

Nevertheless,
we also present results based on the evolution after shock formation.
For this, we make the assumption that the main difference
between shock solutions for hot and cold EOS is that
the kinetic energy converted into heat for the correct solution
is simply lost when using the cold EOS,
while the evolution of the density profile remains similar.
Although this is clearly a strong simplification,
the results may serve as a first estimate.

The reason why we do not simply use the ideal gas EOS is purely technical:
the version of our code which is capable of evolving two-parametric
EOSs is using a method of treating the surface which is
not well suited for high amplitude oscillations.
Fortunately, the only results depending on evolutions beyond shock formation
are the damping times of axisymmetric oscillations,
but not the critical amplitudes for the onset of shock formation;
for the nonaxisymmetric oscillations, no significant shocks occur, and hence
the use of the polytropic EOS is perfectly valid there.

\subsection{Estimating the gravitational wave amplitude}
\label{sec_gwmultipole}
To extract gravitational waves, we use the multipole formalism
for Newtonian sources, as described in \cite{Thorne80}.
In this formalism, the radiation field at distance $r$ far away from the source
is expanded in terms of
spin tensor harmonics
\begin{multline}
h^{TT}_{jk} \left(t+r\right) = \frac{1}{r} \sum_{l=2}^\infty \sum_{m=-l}^l \left[
\Aelec_{lm}\left(t\right) T^{E2,lm}_{jk} \right. \\
+ \left. \Amagn_{lm} \left(t\right) T^{B2,lm}_{jk} \right] , \label{eq_gwstrain}
\end{multline}
and the total gravitational luminosity is given by
\begin{align}
L &= \frac{1}{32\pi} \sum^\infty_{l=2} \sum_{m=-l}^l
\left( |\partial_t \Aelec_{lm} |^2 + |\partial_t \Amagn_{lm} |^2 \right) .
\end{align}
In the Newtonian limit, the coefficients of the radiation field multipole expansion
can be expressed
in terms of the mass- and current-multipoles of the source
\begin{align}
\Aelec_{lm} &= \frac{4\sqrt{2\pi}}{(2l+1)!!}\sqrt{\frac{(l+1)(l+2)}{(l-1)l}(2l+1)} \: \partial_t^l q_{lm} ,\\
\Amagn_{lm} &= \frac{32\pi}{(2l+1)!!} \sqrt{\frac{l+2}{2(l-1)}} \: \partial_t^l J_{lm}   ,
\end{align}
with the multipole moments given by
\begin{align}
q_{lm} &= \sqrt{\frac{4\pi}{2l+1}}\int \rho r^l Y^{lm*} \,d^3x , \label{eq_mass_mult}\\
J_{lm} &= \int \rho r^l \vec{v} \cdot \vec{Y}^{lm*}_B \,d^3x . \label{eq_curr_mult}
\end{align}
The above integrals are defined in the inertial frame, and
$\vec{Y}^{lm}_B$ denotes the vector spherical harmonics of magnetic type.

The error induced by using the above formulas for sources as relativistic
as neutron stars is difficult to estimate analytically.
In \cite{Shib03}, results using the quadrupole formula are compared
with direct wave extraction methods for the case of a pulsating
neutron star model, finding an error around 10-20\usk\% in the strain amplitude.

However, this error does not directly carry over to our results.
The reason is that besides neglecting relativistic corrections,
another error arises from the ambiguity of choosing a coordinate system
for strongly curved spacetimes.
This depends not only on the model,
but also on the gauge choices made when computing the initial data.
To get a rough estimate for the deviation from Euclidean geometry, we compute
\begin{align}
  \eta_1 &= \frac{R_{pe}}{R_e}, &
  \eta_2 &= \frac{R_{pp}}{R_p} , \\
  \eta_3 &= \frac{R_\text{circ}}{R_e}, &
  \eta_4 &= \frac{ \int_0^{R_e} \Vc x^2 \,dx }{ \int_0^{R_e} x^2 \,dx } ,
\end{align}
where $R_e$ and $R_p$ are the equatorial and polar coordinate radius,
while $R_{pe}$, $R_{pp}$, and $R_\text{circ}$ are the proper equatorial radius, proper polar radius,
and equatorial circumferential radius.

From this average quantities we estimate the error of the multipole moments
by assuming that
the radius $r$ in \Eref{eq_mass_mult} and \Eref{eq_curr_mult}
is wrong by a constant factor $\eta_r$,
in the sense that the correct value is in the range $[r/\eta_r,r \eta_r]$,
and that the volume element $d^3x$ is wrong by a factor of $\eta_4$.
We further set $\eta_r$ to the maximum over $\eta_{1\dots 3}$ and the reciprocal values.

In consequence,
the strain amplitude (which is mainly due to the $l=2$ mass multipoles)
would be wrong by a factor $\eta_t=\eta_r^2 \eta_4$.
While for most of our models, $\eta_t$ is in the range 1.2--2,
it is as big as 13 in one case.
This shows how dangerous it is to generalize error bounds for the quadrupole formula
computed for a ``typical'' neutron star model.

Our estimate does not take into account the possibility that large cancellations
occur in the integrals for the multipole moments.
In this study, this affects the quasiradial \Prad-mode oscillations,
in particular for slowly rotating stars.
For those, we do not compute the error;
to obtain robust results,
fully relativistic studies are needed.

To implement the above formalism,
we compute the multipole moments in the corotating frame used in our simulations.
For the current multipoles,
we nevertheless use the inertial frame velocity in the integrals.
It is straightforward to show that the multipole moments in the inertial frame 
are given by
\begin{align}
  q_{lm} &= e^{-im\Omega t} q_{lm}' , \\
  J_{lm} &= e^{-im\Omega t} J_{lm}' ,
\end{align}
where $q_{lm}'$ and $J_{lm}'$ are the multipole moments computed in the corotating frame
defined by $\phi'=\phi - \Omega t$,
with $\Omega$ being the angular velocity of the star.

The multipole moments are computed during the evolution
with a sampling rate of $75\usk\kilo\hertz$.
When evaluating time derivatives of 2nd order or higher,
special care has to be taken to avoid amplification
of high frequency numerical noise.
For this, we first apply numerical smoothing by convolution
of the time series with a Blackman window function.
Derivatives up to 3rd order are then computed using cubic splines.
For derivatives of 4th and 5th order,
we first compute the 3rd derivative and than apply the whole scheme again.

We computed the frequency response function of the resulting scheme,
which effectively cuts off contributions to 
the gravitational wave (GW) signal above $10\usk\kilo\hertz$.
In the frequency range of the actual signal,
the loss of strain amplitude due to the smoothing stays below 15\usk\%.

Although only the $l=2$ mass multipoles are important for our problems,
we generally compute all mass multipole moments up to $l=4$, and the
current multipoles with $m \ge 2, l\le4$,

The second important error for the strain, beside the use of the multipole formula,
is due to the Cowling approximation,
which is known to cause a significant error in the oscillation frequencies.
Given an estimate for the inertial frame frequency $\hat{f}_i$
of a harmonic oscillation in full GR,
one can approximate the strain amplitude $\hat{A}^{E2}_{20}$ by
\begin{align}
  \hat{A}^{E2}_{20} &\approx \Aelec_{20} \left(\frac{\hat{f}_i}{f_i}\right)^2 ,
\end{align}
where $f_i$ and $\Aelec_{20}$  are frequency and strain amplitude in the Cowling
approximation.

\subsection{Measuring dissipation}
In order to measure the damping of oscillations,
we monitor the evolution of an average corotating velocity defined by
\begin{align}
  \bar{v} &= \sqrt{\frac{1}{M} \int \Crmd v_c^i v_c^j g_{ij} \, d^3x} ,\\
  M       &= \int \Crmd \, d^3x ,\\
  v_c^i   &= \frac{u^i}{\Lapse u^0} = v^i - \frac{\beta^i}{\Lapse} ,
\end{align}
where $\Lapse$ and $\beta^i$ are lapse function and shift vector.
Since we are working in the corotating frame, $v_c^i$ is a corotating velocity.

This measure is zero if and only if the fluid velocity is everywhere
the same as for the unperturbed stationary star.
For the nonrotating case in the Newtonian limit, $\bar{v}$ is also directly related
to the total kinetic energy.
As long as there is only one dominant oscillation mode, the decay
of $\bar{v}$ is a measure for the decay of the mode amplitude.
On the other hand,
it is not sensitive to energy transfer from one oscillation mode to another.

From the evolution, we extract three quantities.
\begin{itemize}
\item
The initial amplitude $A_I=\bar{v}(0)$.
\item 
The final amplitude. Since $\bar{v}$ is oscillating,
we use the maximum amplitude during the last oscillation cycle,
\begin{align}\label{eq_def_afinal}
A_F &= \max\{\bar{v}(t) | T_e - T_0 \le t \le T_e \} ,
\end{align}
where $T_e$ is the time over which we evolved the system and
$T_0$ is the period of the oscillation mode used to perturb the system.
\item
The timescale $\tau$ of the initial decay of $\bar{v}$, which 
we define as
\begin{align}\label{eq_def_tau}
  e^{-\frac{\Delta T}{\tau}} &= \frac{\bar{v}_a(\Delta T)}{\bar{v}_a(0)} ,
  &
  \bar{v}_a(t) &= \sqrt{\int_t^{t+T_0} \bar{v}^2 \,dt} ,
\end{align}
where $T_0$ is the oscillation period of the mode used for perturbation,
and $\Delta T = 4 T_0$.
\end{itemize}

To detect the presence of shocks,
we make use of the fact that there exists a conserved energy
when using the Cowling approximation, given by
\begin{align}
  E_c    &= \int \rho_e \Vc \,d^3x , \label{eq_def_econs}\\
  \rho_e &= -\alpha  T^0_0 - \Lf\Rmd .
\end{align}
More precisely, $E_c$ is conserved for smooth and weak solutions of
\Eref{eq_div_tmunu}.
As mentioned,
shock solutions of the reduced system of equations used with a
one-parametric EOS violate the local energy conservation,
and thus the conservation of $E_c$.
Thus, any change of $E_c$ points to the existence of shock waves.

\subsection{Computing eigenfunctions and eigenfrequencies}
To extract eigenfunctions, we use the mode recycling method described in
\cite{DSF2006}.
In short, we perturb the star using a generic perturbation to excite different
modes and extract their frequencies using Fourier analysis.
Then we extract a first guess of the eigenfunction by evaluating
at each point of the numerical grid
the Fourier integral of specific energy and velocity perturbations
at the frequency of the desired mode.

The estimate of the eigenfunction obtained in this way is in general still
contaminated with other modes, due to the finite evolution time.
Therefore, additional simulations are performed, using the eigenfunction obtained
in the previous step as initial perturbation,
until only the desired oscillation mode is present in the evolution.

We further improved this scheme by making use of the fact that for any axisymmetric star,
the eigenfunctions of $\Sed, v^d$, $v^z$, and $v^\theta$ can be written in the
form
\begin{align}
  \delta \Sed(d,z,\phi) = e^{i\beta} \delta\hat{\Sed}(d,z) e^{im\phi} .
\end{align}
The two-dimensional eigenfunction $\delta\hat{\Sed}(d,z)$
is real-valued and has a well defined $z$-parity.
The constant $\beta$ is an irrelevant complex phase.

The Fourier analysis of the numerical evolution yields an
estimate for the complex-valued $\delta\Sed(d,z,\phi)$.
To obtain $\delta\hat{\Sed}$, we first compute the integral
\begin{align}\label{eq_ef_phi_avg}
  \delta\hat{\Sed}_c &=
  \int_0^{2\pi} \delta \Sed(d,z,\phi) e^{-im\phi}\, d\phi .
\end{align}
Since we use Cartesian coordinates for nonaxisymmetric problems,
this step involves interpolation.
For this, a multidimensional cubic spline interpolation method is applied.
Note we also increase the resolution by evaluating the above integral
at finer intervals than the original grid spacing.

Errors due to the presence of
unwanted modes with different $\phi$-dependency are strongly suppressed
during the computation of $\delta\hat{\Sed}_c$.
This step is absolutely necessary to separate co- and counter-rotating modes
for slowly rotating stars, since their frequencies approach each other.

It seems plausible to take the real or imaginary part of $\delta\hat{\Sed}_c$ as estimate
for the eigenfunction.
However, this is a bad idea since the phase is arbitrary,
which can lead to strong suppression of the actual eigenfunction
with respect to numerical errors.
It is preferable to remove the complex phase factor,
using the expression
\begin{align}\label{eq_phase_avg}
   e^{2 i \beta} &=
    \frac{\int \left(\Rmd \delta\hat{\Sed}_c\right)^2 \,d^2x}{\int \left|\Rmd \delta\hat{\Sed}_c\right|^2 \,d^2x} ,&
   \delta \hat{\Sed} &= e^{-i\beta} \delta\hat{\Sed}_c .
\end{align}
It is easy to see that this equation holds for the correct eigenfunction,
which has constant complex phase.
In case of numerical errors, the above expression yields an averaged
phase which is insensitive to the unavoidable phase errors near the nodes
of the eigenfunction, as well as the errors at the stellar surface.

Since analytically the complex phase should be constant, we can
convert it's variation into a measure for the quality
of the numerical eigenfunction
\begin{align}
  q &= \frac{\int \left| C^2 - |C|^2 \right| \,d^2x}{2\int |C|^2 \,d^2x}, &
  C &= \Rmd  \delta\hat{\Sed} . \label{eq_ef_qual}
\end{align}
It is easy to see that $q=0$ for an exact eigenfunction and $q=1$ in the worst case.
Due to the density weight, $q$ is a measure for the error of bulk motion,
while errors at the surface are ignored.
The complex phase of the velocity components is removed using the phase computed from $\delta\hat{\Sed}_c$,
taking into account the phase shift by $\pi/2$ of $\delta v^d, \delta v^z$
with respect to $\delta \Sed, \delta v^\phi$.
The quality measure is computed for each of the velocity components separately.

Axisymmetric eigenfunctions are extracted from two-dimensional simulations,
in which case we do not need to evaluate \Eref{eq_ef_phi_avg}.
Eigenfunctions of radial modes of spherical stars are computed using one-dimensional
simulations at high resolutions.
The eigenfunctions become one-dimensional and the integrals (\ref{eq_phase_avg}),
(\ref{eq_ef_qual}) are replaced by integrals over the radial coordinate.

The eigenfunctions used in this study are only extracted once with a good resolution,
and then used in all simulations, regardless of grid resolution or dimensionality.
For this, linear interpolation is used to map the numerical eigenfunction to the
desired resolution and coordinate system.
This way, we can do convergence tests and comparisons between 2D and 3D simulations
without worrying about differences in the eigenfunctions itself.

The frequencies are extracted using the time evolution
of density and velocity at some sample point in the corotating frame,
and of the multipole moments in the inertial frame defined in \Sec{sec_gwmultipole}.
For this, we fit an exponentially damped sinusoidal,
which is usually more accurate than using the Fourier transform.
The comparison between inertial and corotating frame
is useful to identify $m$ for unknown modes,
and serves as a consistency check.

The frequencies in the inertial and corotating frame are related by
\begin{align}
  f_i = \left| f_c - m F_R \right| ,
\end{align}
where $f_c$ is the frequency in the corotating frame, $f_i$ in the inertial frame,
and $F_R$ is the rotation rate of the star.
All frequencies are defined as positive and measured with respect to coordinate time.
For our setups,
$F_R$ and $f_i$ are also identical to
the rotation rate and oscillation frequency observed at infinity,
see \Sec{sec_initial_data}.
We define $m$ such that $m>0$ if the wave-patterns of the mode appear
counterrotating in the corotating frame,
and $m<0$ for modes which are corotating in the corotating frame.
Note a mode counter-rotating in the corotating frame appears corotating
in the inertial frame if $f_c  < m F_R$.

\subsection{Setting up initial data}
\label{sec_initial_data}
Our stellar models are rigidly rotating (and nonrotating)
stationary configurations of an ideal fluid in general relativity,
with a polytropic EOS defined by \Eref{eq_eos_poly}.
The models are characterized uniquely by central density $\Rmd_c$,
rotation rate $F_R$, polytropic exponent $\Gamma$ and polytropic density
scale $\Rmdp$.

To compute the spacetime describing a rigidly rotating relativistic star,
we use the code described in \cite{Ansorg03,Ansorg08}.
This code is able to solve the full set of stationary Einstein and hydrostatic equations
with high precision even for models rotating near the Kepler (mass shedding) limit.

For all simulations of a given model,
we use one and the same spacetime which is computed once with a resolution of at
least 100 points per stellar radius.
The data is mapped onto the computational grid used in our simulations
using linear interpolation.
The shift vector is initialized such that we obtain coordinates corotating with
the (rigidly) rotating star.
For three-dimensional simulations, we apply the standard transformation
from cylindrical to Cartesian coordinates.

To set up spherical stars,
we use our own code to solve the ordinary differential equations
derived by \cite{OV39} (TOV-equations).
For two- or three-dimensional simulations,
a standard transformation from spherical coordinates
to cylindrical or Cartesian coordinates is applied.

We note that the line element found by the two methods
is not exactly the same for a given nonrotating star due to different
gauge choices.
The line elements for the rotating model can be found in \cite{Ansorg03,Ansorg08}.
For the simulation itself the choice of coordinates doesn't matter,
since it is gauge invariant (up to numerical errors).
It will however have an impact on the error of the GW extraction,
as discussed in \Sec{sec_gwmultipole}.

The time coordinate, which is only fixed up to a global factor,
is normalized by the initial data codes such that $\Lapse=1$ at infinity.
Since the spacetime is stationary,
any frequency measured with respect to coordinate time
at a fixed point in the inertial coordinate frame
is identical to the frequency observed at infinity.

To excite oscillations, we perturb specific energy $\Sed$ and 3-velocity $v^i$.
Since we are using the Cowling approximation during evolution,
we neither perturb the metric nor reinforce the constraint equations.
In case the density becomes negative, which can happen near the surface,
it is simply reset to zero.
When perturbing with axisymmetric eigenfunctions,
we have the freedom to choose the phase of the oscillation
such that the initial density perturbation is zero,
and perturb only the velocity.
For nonaxisymmetric modes,
both specific energy and velocity are perturbed.

\section{Problem setup}
\label{sec_models}
In order to study nonlinear effects,
we perturb stationary neutron star models with various eigenfunctions,
which are scaled to amplitudes ranging from the linear regime
to the strongly nonlinear one, and let the system evolve long enough to observe
strong damping effects.
In detail, we study the \Prad, \Paxi, \Pctr, and \Pcor~modes.

We use models with three different polytropic EOSs, which are summarized in \Tab{tab_eos}.
The motivation behind our choice is to cover a wide range of stiffness,
in order to demonstrate it's influence.
EOS A was introduced in \cite{2004PhRvD..70h4026S} as a rough
polytropic approximation to the more realistic Pandharipande EOS as tabulated
in \cite{Arnett77}. It is the stiffest of our EOSs.
The models with EOS B and C (not to be confused with B and C in \cite{Arnett77}) are
generic toy models; EOS C is very soft, while EOS B is of medium stiffness.
EOS B is often used as a reference point in numerical studies.

For each EOS, we investigate a nonrotating model as well as
rigidly rotating models with various rotation rates,
some close to the Kepler limit.
Our models have gravitational (ADM) masses 1.4--1.9$\usk M_\odot$,
which is in the range of observed neutron star masses.
We checked that the nonrotating models are on the stable branch of the mass-radius diagram,
and expect the same for the rotating ones.
The central sound speed is $0.75\usk c$ for model MA100, $0.45\usk c$ for MB100, and $0.21 \usk c$ for MC100. 
All our models are summarized in \Tab{tab_models}.

Model MA65 is of particular astrophysical interest.
As shown in \Sec{sec_freq_ef},
the counter-rotating \Pctr~mode is most probably
subject to the CFS instability in full GR.
We stress that the CFS instability is not active in the Cowling approximation.
Even in full GR, it's growth timescale is on the order of seconds,
and therefore irrelevant on the timescales of our simulations.
However, it could provide a mechanism of exciting the high amplitudes
investigated in this study,
provided the instability is not suppressed already at much lower amplitudes
due to viscosity or mode coupling effects.

The maximum amplitude of the initial perturbation is chosen such that during the evolution,
the fluid stays inside a given region,
which is usually twice as big as the bounding box of the star.
Only for 3D simulations of model MA65,
we had to restrict ourselves to an expansion factor of 1.6,
because otherwise the corners of the corotating coordinate system would move with superluminal speed.

To excite high amplitude oscillations, we linearly scale the eigenfunctions
of specific energy and 3-velocity, and add them to the background model.
For axisymmetric simulations,
the sign of the perturbation is chosen such that the first maximum of the $x$-velocity 
along the $x$-axis  is positive.
For nonaxisymmetric perturbations, the sign is irrelevant
since changing the sign corresponds to a rotation.
In the rest of this paper we refer to such a high amplitude perturbation
based on the eigenfunction e.g. of the \Prad~mode simply as an \Prad-mode perturbation.

Note that this choice is not unique.
For example, one could  scale the momentum density instead of the velocity,
or change the sign of the perturbation.
In the nonlinear regime, this will lead to small differences in the results.
Also, our setups probably differ slightly from what one would obtain by letting a mode grow
to high amplitudes by means of some physical instability or artificial backreaction force.

\begin{table}
\caption{\label{tab_eos}
The polytropic EOS used in our models, specified by polytropic constant $\Gamma=1+1/n$
and polytropic density scale $\rho_p$.
The more commonly used polytropic
constant $K = \Rmd_p^{1-\Gamma}$ is given in geometric units $G=c=M_\odot=1$.
}
\begin{ruledtabular}
\begin{tabular}{lllll}
Name & $n$       & $\Gamma$ & $\rho_p / \gram\usk\centi\meter\rpcubed$ & $K$ \\\hline
C    & 2         & 1.5      & 7.000\e{16}            & 2.970 \\
B    & 1         & 2        & 6.176\e{15}        & 100.0  \\
A    & 0.6849    & 2.46     & 4.070\e{15}        & 11.65 \\
\end{tabular}
\end{ruledtabular}
\end{table}

\begin{table}
\caption{\label{tab_models}
Details of the stellar models. 
$M$ is the gravitational mass,
$F_R$ the rotation rate as observed from infinity,
$R_c$ the equatorial circumferential radius, 
$R_p$ and $R_e$ the polar and equatorial coordinate radius,
and $\rho_c$ the central rest mass density.
}
\begin{ruledtabular}
\begin{tabular}{lllllll}
Name  & EOS & $M / M_\odot$  & $F_R / \hertz$ & $R_c / \kilo\meter$ & $R_p/R_e$ & $\rho_c / \gram\usk\centi\meter\rpcubed$ \\\hline
MA100 & A   & 1.615          & 0              & 9.529      & 1         & 2.065\e{15}        \\
MA65  & A   & 1.910          & 1687           & 11.61      & 0.65      & 2.065\e{15}        \\
MB100 & B   & 1.400          & 0              & 14.16      & 1         & 7.905\e{14}        \\
MB85  & B   & 1.503          & 590.9          & 15.38      & 0.85      & 7.905\e{14}        \\
MB70  & B   & 1.627          & 792.1          & 17.27      & 0.70      & 7.905\e{14}        \\
MC100 & C   & 1.400          & 0              & 47.60      & 1         & 7.618\e{13}        \\
MC95  & C   & 1.400          & 56.97          & 51.17      & 0.95      & 6.660\e{13}        \\
MC85  & C   & 1.400          & 83.47          & 59.41      & 0.85      & 5.222\e{13}        \\
MC65  & C   & 1.400          & 93.11          & 80.80      & 0.65      & 4.007\e{13}        \\
\end{tabular}
\end{ruledtabular}
\end{table}

\section{Numerical results}
\label{sec_results}
\subsection{Mode frequencies and eigenfunctions}
\label{sec_freq_ef}
As a prerequisite for our studies, we extracted the frequencies and eigenfunctions
of various low-order axisymmetric and nonaxisymmetric pressure modes.
All frequencies can be found in \Tab{tab_freq_modes}.
Based on convergence tests, including those in \cite{Kastaun06},
we are confident that the numerical error of the frequency is generally below 2\usk\%.

We also compared our results to those obtained by \cite{2008PhRvD..78f4063G},
where a linearized evolution code and Cowling approximation was used.
The differences are below 1.2\usk\% for models MA100, MA65,
and below 0.6\usk\% for models MB100, MB85, MB70,
which is an excellent agreement.
For the nonrotating models MC100 and MA100,
the frequencies we extracted for the \Paxi and \Pctr~modes
(which are exactly the same analytically)
agree better than 0.1\usk\%.

For the extraction of nonaxisymmetric modes,
the mode-recycling method is computationally very expensive,
since it requires repeated 3D simulations over many oscillation periods.
Therefore, we did not repeat the recycling step as often as for
axisymmetric modes, and used a resolution of only 50 points per stellar radius,
in contrast to 100--200 points for 2D simulations.

As a consequence,
when using the nonaxisymmetric numeric eigenfunctions as a perturbation,
other modes are sometimes excited as well,
at amplitudes up to a few \% of the desired mode.
We believe that regarding the dynamics of the evolution,
this error can be neglected in comparison to other numerical errors.
When interpreting the GW signal on the other hand,
it has to be taken into account.

In agreement with \cite{2008PhRvD..78f4063G},
we find that the counter-rotating \Pctr~mode of model MA65
becomes corotating in the inertial frame.
Although we are working in the Cowling approximation,
this should be the case in full GR as well.
As shown in \cite{2010PhRvD..81h4055Z} for similar models,
dropping the Cowling approximation seems to lower the
neutral point,
i.e. the critical rotation rate where the counter-rotating $f$~mode becomes corotating
in the inertial frame.
Therefore, the counter-rotating \Pctr~mode of model MA65
is most probably CFS-unstable.

\begin{figure}
  \includegraphics[width=\columnwidth]{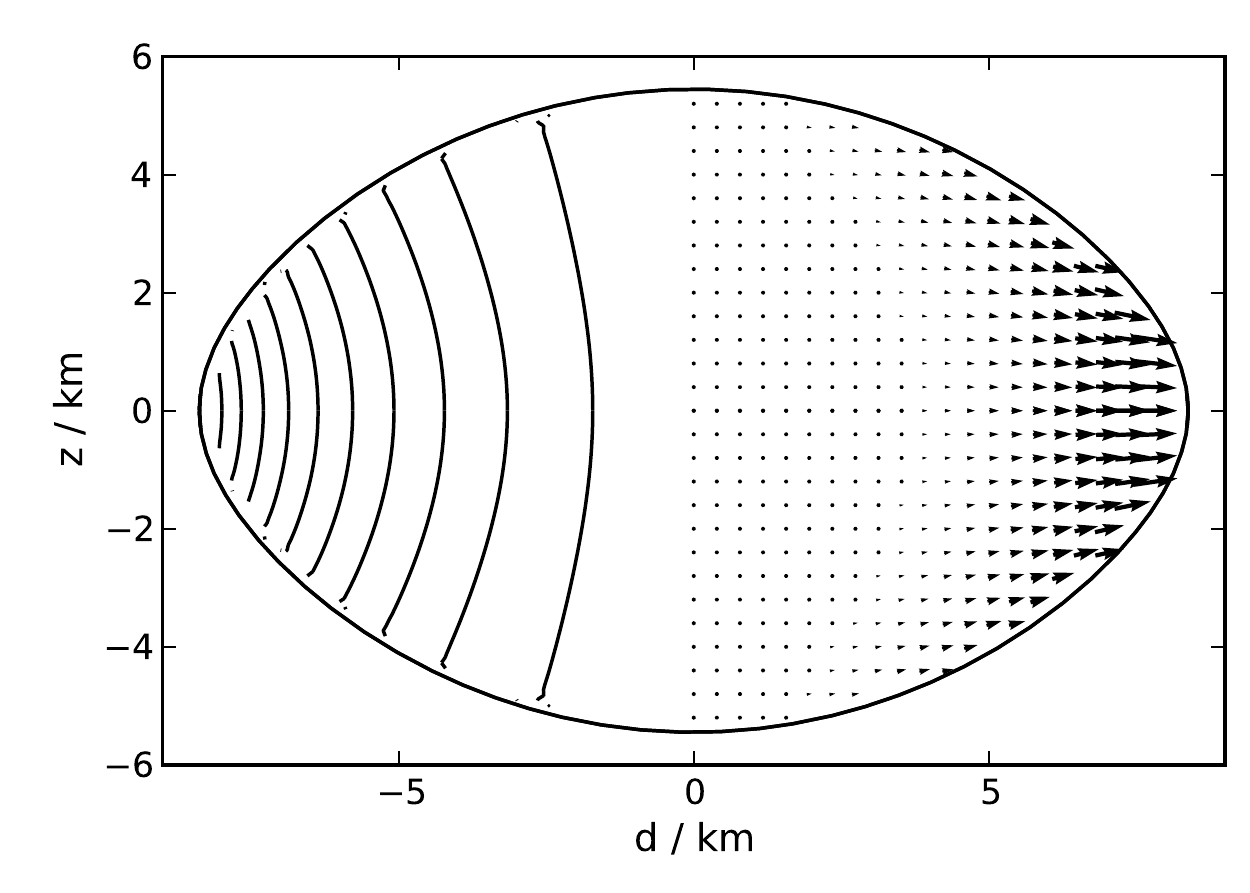}
  \caption{\label{fig_ef_jega65_fl2m2}
Eigenfunction of the counter-rotating \Pctr~mode of model MA65.
The left half shows the two-dimensional eigenfunction $\delta\hat{\Sed}(d,z)$ of specific energy
as a contour plot. The eigenfunction is zero on the rotation axis.
The arrows in the right half correspond to the eigenfunction of the velocity in the meridional plane.
  }
\end{figure}

\begin{figure}
  \includegraphics[width=\columnwidth]{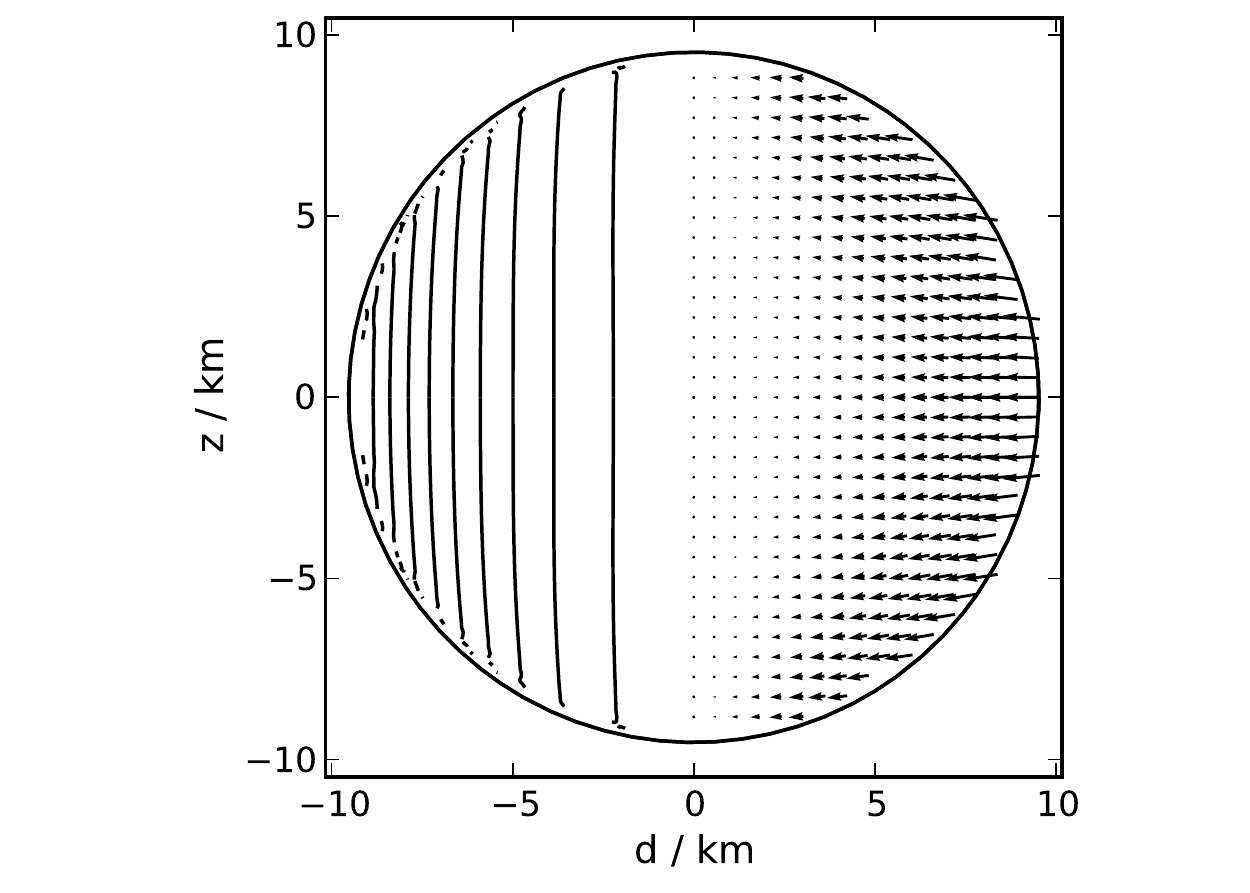}
  \caption{\label{fig_ef_ega1_fl2m2}
Like \Figure{fig_ef_jega65_fl2m2}, but showing
the \Pctr~mode of nonrotating model MA100.
The specific energy eigenfunction is zero only on the rotation axis.
  }
\end{figure}

The eigenfunction of the \Pctr~mode is shown in \Figure{fig_ef_jega65_fl2m2}.
Although model MA65 is rotating quite rapidly,
the eigenfunction is only moderately deformed
in comparison to the nonrotating model MA100 shown in \Figure{fig_ef_ega1_fl2m2}.
The main difference is that,
compared to the inner region,
the oscillation amplitude near the equator is increased.
This is to be expected since the material is bound less strongly with
increasing centrifugal force.

The eigenfunctions we found for stars with moderate rotation rates are
well described by the slow-rotation limit,
where the eigenfunctions are given by a spherical harmonic
times some radial function.
For the purpose of our study,
it is more interesting to look at the rapidly rotating models.

\begin{figure}
  \includegraphics[width=\columnwidth]{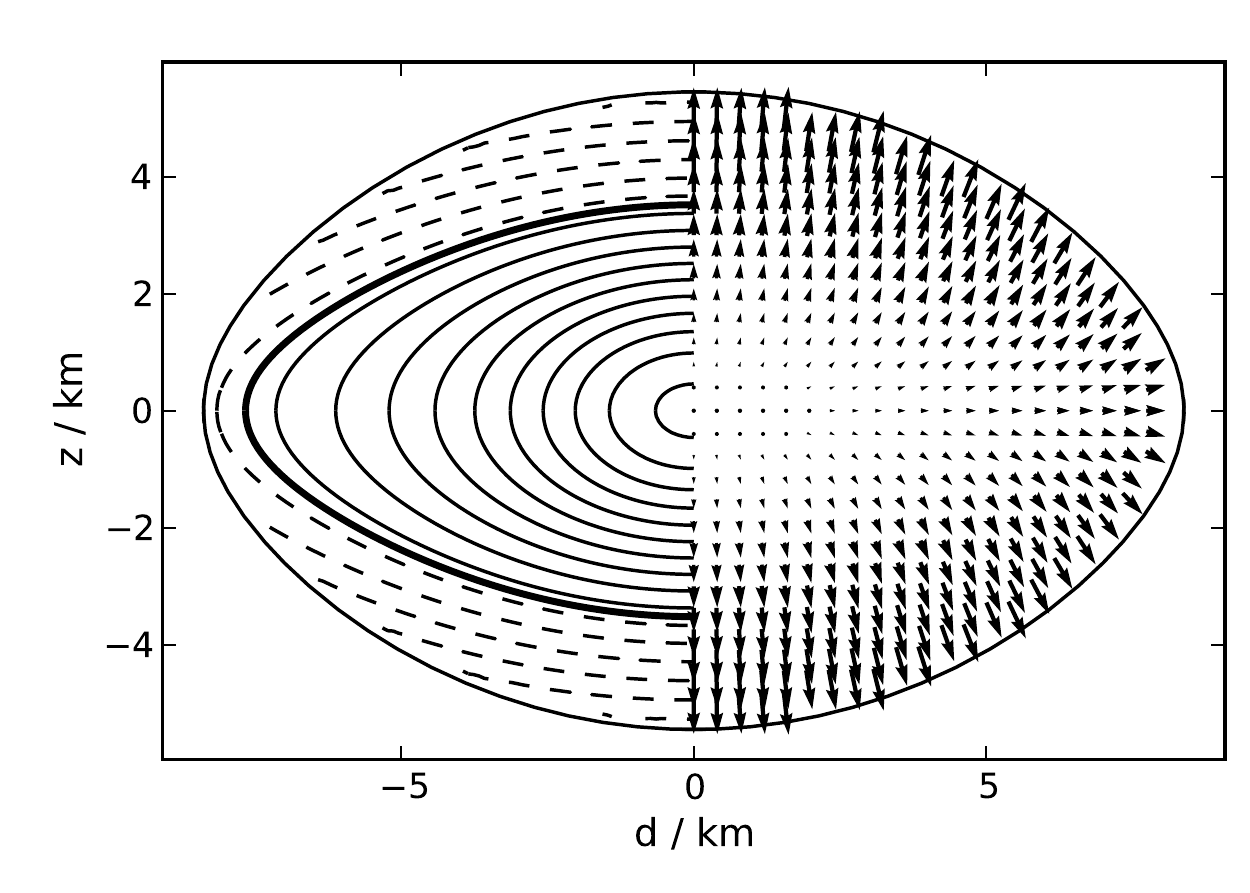}
  \caption{\label{fig_ef_jega65_fl0m0}
Eigenfunction of the quasiradial \Prad~mode of model MA65,
plotted like in \Figure{fig_ef_jega65_fl2m2}.
The additional thick solid line marks the node.
  }
\end{figure}

One important observation with regard to GW emission
is that for rapidly rotating models,
the quasiradial \Prad~mode possesses a considerable quadrupole moment
due to the oblateness of those stars.
As an example, the quasiradial \Prad~mode of model MA65
is shown in \Figure{fig_ef_jega65_fl0m0}.
We have to stress that the figure shows the eigenfunction in the coordinate
system set up by the initial data code.
It would be easy to construct an asymptotically Euclidean coordinate system in which
even a spherical star looks oblate.
However, given that MA65 is rotating near the Kepler limit,
we are confident that the deformation of the star and the eigenfunction
is not mainly a coordinate artifact.

\begin{figure}
  \includegraphics[width=\columnwidth]{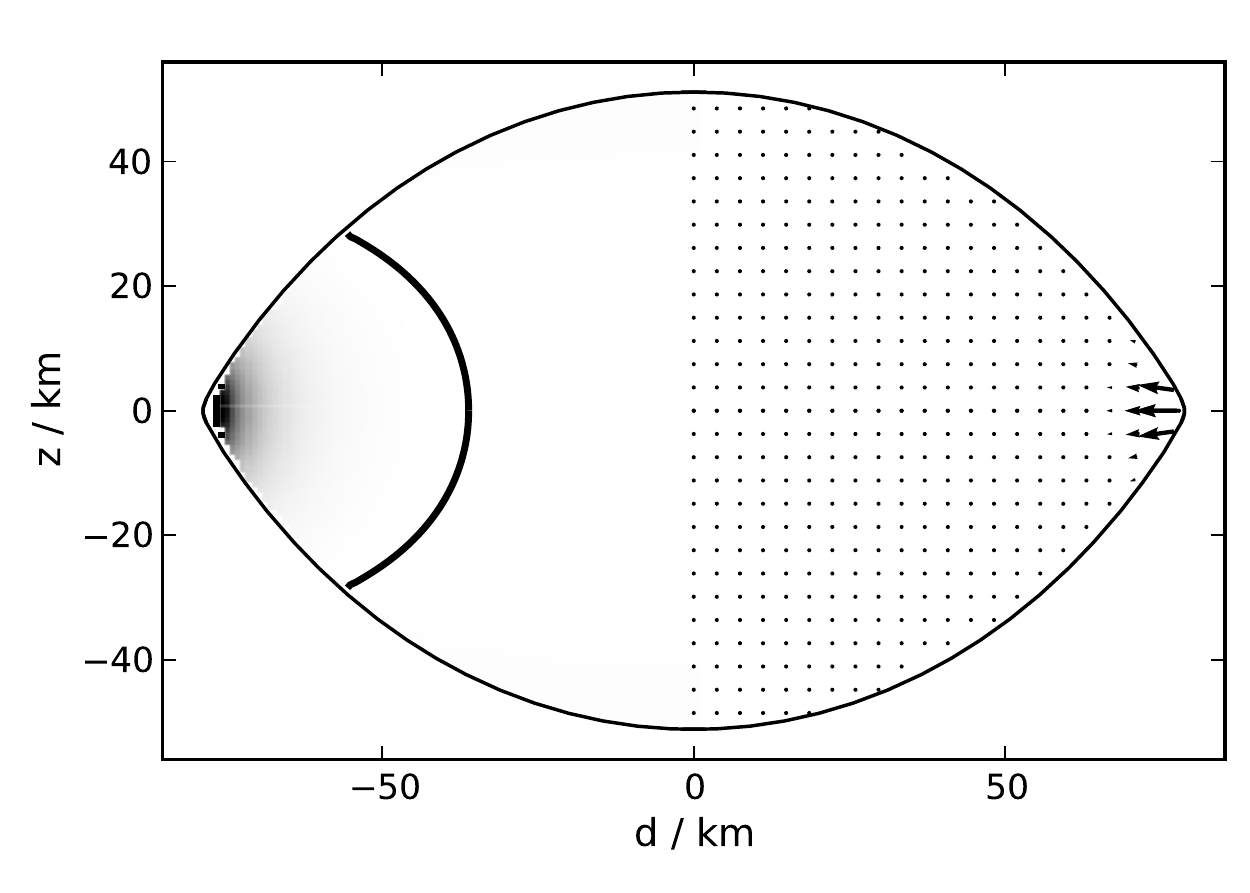}\\
  \includegraphics[width=\columnwidth]{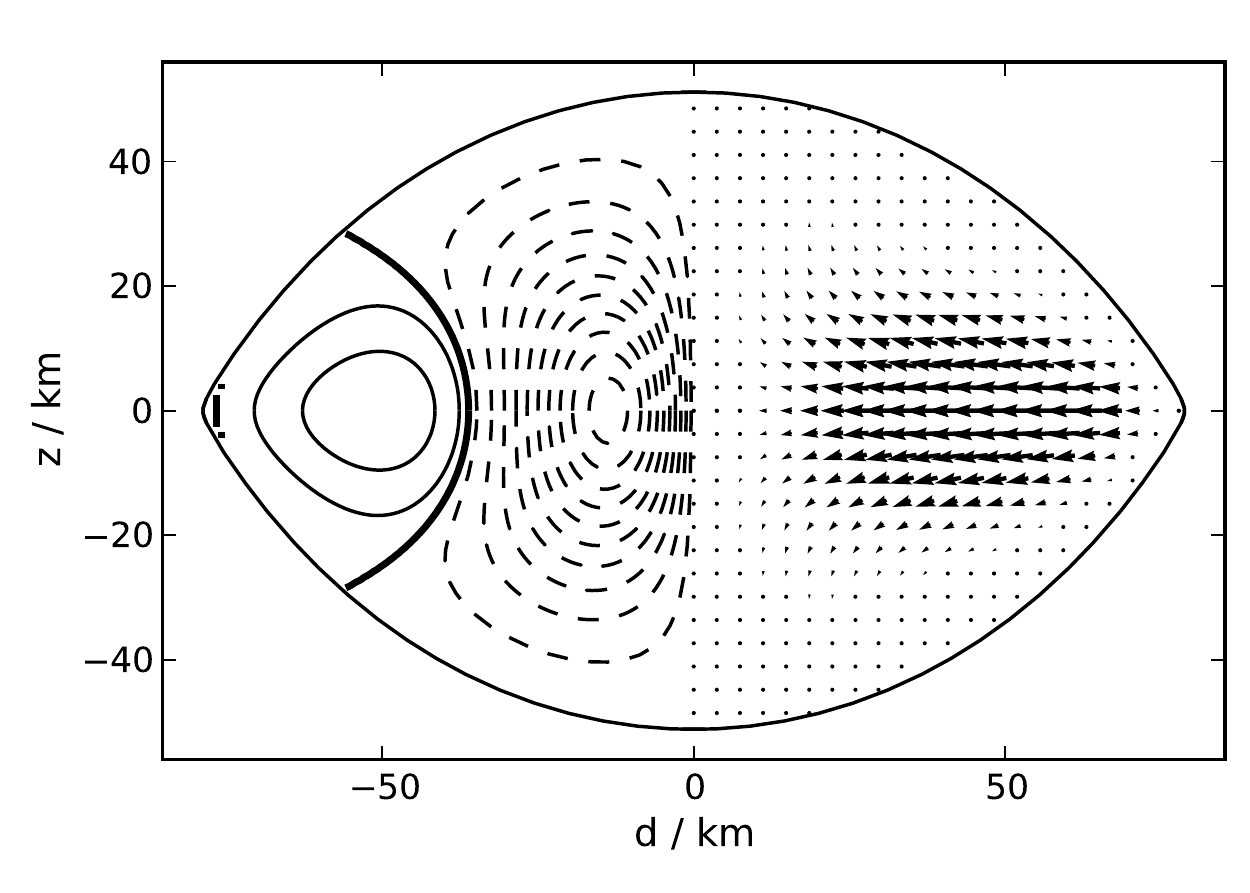}
  \caption{\label{fig_ef_jsft7_fl2m0}
Eigenfunction of the axisymmetric \Paxi~mode of model MC65.
TOP: Eigenfunction $\delta\hat{\Sed}(d,z)$ as gray-scale plot.
The thick solid line marks the node position.
The arrows show the velocity eigenfunction.
BOTTOM: Same eigenfunctions scaled by $\rho d$, to visualize the bulk properties of the mode.
The node near the equator is very likely a numerical artifact.
  }
\end{figure}

Another effect becomes important very close to the Kepler limit.
As shown in \Figure{fig_ef_jsft7_fl2m0},
the \Paxi-mode eigenfunction of model MC65 is not only strongly deformed
in comparison to the slow-rotation limit,
it also develops very pronounced peaks near the equator.
Looking at the bulk properties of mode,
we find that the dynamics of the oscillation is still determined by
the inner regions of the star,
but a given amplitude in the interior induces much higher amplitudes
near the equator.
This is not surprising since the material there is only marginally bound
for this model.
As a consequence, oscillation modes of MC65 can store only little energy before
nonlinear effects set in, as will be shown in \Sec{sec_damp_axi}.
In fact, we did not even succeed to obtain a clean eigenfunction of the \Prad~mode
of this model, due to strong nonlinear couplings.

Since the GW luminosity depends strongly on the frequency,
we have to estimate the influence of the Cowling approximation.
Frequencies in Cowling approximation and full GR,
computed by means of fully relativistic simulations,
are given in \cite{2010PhRvD..81h4055Z},
for model MB100 and models similar to MA100 and MA65, amongst others.
After adding an additional safety margin to account for the different models,
we assume that we over-estimate the frequencies of the \Prad~modes
by a factor of less than 2.5 and the \Paxi~modes by a factor less than 1.5.
Further, we assume that the frequency of the \Pctr~mode of 
in the \textit{corotating} frame is over-estimated by a factor $< 1.3$.
For model MA65, this implies that the frequency in the inertial frame
is \textit{under-estimated} by a factor up to 2.3.
For model MB70, we find that the \Pctr~mode could become CFS-unstable
in full GR, with an inertial frame frequency in the range 0--90\usk\hertz.

\begin{table}
\caption{\label{tab_freq_modes}
Frequencies $f_i$ of the different oscillation modes
as observed from infinity in the inertial frame, and corresponding frequencies
$f_c$ in the corotating frame.
$q$ is the quality factor of the numerical eigenfunction
for $\delta \hat{\epsilon}$ defined in \Eref{eq_ef_qual}.
}
\begin{ruledtabular}
\begin{tabular}{lllll}
Model   & Mode   &$f_c / \hertz$ & $f_i / \hertz$ & q  \\\hline
MC100   & \Prad  &472            &472             & 2\e{-4}     \\
MC100   & \Paxi  &418            &418             & 5\e{-4}    \\
MC100   & $\Pctr \equiv \Pcor$ &418            &418     & 4\e{-4} \\
MC95    & \Prad  &437            &437             & 8\e{-5}       \\
MC95    & \Paxi  &392            &392             & 2\e{-3}      \\
MC85    & \Prad  &383            &383             & 5\e{-5}      \\
MC85    & \Paxi  &337            &337             & 2\e{-4}      \\
MC85    & \Pctr  &371            &205             & 3\e{-3}      \\
MC65    & \Paxi  &239            &239             & 2\e{-4}      \\
MB100   & \Prad  &2686           &2686            & 3\e{-5}       \\
MB100   & \Paxi  &1883           &1883            & 5\e{-4}       \\
MB85    & \Paxi  &1893           &1893            & 4\e{-4}      \\
MB70    & \Paxi  &1785           &1785            & 3\e{-4}      \\
MB70    & \Pctr  &1948           &364             & 4\e{-3}      \\
MA100   & \Prad  &4606           &4606            & 4\e{-5}      \\
MA100   & \Paxi  &3038           &3038            & 5\e{-4}      \\
MA100   & $\Pctr \equiv \Pcor$  &3037         &3037  & 2\e{-3}  \\
MA65    & \Prad  &3997           &3997            & 3\e{-4}      \\
MA65    & \Paxi  &2662           &2662            & 2\e{-3}      \\
MA65    & \Pctr  &2856           &518             & 6\e{-3}     \\
MA65    & \Pcor  &1179           &4553            & 2\e{-2}     \\
\end{tabular}
\end{ruledtabular}
\end{table}

\subsection{Damping of axisymmetric modes}
\label{sec_damp_axi}
For all axisymmetric oscillations of all models, we find a common behavior
at high amplitudes:
if the initial amplitude exceeds a certain threshold,
shock waves form in the outer layers of the star, which dissipate energy
during a few oscillation cycles until the oscillation amplitude falls below
the threshold again.
After this phase, we usually observe only the numerical damping.
In some cases, we also see mode coupling effects,
which are however small compared to the strongest damping due to shocks.

\begin{figure}
  \includegraphics[width=\columnwidth]{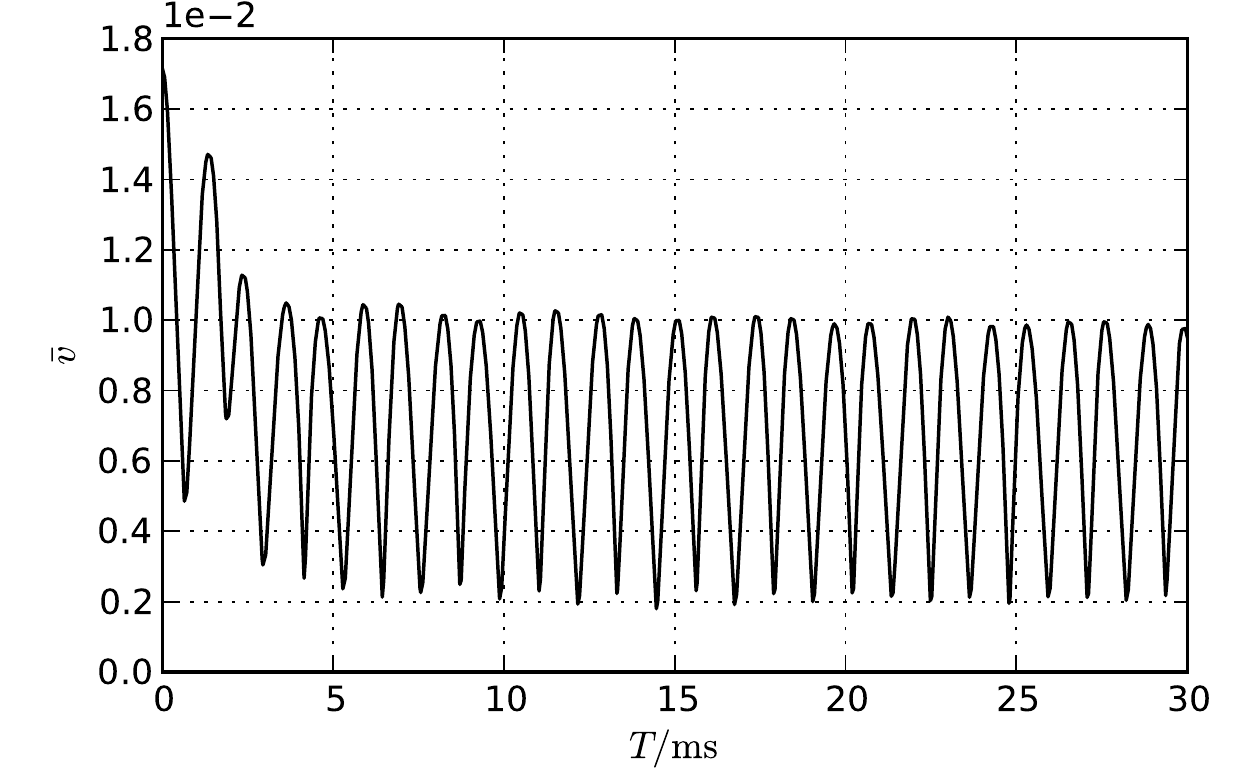}
  \caption{\label{fig_shock_vm2}
Evolution of mean velocity $\bar{v}$ for an \Prad-mode perturbation
of model MC95 at high amplitudes.
  }
\end{figure}

The final amplitude can be smaller than the threshold,
in particular for strong initial excitation.
The final amplitude as a function of the initial one thus has a maximum for some modes.
\Figure{fig_shock_vm2} shows a typical evolution of the mean velocity in presence of shocks.

\begin{figure}
  \includegraphics[width=\columnwidth]{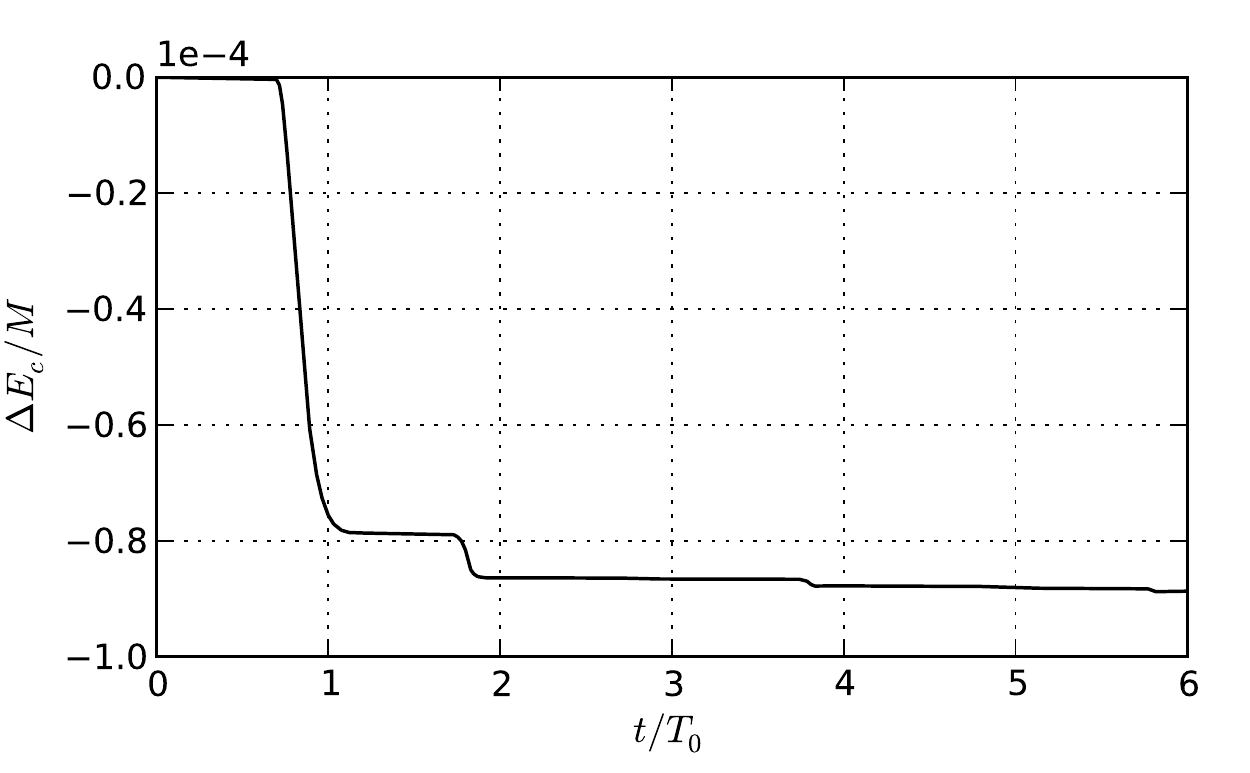}
  \caption{\label{fig_shock_econs}
Violation of energy conservation due to shocks,
for an \Prad-mode perturbation
of model MC95.
Plotted is the loss of conserved energy $E_c$ in units of the total stellar mass versus the time in units
of the oscillation period $T_0$.
Only the phase of strong decay is shown.
  }
\end{figure}

To detect shocks, we monitor the conserved energy $E_c$ defined by \Eref{eq_def_econs}.
\Figure{fig_shock_econs} shows an example of energy dissipation in the presence of shocks.
To investigate the details of shock formation,
we produced movies from several of our simulations,
showing the evolution of the density and velocity in the meridional plane.
The shock formation was clearly visible for simulations with significant energy loss.
For one-dimensional simulations, we also verified that shock formation and energy
dissipation occur simultaneously.
The shocks we observed formed in the outer layers of the star,
but not directly at the surface.

\begin{figure}
  \includegraphics[width=\columnwidth]{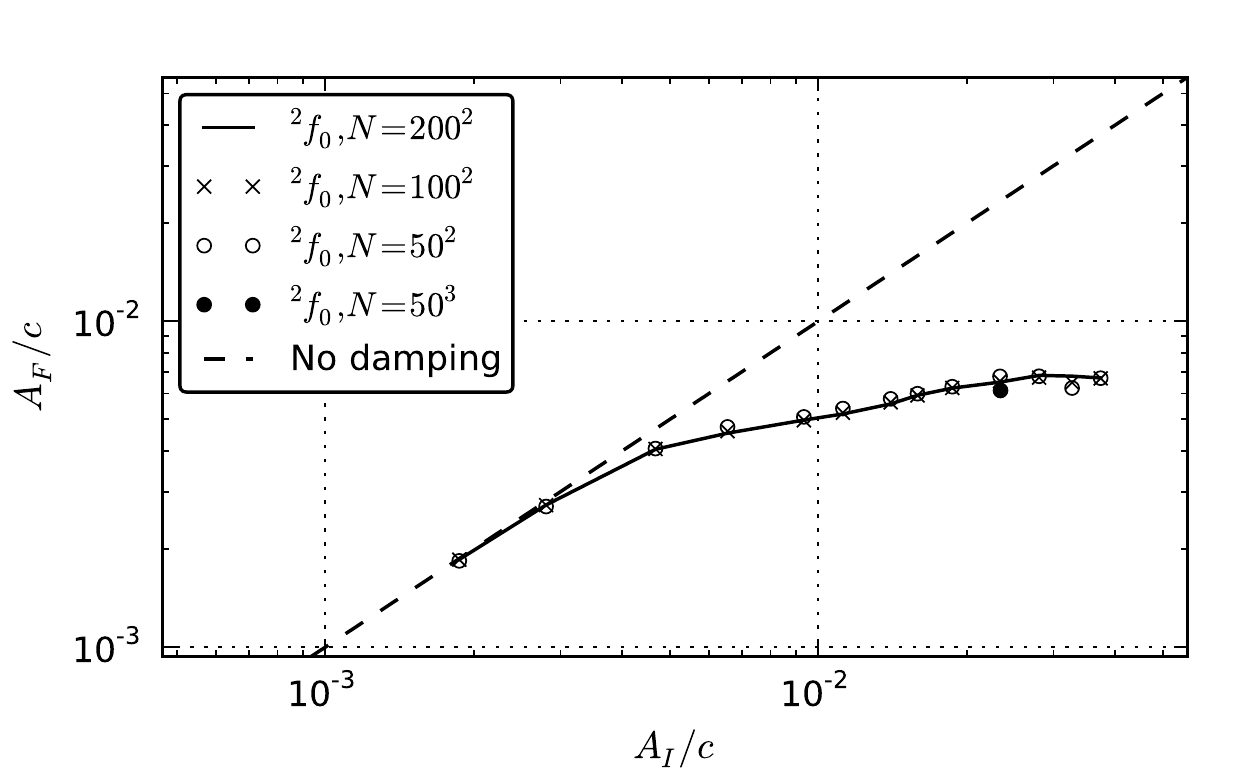}
  \caption{\label{fig_vm_jsft3_fl2m0}
Nonlinear damping of the \Paxi~mode of model MC85.
Plotted is the final amplitude $A_F$ after an evolution time of $20\usk\milli\second$
over the initial amplitude $A_I$.
We show results of axisymmetric simulations with resolutions
of $N=50$, 100, and 200 points per stellar radius, as well as a three-dimensional simulation
with resolution $N=50$.
  }
\end{figure}

To get an overview on the magnitude of nonlinear damping effects,
we plot the final amplitude $A_F$ defined in \Eref{eq_def_afinal}
versus the initial amplitude $A_I$.
As an example, \Figure{fig_vm_jsft3_fl2m0} shows the results for the
\Paxi~mode of model MC85.

In general, we choose different evolution times for different sequences,
such that the strong damping phase at high amplitudes
(compare \Figure{fig_shock_vm2}) is included, but not much longer.
We choose short evolution times because we are interested mainly
in strong damping effects.
In any case,
we can only accurately measure damping effects acting on timescales
significantly shorter than the timescale of numerical damping.

For each sequence,
we extract the amplitude $A_D$ at which nonlinear damping effects
become stronger than the numerical damping, given in \Tab{tab_onset_nl}.
This amplitude is independent
from
the exact value of the evolution time.
As discussed in \Sec{sec_polyshock}, the onset of shock formation is
predicted accurately despite the use of a cold (one-parametric) EOS.
Therefore, the onset of nonlinear damping is not affected by this
approximation neither.

The values we found for the \Prad and \Paxi~modes of the different models
cover a wide range $A_D=0.003\dots 0.13 \usk c$.
We will discuss the dependence on the model parameters in sections~\ref{sec_damp_eos}
and \ref{sec_damp_rot}.
Naturally, the values of $A_D$ provide upper limits for the applicability of linear
perturbation theory.

\begin{figure}
  \includegraphics[width=\columnwidth]{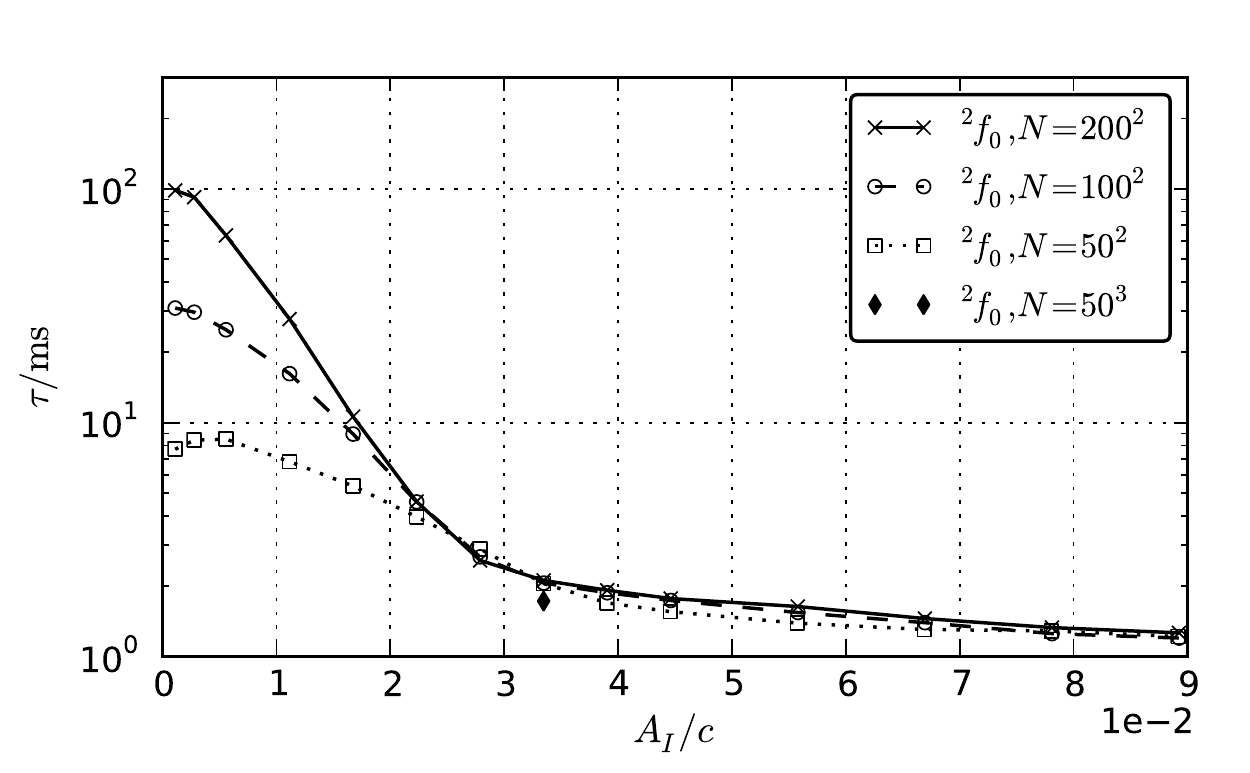}
  \caption{\label{fig_tau_jega65_fl2m0}
Timescale of initial decay $\tau$ due to shock damping versus the
initial amplitude $A_I$, for an \Paxi-mode perturbation
of model MA65.
Shown are results from axisymmetric simulations with three different resolutions
as well as one 3D-simulation of the same problem.
The resolution $N$ is given in points per stellar equatorial radius.
  }
\end{figure}

The final amplitude contains only information on how much the system is damped,
but not how fast.
To quantify the damping speed,
we use the timescale $\tau$ of the initial decay defined in \Eref{eq_def_tau}.
\Figure{fig_tau_jega65_fl2m0} shows $\tau$ versus the initial amplitude $A_I$
for the case of the axisymmetric \Paxi~mode of model MA65.
We stress that the values of $\tau$ \textit{are} affected by the use of a cold EOS,
as discussed in \Sec{sec_polyshock}, and should be regarded as an educated guess.

Not surprising, the damping becomes faster with increasing amplitude.
By definition, $\tau$ is a measure for time averaged decay.
More detailed investigation reveals that for the highest amplitudes,
the larger part of the energy is dissipated on timescales smaller
than the oscillation period during the formation of a strong,
but short-lived  shock.

The decay timescale can be used to estimate
the required growth timescale of a hypothetical instability
saturating at a given amplitude.
We are not aware of any axisymmetric instability acting on timescales
shorter than the numerical damping.
For any process with longer growth times,
the value $A_D$ provides an upper limit for the saturation amplitudes.

To estimate the numerical errors,
we computed several sequences using 3 different resolutions.
Note it is important to choose the sign of the perturbation consistently,
since we observed differences in the nonlinear regime comparable to
the numerical error at low resolution.

By far the greatest errors are found for the stiff EOS A.
\Figure{fig_tau_jega65_fl2m0} shows the convergence of the
nonlinear damping timescale
of the \Paxi~mode of model MA65 as a representative example.
To explain the behavior at low amplitudes,
we note that when approaching the linear regime,
the timescale of physical damping goes to infinity.
In that case, we only see the numerical damping,
which becomes weaker with increasing resolution.
The stronger the damping, the more accurate is the value we find for $\tau$.

For the soft EOS C, the numerical damping is extremely low.
For a resolution of 100, we find timescales $\tau_N>800\usk\milli\second$
in the linear regime.
Also the nonlinear results match extremely well,
as shown in \Figure{fig_vm_jsft3_fl2m0}.
The accuracy of models with EOS B is in between the other two cases.
Since the numerical damping is mainly given by surface effects,
as shown in \cite{Kastaun06},
and the steepness of the density gradient at the surface increases with stiffness,
this behavior is not surprising.

To get a rough error estimate for simulations which we only ran at one resolution,
we assume that the numerical damping observed in the linear regime
for single oscillations is also operational at high amplitudes,
acting on the same timescale $\tau_D$.
We then scale our time series by $\exp(t/\tau_D)$ to obtain corrected
values for final amplitude $A_F$ and damping time $\tau$.

To estimate the accuracy of three-dimensional simulations
we also performed 3D simulation of axisymmetric modes at high amplitudes.
We generally found a good agreement  with the axisymmetric results for the same setup;
see \Figure{fig_vm_jsft3_fl2m0} and \Figure{fig_tau_jega65_fl2m0}.

\begin{table}
\caption{\label{tab_onset_nl}
Onset of nonlinear damping.
$A_D$ is the initial amplitude $A_I$ at which nonlinear damping
is strong enough to be distinguishable from numerical damping,
read off from plots of initial versus final amplitude.
}
\begin{ruledtabular}
\begin{tabular}{lll}
Model   &Mode     &$A_D / (10^{-3}\usk c)$ \\\hline
MC100    & \Prad   &$11 \pm 1 $      \\
        & \Paxi   &$4 \pm 1  $      \\
MC95   & \Prad   &$9  \pm 2 $      \\
        & \Paxi   &$5 \pm 2  $      \\
MC85   & \Prad   &$7  \pm 2 $      \\
        & \Paxi   &$4 \pm 1  $      \\
MC65   & \Paxi   &$<0.1  $         \\
MB100     & \Prad   &$53 \pm 7 $      \\
        & \Paxi   &$20 \pm 5$       \\
MA100    & \Prad   &$100 \pm 25$      \\
        & \Paxi   &$45 \pm 15$      \\
MA65  & \Prad   &$40 \pm 15$      \\
        & \Paxi   &$8  \pm 3 $
\end{tabular}
\end{ruledtabular}
\end{table}

\subsection{Influence of the EOS on damping}
\label{sec_damp_eos}
To study the influence of the EOS, we use the nonrotating models MC100, MB100, MA100,
which have similar masses, but EOSs of different stiffness.

\begin{figure}
  \includegraphics[width=\columnwidth]{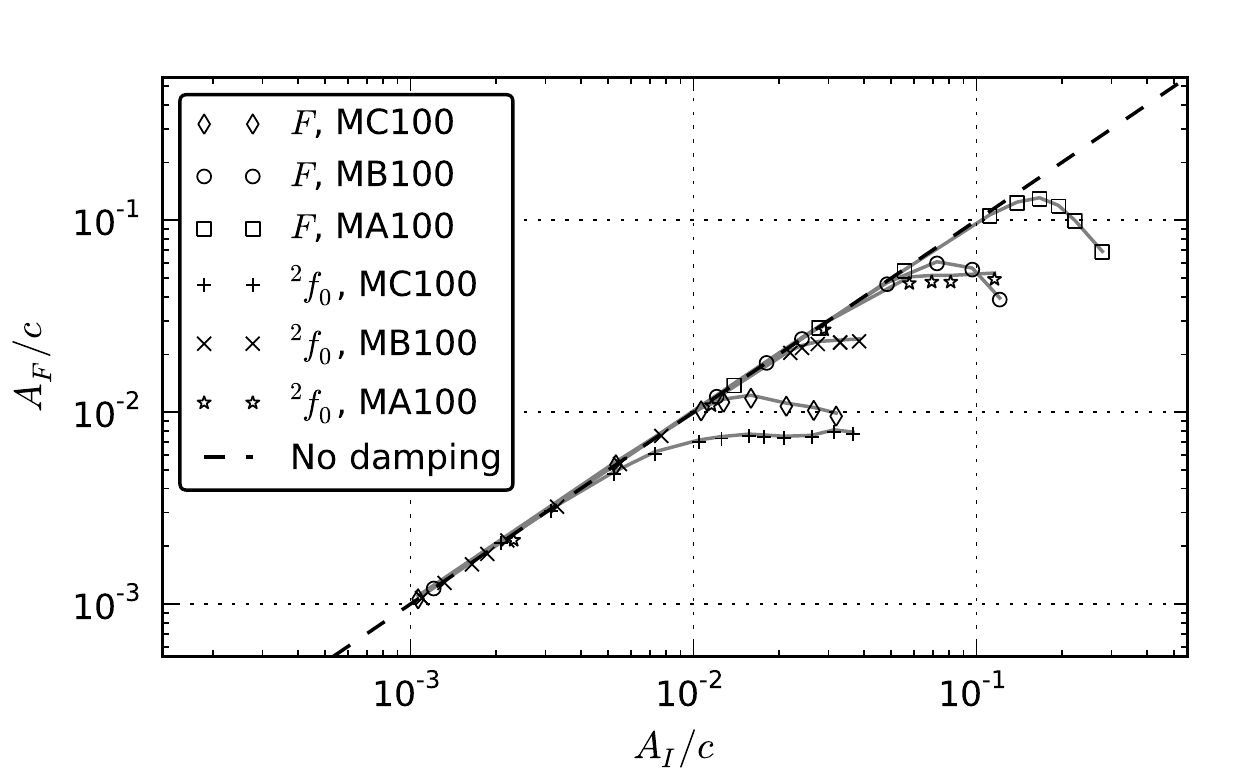}
  \caption{\label{fig_vm_eos}
Onset of nonlinearity for the \Prad and \Paxi~modes of the nonrotating
models MB100, MC100, and MA100.
Plotted is the final amplitude $A_F$ versus the initial one $A_I$.
The solid gray lines are values corrected for the numerical damping.
The \Prad~mode results are extracted from one-dimensional simulations with a
resolution of 200 points per stellar radius, the \Paxi~modes are computed
with two-dimensional simulations at resolution of $100^2$.
  }
\end{figure}

The amount of nonlinear damping for the different models is shown in
\Figure{fig_vm_eos}.
The amplitudes $A_D$ at which nonlinear damping effects
become visible is given in \Tab{tab_onset_nl}.

As one can see, the nonlinearity sets in one order of magnitude earlier for the
EOS C, which is the softest one, than for EOS A, the stiffest one.
Also the maximum final amplitude is the largest for the stiff EOS.
Note that MA100 is 13\usk\% heavier than MC100,
which might account for part of the differences.
Nevertheless, the masses of models MC100 and MB100 are exactly the same,
and still the damping differs strongly.
We conclude that with increasing stiffness of the EOS a neutron star (of a given mass)
can store more energy in axisymmetric oscillations before strong damping sets in.

We believe that the onset of shock formation is determined mostly by the stiffness
of the EOS in the density range of the outer layers of the star,
since this is the region where the shocks form.
However, this requires further investigation.

\begin{figure}
  \includegraphics[width=\columnwidth]{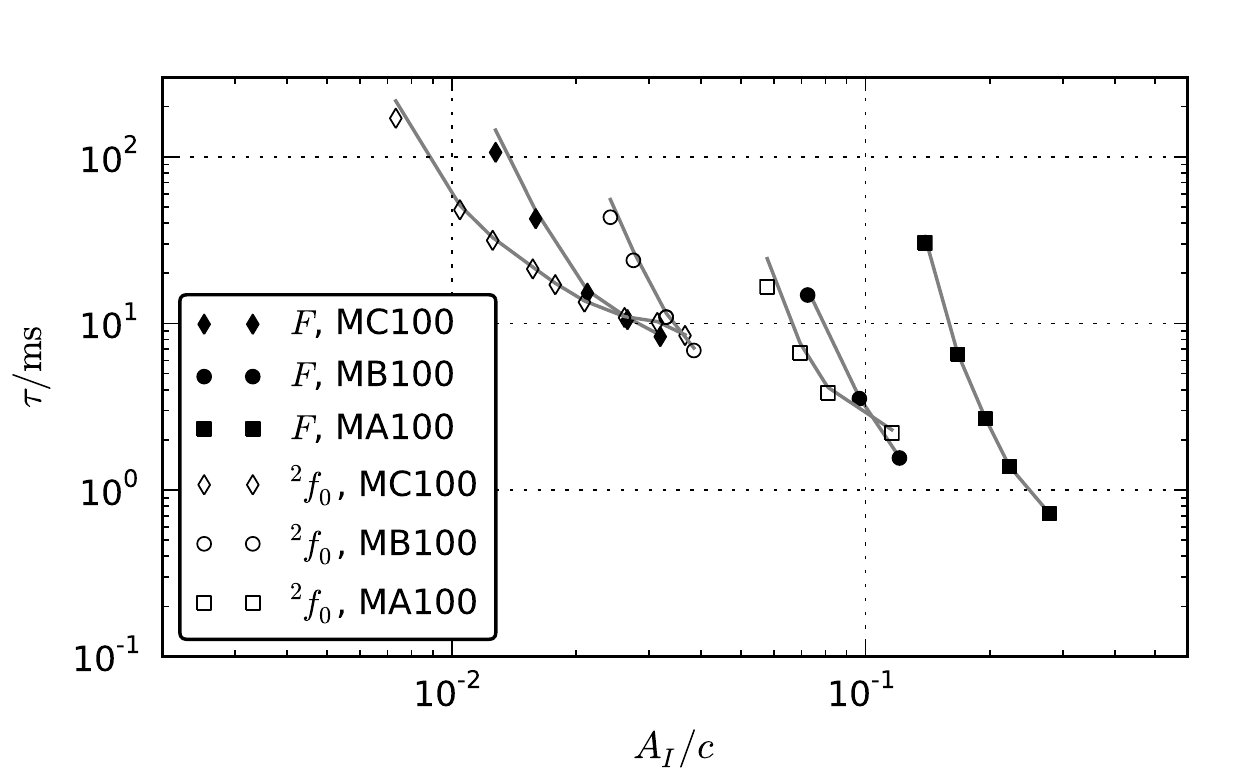}
  \caption{\label{fig_tau_eos}
Initial damping timescale $\tau$ versus initial amplitude $A_I$,
for the same simulations shown in \Figure{fig_vm_eos}.
The solid gray lines are values corrected for the numerical damping.
We only show amplitudes for which the difference is
less than 50\usk\%.
For slower damping, the error of $\tau$ blows up quickly.
  }
\end{figure}

In \Figure{fig_tau_eos}, we plot the damping timescale for the different
nonrotating models.
At a given damping timescale,
the possible oscillation amplitudes differ by roughly one order of magnitude.
Further, the damping timescale at the highest possible amplitudes
(before material starts leaving the computational domain)
decreases with increasing stiffness,
i.e. the damping is faster.

\subsection{Influence of rotation on damping}
\label{sec_damp_rot}
\begin{figure}
  \includegraphics[width=\columnwidth]{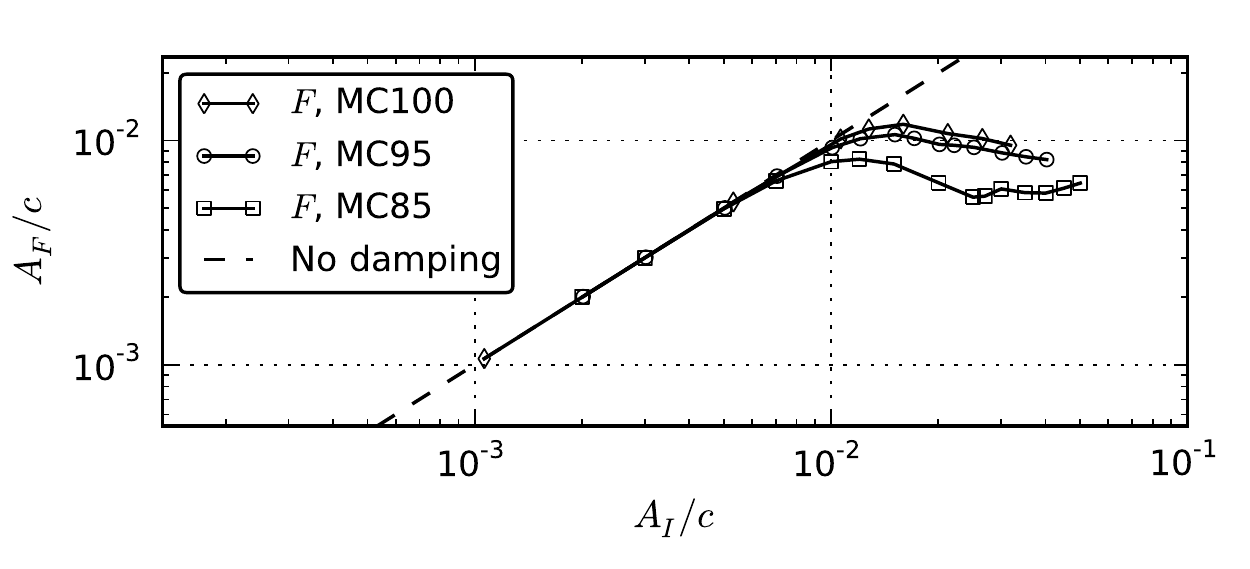}
  \includegraphics[width=\columnwidth]{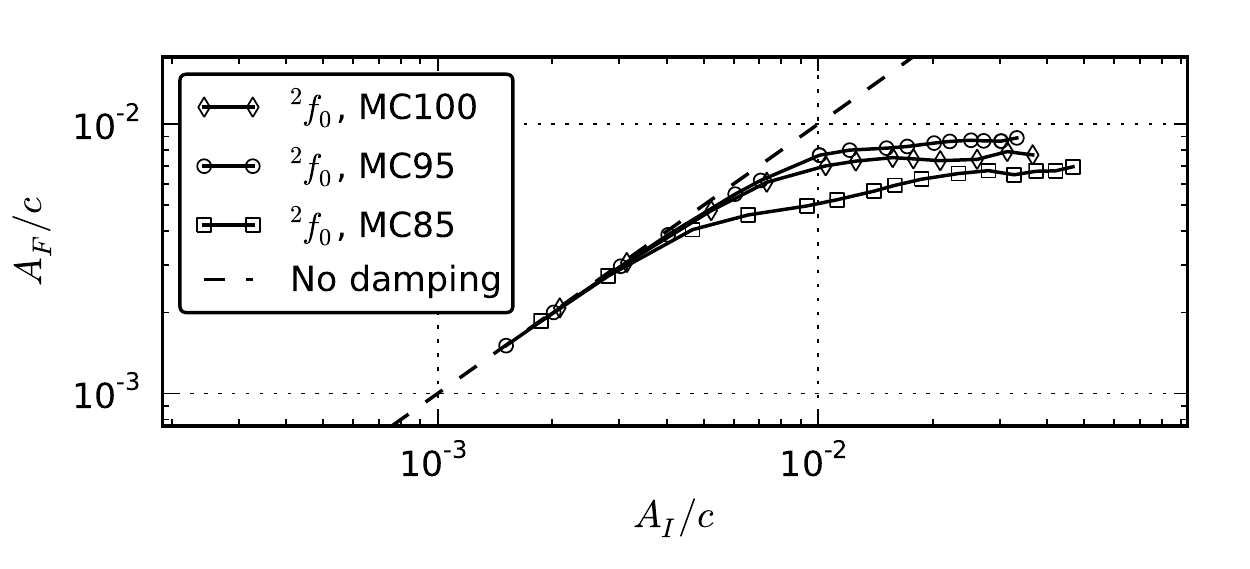}
  \caption{\label{fig_vm_rot}
Influence of rotation on nonlinear damping of axisymmetric modes.
Shown are final versus initial amplitude for a perturbation with the \Prad~mode (TOP)
and the \Paxi~mode (BOTTOM) of models MC100, MC95, MC85.
The evolution time was $20\usk\milli\second$ for the \Paxi~mode and
$15\usk\milli\second$ for the \Prad~mode.
  }
\end{figure}

\begin{figure}
  \includegraphics[width=\columnwidth]{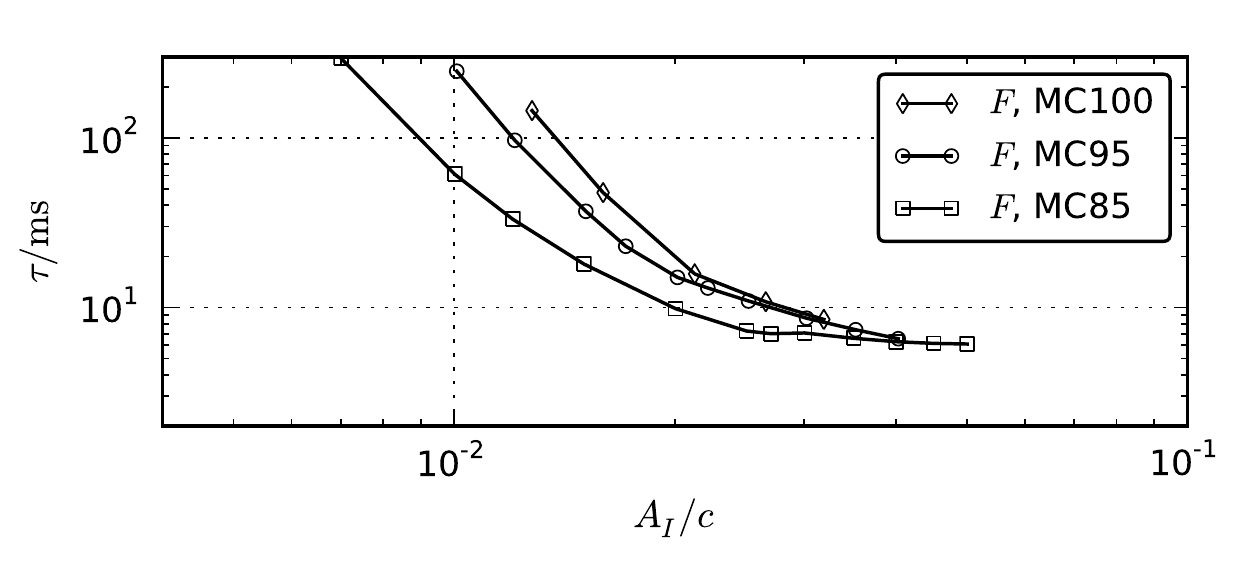}
  \includegraphics[width=\columnwidth]{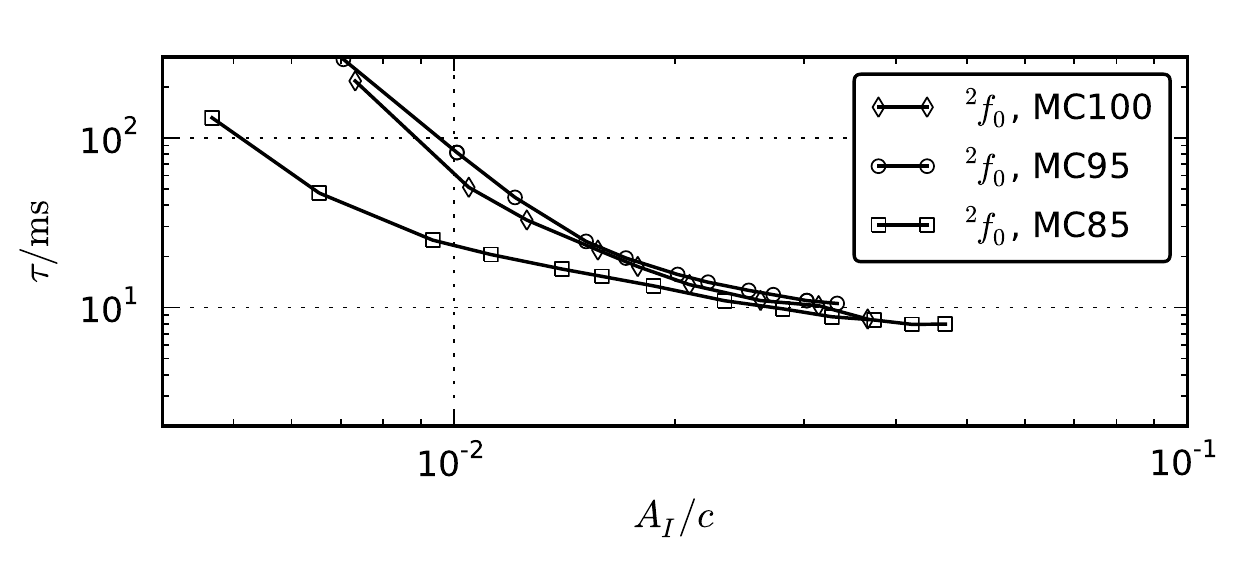}
  \caption{\label{fig_tau_rot}
Influence of rotation on damping timescale $\tau$ of
\Prad~modes (TOP) and \Paxi~modes (BOTTOM).
The values are already corrected for the numerical damping.
Only points are shown for which the correction is less than 50\usk\%.
  }
\end{figure}

\begin{figure}
  \includegraphics[width=\columnwidth]{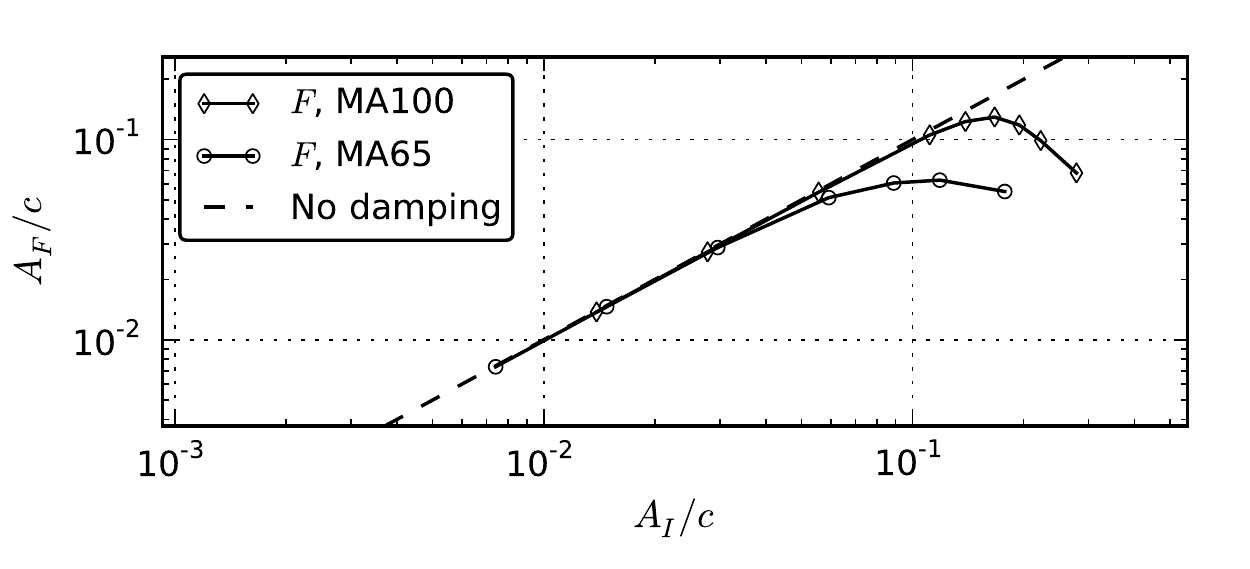}
  \includegraphics[width=\columnwidth]{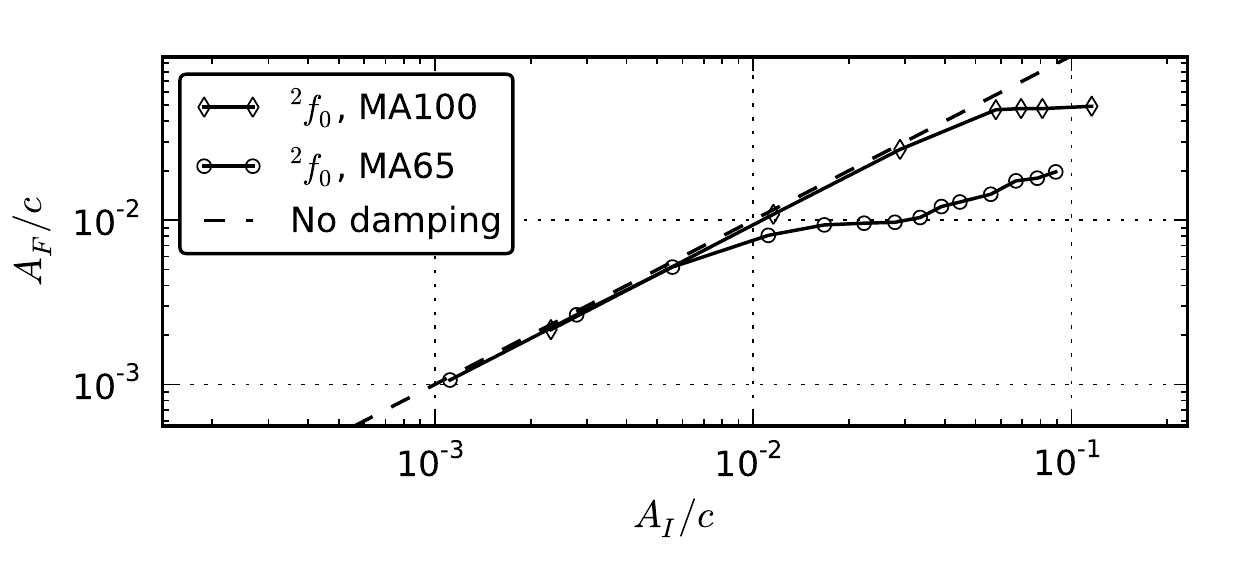}
  \caption{\label{fig_vm_rot_eosa}
Nonlinear damping of \Prad~mode (TOP) and \Paxi~mode (BOTTOM)
for nonrotating model MA100 and rapidly rotating model MA65.
Evolution time was $4\usk\milli\second$.
  }
\end{figure}

To investigate the influence of rotation on axisymmetric modes,
we first compare the models MC100, MC95, and MC85, which have same mass and EOS,
but different rotation rates up to 90\usk\% of the Kepler limit.
The amount of damping for the \Prad and \Paxi~modes
is shown in \Figure{fig_vm_rot}, and the damping timescales in \Figure{fig_tau_rot}.
For the stiff EOS A, we compared nonrotating model MA100 with the rapidly rotating model MA65.
Note however that MA65 is also 18\usk\% heavier.
The results are shown in \Figure{fig_vm_rot_eosa}.

The differences between models MC100 and MC95 are very small.
For MC85 on the other hand,
the nonlinear damping is significantly faster and stronger in comparison.
The onset of nonlinearity occurs at slightly lower amplitudes for the \Prad~mode,
while for the \Paxi~modes we found no significant difference.
Since the fluid is bound less strongly with increasing centrifugal force,
an increase of nonlinearity at given amplitude is to be expected.

An extreme case is given by model MC65, which is rotating almost at breakup velocity.
For this case nonlinear effects start much earlier than for MC85.
For the \Paxi~mode, nonlinear effects are definitely present for $A_I > 10^{-4}$.
At an amplitude as low as $A_I=1.5\e{-3}$ a fraction of $2\e{-5}$ of the total mass
leaves the computational domain.
We obviously observe what is called mass shedding induced damping,
i.e. even small oscillations liberate mass near the equator.
The same mechanism has already been observed in \cite{SAF2004,DSF2006}
for models with EOS B.
Since stars rotating so close to the Kepler limit are unlikely to occur in nature,
we did not investigate this case further.

\subsection{Damping of nonaxisymmetric modes}
\label{sec_damp_m2}
To study the evolution of nonaxisymmetric modes,
we performed three-dimensional simulations of the counter-rotating \Pctr~mode of the rotating models
MA65, MA100, MB70, and MC85.
For model MA65, we also studied the corotating \Pcor~mode.

In contrast to the axisymmetric case,
we did not observe any steep decrease of the conserved energy $E_c$
that would point to formation of strong shocks.
We did however observe a continuous decrease of the mean velocity and
the $l=m=2$ multipole moment.
The dissipation of energy was considerably slower than
the one observed in case of shock formation in axisymmetric oscillations,
even for high perturbation amplitudes.

\begin{figure}
  \includegraphics[width=\columnwidth]{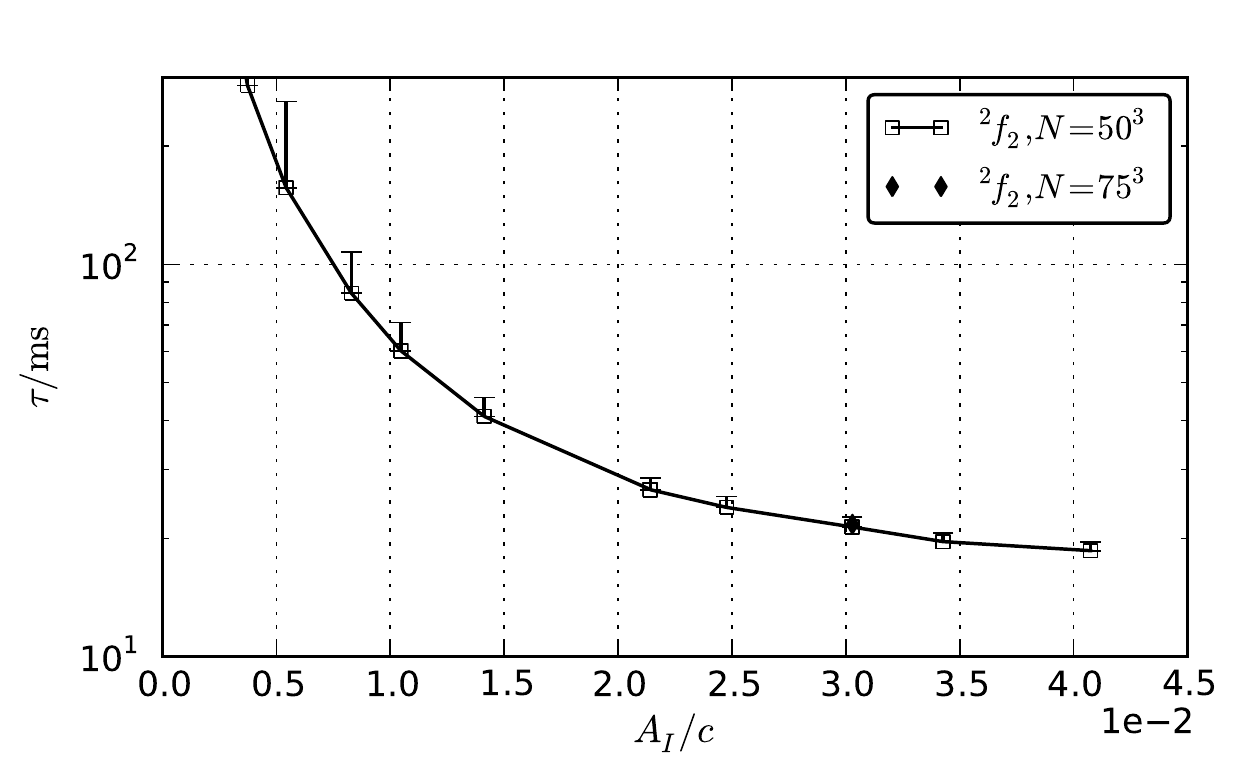}
  \caption{\label{fig_tau_jsft3_fl2m2}
Timescale $\tau$ of initial damping for the \Pctr~mode of model MC85.
Shown are numerical results obtained at resolutions 50 and 75
points per stellar equatorial radius.
The error bars are our estimate for the influence of numerical damping.
  }
\end{figure}

\begin{figure}
  \includegraphics[width=\columnwidth]{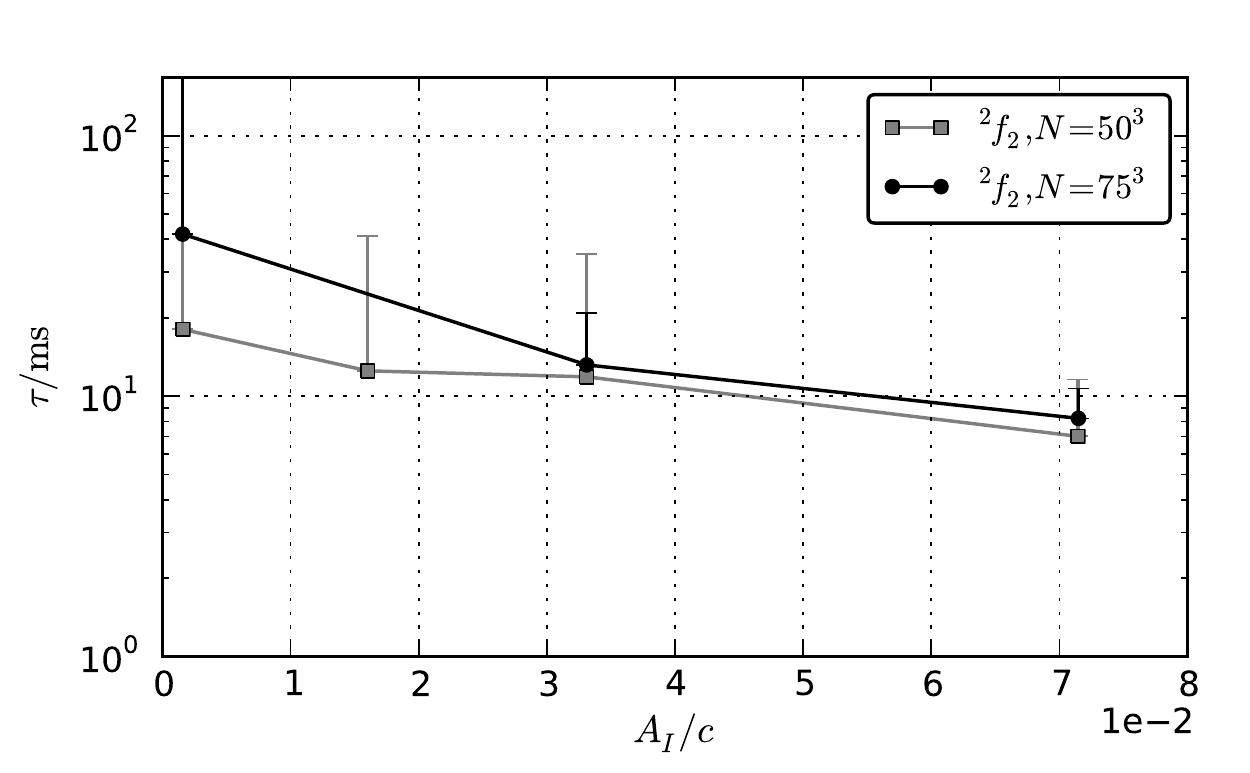}
  \caption{\label{fig_tau_jbu6_fl2m2}
Like \Figure{fig_tau_jsft3_fl2m2}, but for the \Pctr~mode of model MB70.
Error bars leaving the plot are of infinite size,
i.e. the numerical results are compatible with no physical damping.
  }
\end{figure}

\begin{figure}
  \includegraphics[width=\columnwidth]{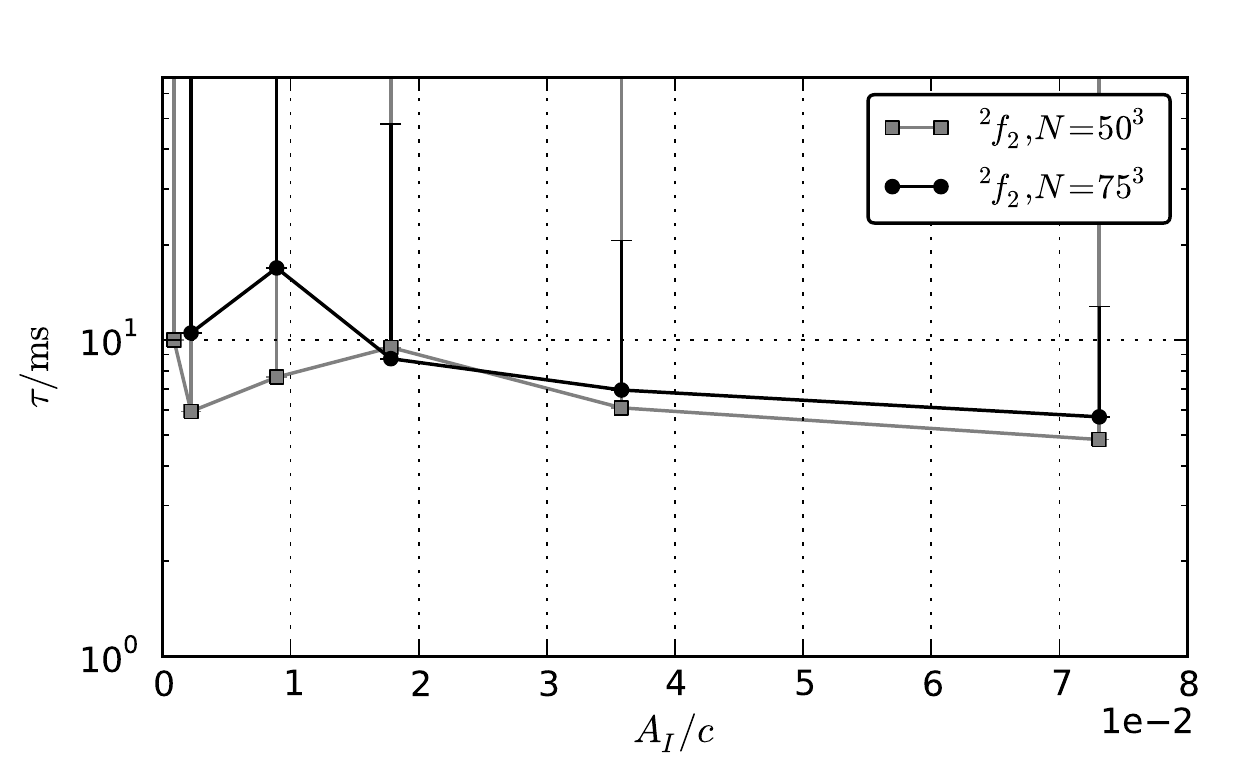}
  \caption{\label{fig_tau_jega65_fl2m2}
Like \Figure{fig_tau_jsft3_fl2m2}, but for the \Pctr~mode of model MA65.
  }
\end{figure}

\begin{figure}
  \includegraphics[width=\columnwidth]{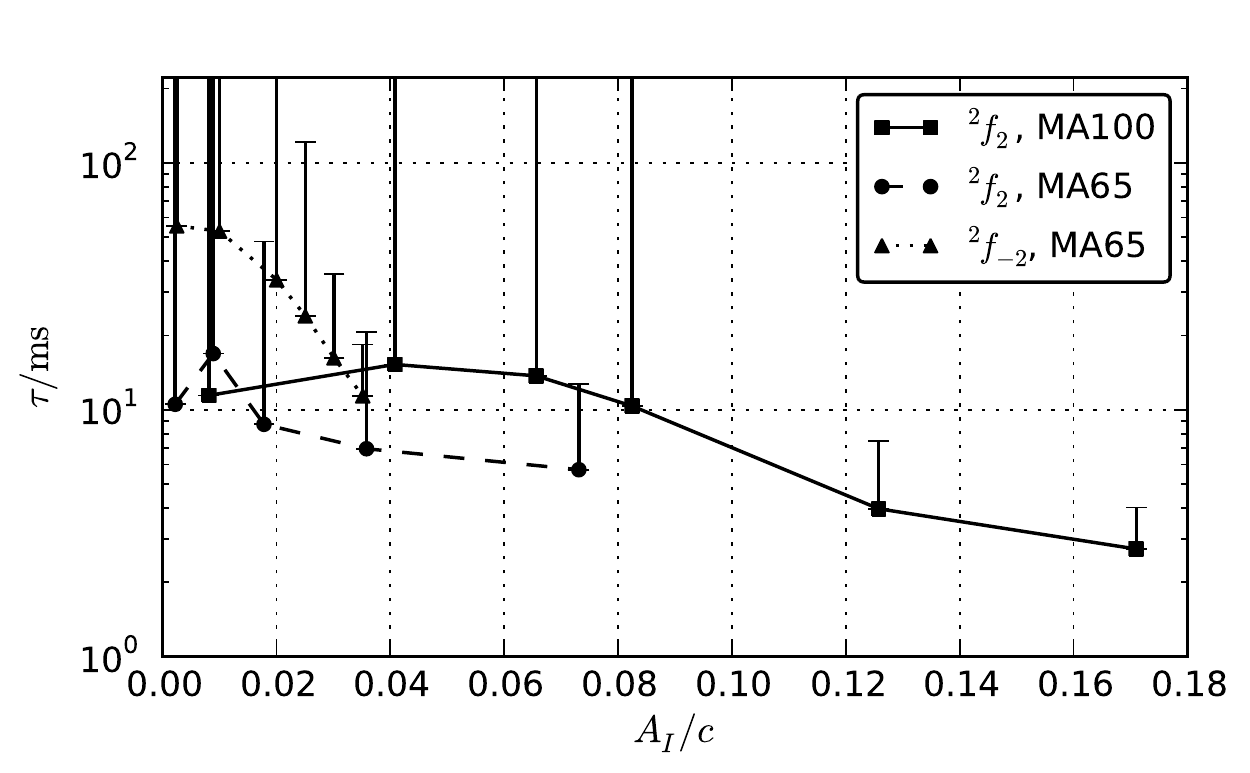}
  \caption{\label{fig_sat_rot_m2}
Nonlinear damping of nonaxisymmetric modes for nonrotating model MA100 and rapidly rotating model MA65.
  }
\end{figure}

Figures \ref{fig_tau_jsft3_fl2m2}--\ref{fig_sat_rot_m2} show the damping timescale.
For models MA65 and MB70,
it proved difficult to disentangle physical and numerical damping,
which are of comparable strength at the resolutions we could afford.

To estimate the error of the damping,
we assume that the timescale of the numerical damping in the nonlinear regime
is equal to the one observed in the linear regime.
For the axisymmetric case described in \Sec{sec_damp_axi},
this estimate agreed well with the convergence test results.
For axisymmetric setups, we also proved that the code is as accurate in 3D as in 2D.
As a further check, we compared resolutions of 50 and 75 points per stellar radius.
The results are shown in \Figure{fig_tau_jsft3_fl2m2} to \ref{fig_tau_jega65_fl2m2}.

\begin{figure}
  \includegraphics[width=\columnwidth]{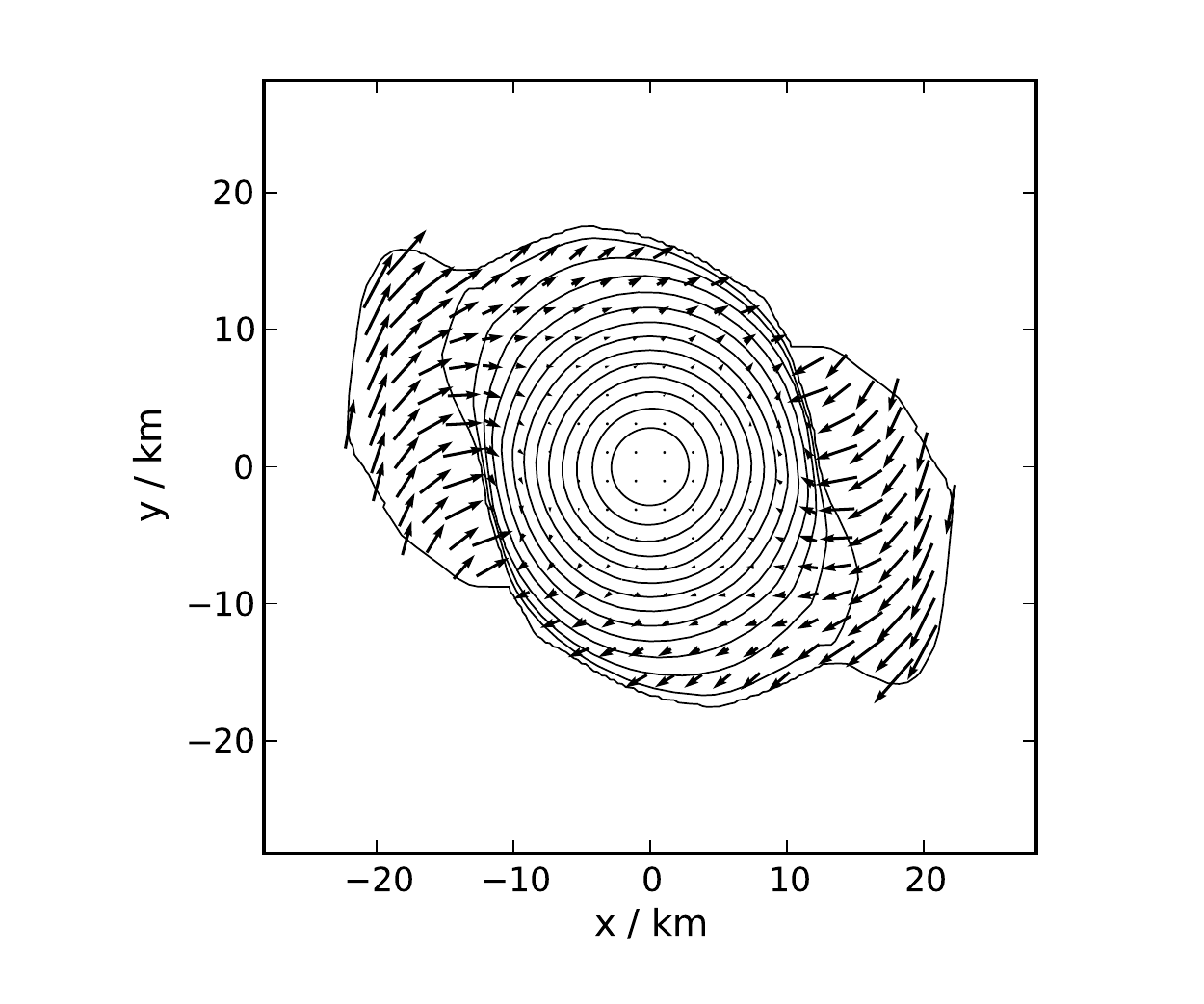}
  \caption{\label{fig_wave_jbu6}
Snapshot of the evolution of density and velocity in the equatorial plane,
for an \Pctr-mode perturbation of model MB70 with amplitude
$A_I=0.04$.
Shown are the lines of constant density $\Rmd$,
spaced at regular intervals of $\sqrt{\Rmd}$.
The outermost line corresponds to a density of $10^{-4}$ times the central density.
The arrows correspond to the velocity in the corotating frame.
The wave-patterns are rotating clockwise, leaving behind low density
regions which fall back. Note those regions are not well resolved numerically.
  }
\end{figure}

\begin{figure}
  \includegraphics[width=\columnwidth]{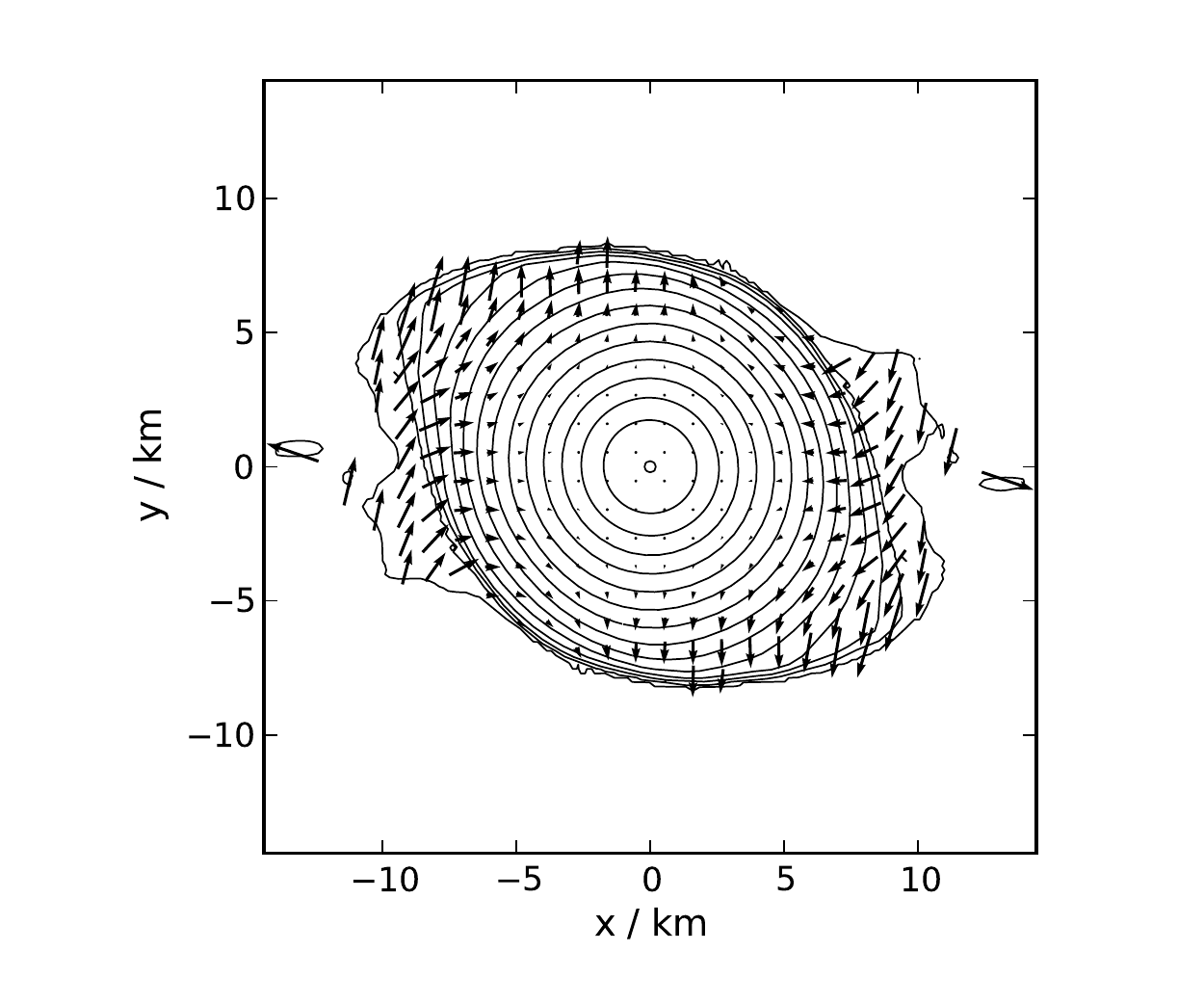}
  \includegraphics[width=\columnwidth]{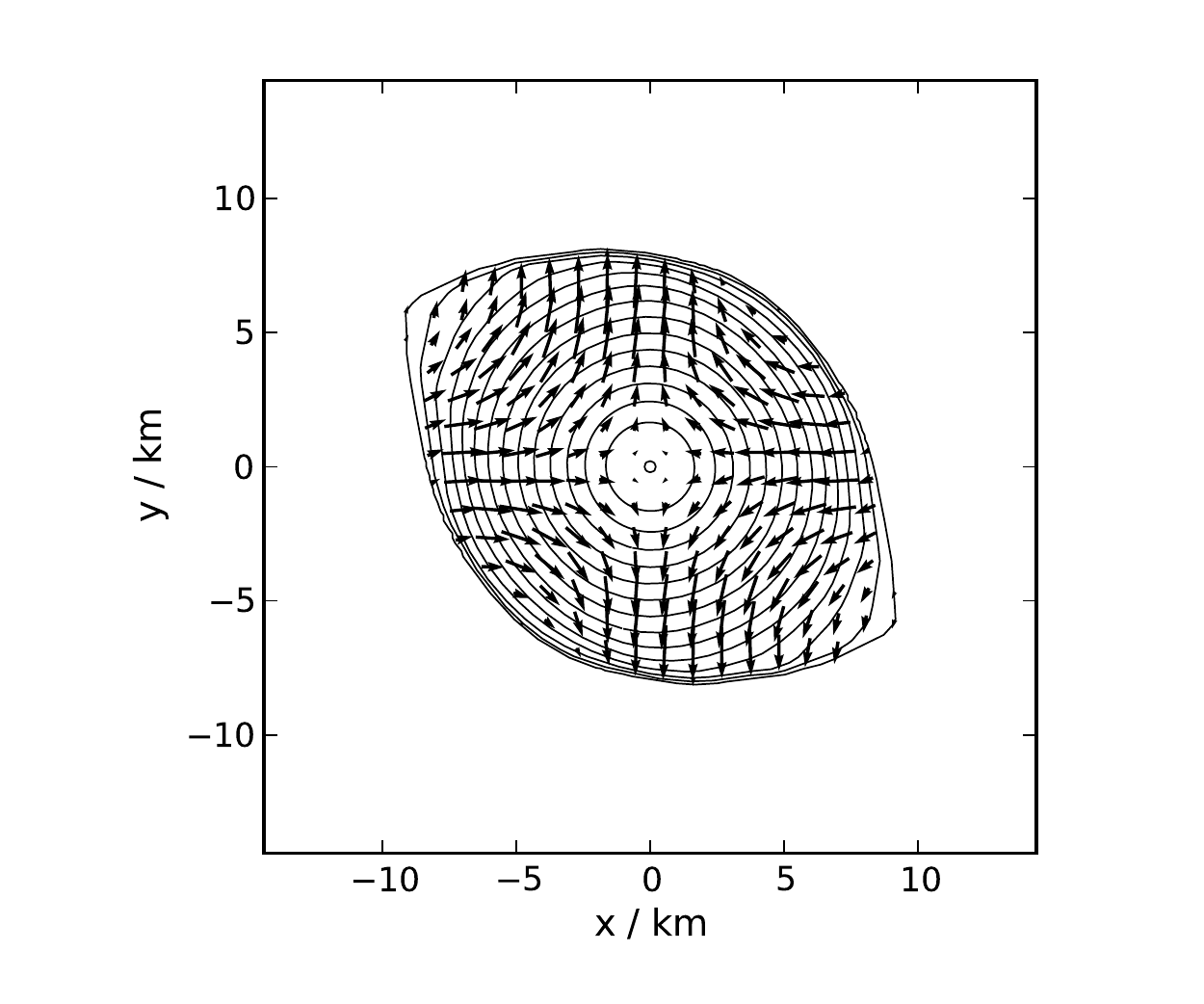}
  \caption{\label{fig_wave_jega65}
TOP: Like \Figure{fig_wave_jbu6}, but for the \Pctr~mode of model MA65.
BOTTOM: Same data, but only densities above $10^{-2}$ times the central one are shown,
and the arrows correspond to the momentum density instead of velocity.
  }
\end{figure}

To identify the damping mechanism, we produced animations
showing the density and velocity evolution
in the equatorial plane as well as the meridional planes.
For high amplitudes,
we found that the wave-patterns traveling around the star become
more and more nonsinusoidal towards the surface,
where we observe effects similar to wave breaking in ocean waves.
A snapshot is shown in \Figure{fig_wave_jbu6} and \ref{fig_wave_jega65}.
We do not find any formation of shocks in the interior of the star,
only at the surface in regions with densities around $10^{-4}\dots 10^{-3}$
times the central one.
Note this also justifies the approximation of adiabatic evolution
implied by the use of a polytropic EOS.
For low amplitudes, the wave-patterns looked exactly like
the eigenfunctions we used to excite the oscillation.
One important dissipative mechanism at high amplitudes thus seems to be wave breaking
at the surface.
As will be shown in the next section, the \Pctr~modes can also
participate in mode coupling.

Since wave breaking is very difficult to resolve numerically,
our results are probably not as accurate as the axisymmetric ones.
More important, the surface of realistic neutron stars does not consist
of a boundary between a cold fluid and vacuum, like in our models;
there is either a solid crust or a hot envelope, and also a magnetic field.
To compute the damping of the nonaxisymmetric modes correctly,
it will be necessary to add a more detailed description of the surface
to the numerical models.

Taking into account the strong deformation of the star,
one also has to question the accuracy of the Cowling approximation.
On the other hand,
the gravitational field is mainly determined by the denser regions of the star,
for which we observe much smaller deformations.
We therefore assume that the error on the bulk motion of the star
is comparable to the linear regime.
This should be verified by means of a fully relativistic study.


\subsection{Mode coupling}
\label{sec_mode_coupling}
For the \Pctr~modes of models MA65 and MB70,
we found that beside wave breaking,
mode coupling has a significant influence on the damping.

\begin{figure}
  \includegraphics[width=\columnwidth]{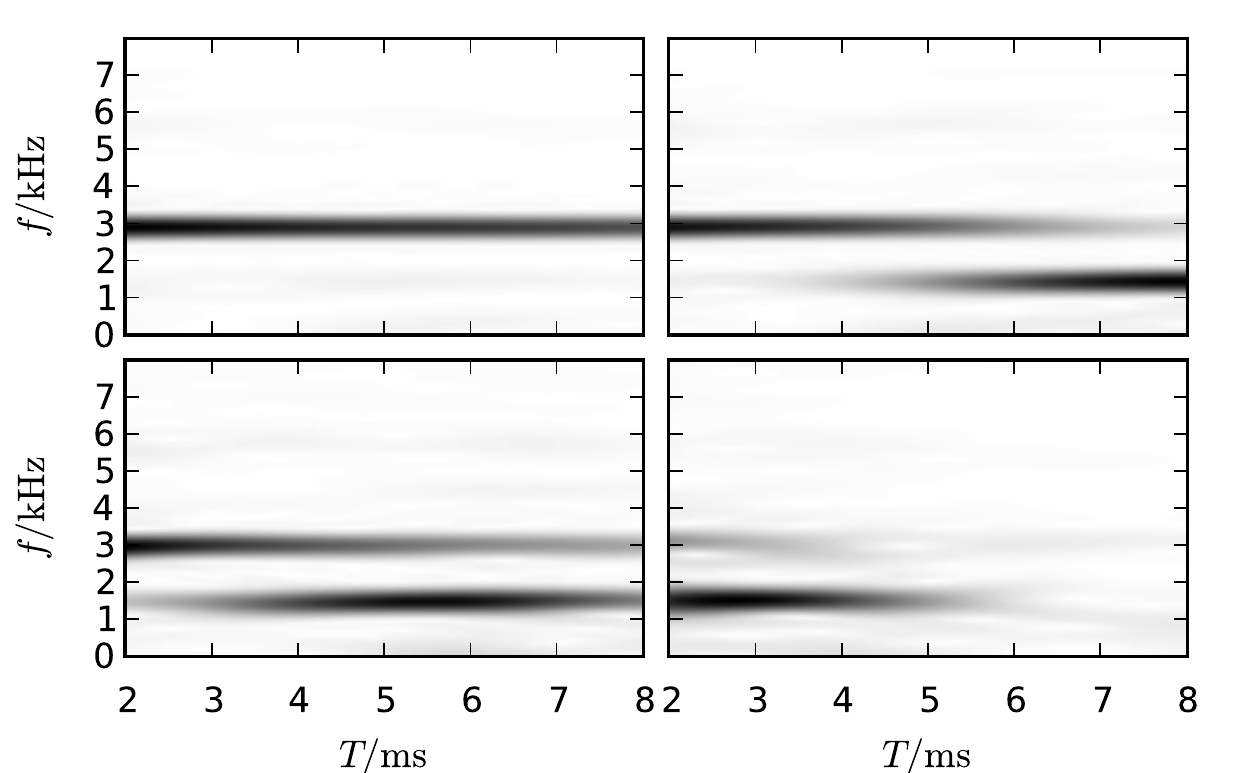}
  \caption{\label{fig_spg_jega65}
Mode coupling at different amplitudes
of the \Pctr~mode of model MA65 to an unknown mode at
half the frequency.
Shown is time evolution of the Fourier spectrum of the $\phi$-velocity in the corotating frame,
at $r=\frac{1}{2}R$ on the space diagonal.
The spectra were computed using a running window of width $4\usk\milli\second$.
The perturbation amplitudes $A_I$ are 0.01 (upper left), 0.02 (upper right), 0.04 (lower left),
and 0.08 (lower right).
The frequency resolution is $250\usk\hertz$.
  }
\end{figure}

In the spectrogram shown in \Figure{fig_spg_jega65},
one can easily spot mode coupling between the \Pctr~mode
and some unidentified mode, called mode X in the following.
For low amplitudes, only the \Pctr~mode is excited.
With increasing amplitude, mode X is growing
faster.
At the highest amplitude, it becomes dominant
in the $v^\phi$-spectrum on a timescale as low as $4\usk\milli\second$.

\begin{figure}
  \includegraphics[width=\columnwidth]{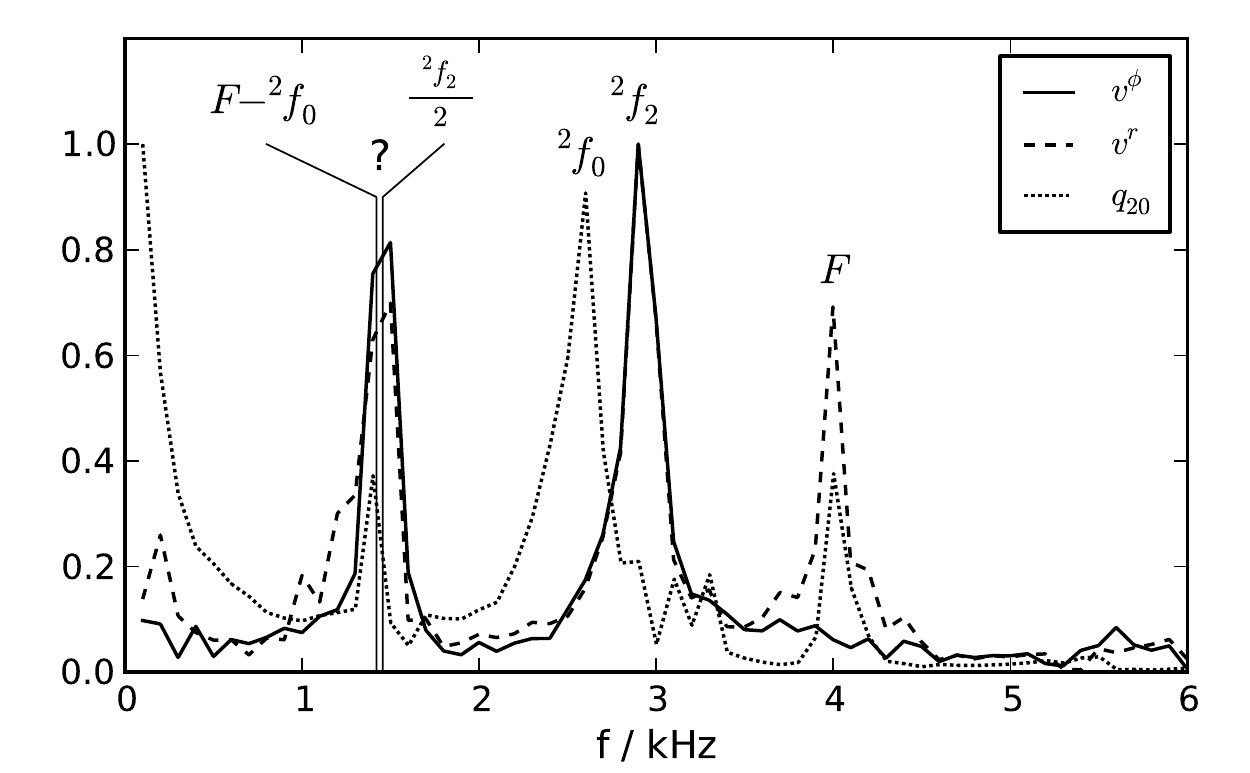}
  \caption{\label{fig_spec_jega65}
Fourier spectra of $v^\phi$, $v^r$, and $q_{20}$, in arbitrary units,
for a \Pctr-mode oscillation of model MA65, at amplitude $A_I=0.04$.
All frequencies are given in the corotating frame.
  }
\end{figure}

The Fourier spectrum for amplitude $A_I=0.04\usk c$ is shown in \Figure{fig_spec_jega65}.
Mode X is located at $1458 \pm 50\usk\hertz$,
the \Prad~mode at $3997\pm 50\usk\hertz$, the \Paxi~mode at $2590\pm 50\usk\hertz$,
and the \Pctr~mode at $2912\pm 50\usk\hertz$.

From this, we find two interesting resonances for the unknown mode.
First, mode X is located at half the frequency of the \Pctr~mode,
measured in the corotating frame.
Second, the frequency also equals (up to the resolution of the Fourier spectrum)
the difference between the \Prad and the \Paxi~mode.
The latter could be interpreted as a hint that the unidentified peak does not belong to
a proper oscillation mode, but is a nonlinear harmonic instead,
i.e. a typical feature of Fourier transforms in case of nonharmonic waveforms
and nonlinear superpositions.
However, the \Prad~mode is not present in the spectrum of $v^\phi$,
and the \Paxi~mode is neither present in the spectrum of $v^\phi$ nor of $v^r$,
while the unknown peak is present in the spectra of $v^\phi$ and $v^r$.
This supports the interpretation that the peak belongs to an actual oscillation mode,
because the amplitude of a nonlinear harmonic in the Fourier spectrum of some quantity
scales with the product of the amplitudes (in the same spectrum) of the modes
producing the harmonic.
However, it is still possible that the unidentified peak in the spectrum
of the multipole $q_{20}$ is a nonlinear harmonic of the \Prad and \Paxi~modes,
and thus unrelated to mode X.

Since the frequency of the unidentified mode is in the
inertial mode range,
and well below the frequencies of the lowest order axisymmetric pressure modes,
it is most probably an inertial mode.
In order to identify the modes appearing
for models MB70 and MA65, we first attempted mode recycling
with axisymmetric simulations.
However, the method did not converge.
This does not necessarily mean that the mode is nonaxisymmetric.
As described in \cite{Kastaun08},
extracting inertial modes using mode recycling is more difficult than for pressure modes.
Nevertheless, several low-order inertial modes of model MB70
were identified in \cite{Kastaun08};
none of them matches the frequency of the unknown mode.
Since extracting inertial modes typically requires more recycling steps
than pressure modes (due to the dense spectrum),
and because the frequencies are lower,
extracting nonaxisymmetric inertial modes is computationally expensive.
Therefore, we only performed the first step,
using an \Pctr-mode perturbation of model MA65 at amplitude $A_I=0.04$,
for which we know that the unknown mode is strongly exited.
The resulting three-dimensional numerical eigenfunction had predominantly a $|m|=2$ $\phi$-dependency.
We tried extracting a two-dimensional eigenfunction assuming either
$m=+2$ or $m=-2$, but obtained a big phase error ($q\approx 0.5$),
which typically points to a superposition of different modes
of almost the same frequency.
In our case, it seems that the unidentified peak actually consists of two
high order inertial modes with $m=\pm 2$,
although a clear identification will require additional mode recycling steps.

We note that the mode coupling is significantly stronger for resolution 75 than
for resolution 50, which means the process is not well resolved numerically.
In the aforementioned low-quality numerical eigenfunctions,
we found structures smaller than one fifth of the stellar radius.
Therefore, the excited inertial modes probably suffer from strong numerical damping,
which suppresses the coupling at resolution $N=50$.
Further, the time evolution in presence of mode coupling effects typically
depends strongly on the initial amplitudes of both modes,
even if one amplitude is small.
In our case, the unknown mode is probably seeded only by numerical errors;
in real stars, the situation might be different.
It is hence possible that the \Pctr~mode  decays even faster
in reality than in our simulations. 

Beside the inertial modes, the \Pctr~mode seems to couple to the \Pcor~mode
as well, albeit less dramatic.
This coupling will be discussed in \Sec{sec_gw_lum},
since it interferes with the estimation of the GW strain.
It is still unclear what the exact nature of the observed couplings is,
and whether the mode coupling picture is adequate at such high amplitudes at all.

To measure the decay in presence of mode coupling,
$\bar{v}$ is not useful because it is quite insensitive to
transfer of energy between modes.
However, we can use the dominant multipole moment of the mode,
which for the \Pctr~modes is the $l=m=2$ mass multipole.
In the Fourier spectra of $q_{22}$, the dominant contribution is due to the \Pctr~mode.
By fitting a function $q^0 \exp(-t/\tau_m + i \omega t)$ to the initial evolution of
the multipole moment $q_{22}$, we obtain another damping timescale $\tau_m$.

\begin{figure}
  \includegraphics[width=\columnwidth]{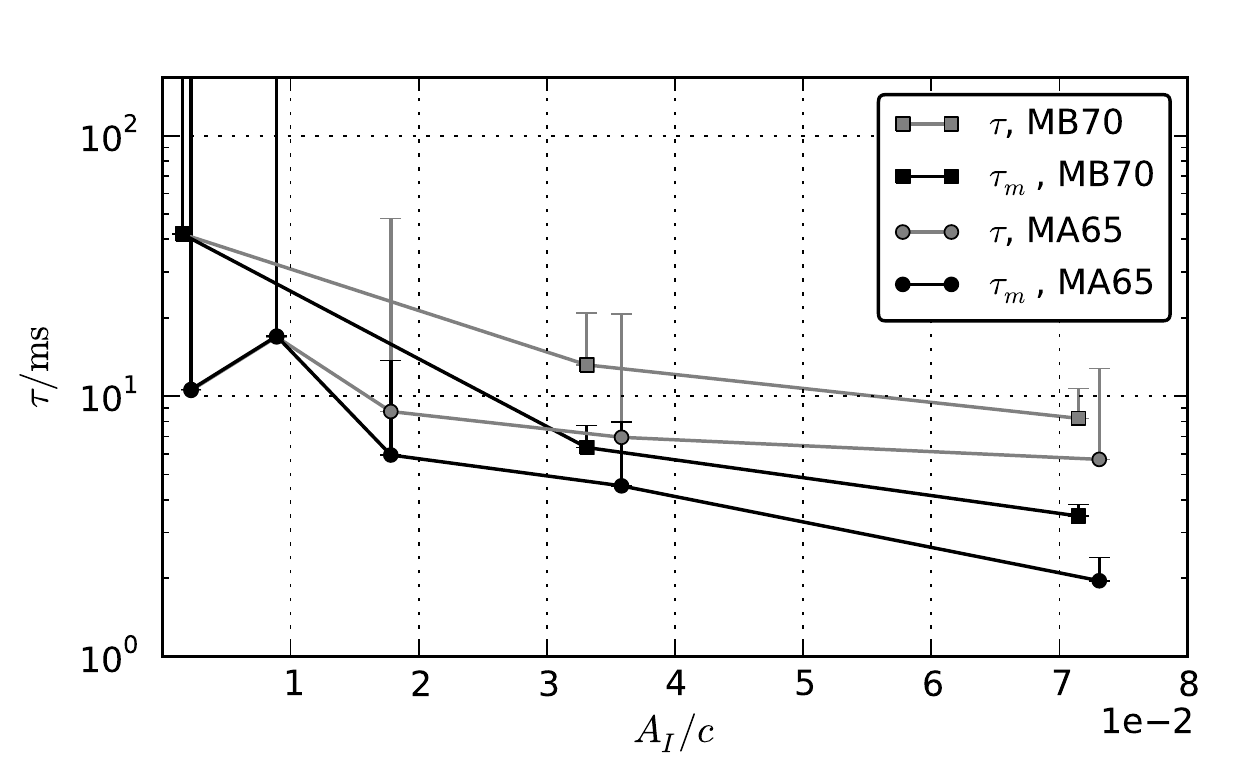}
  \caption{\label{fig_tau_multi}
Decay timescales $\tau$ and $\tau_m$ of $\bar{v}$ and $q_{22}$
versus initial amplitude $A_I$, for models MB70 and MA65
perturbed with the \Pctr~mode.
The results are obtained at resolution 75.
The error bars estimate the influence of the numerical damping.
  }
\end{figure}

While $\tau$ is a good  measure for the overall energy dissipation,
$\tau_m$ is a measure for the decay of the \Pctr~mode.
\Figure{fig_tau_multi} shows a comparison between the two.
As one can see, mode coupling is even more important for the decay of the \Pctr~mode
than dissipative effects, e.g. wave breaking.

At least this is the case for models MA65 and MB70;
for the \Pctr~mode of model MA100
and the \Pcor~mode of model MA65, we do not observe mode coupling,
while for model MC85 we observe weak coupling to a mode at one quarter of
the (corotating) frequency of the \Pctr~mode.
What is the difference?
In model MB70 and MA65, the coupled mode is an inertial mode
which oscillates at $\frac{1}{2} f_c$,
with  $f_c$ being the frequency of the \Pctr~mode in the corotating frame.
Since the inertial mode range is bounded by $2 F_R$ (see \cite{Kastaun08}),
where $F_R$ is the rotation rate of the star,
this is only possible if
\begin{align}
  4 F_R > f_c  .
\end{align}
Model MC85 (and MA100 anyway)  rotates too slowly to satisfy this condition.
The 1:4 resonance to an inertial mode on the other hand is possible for model MC85.

Note the above condition automatically holds for CFS-unstable \Pctr~modes,
for which $f_c<2F_R$.
It seems plausible that the coupling is also present at lower amplitudes,
but evolves at longer timescales.
To compute this timescale and thus a more stringent limit
on the saturation amplitude of  CFS-unstable \Pctr~modes,
perturbative studies are required.

The upper limits for the onset of nonlinearity and the corresponding
damping timescales for the nonaxisymmetric modes are summarized
in \Tab{tab_onset_nl_m2}. 
The mode coupling effects for models MA65 and MB70 are taken into account.
Since the CFS instability of the \Pctr~mode of model MA65
has a growth time in the order of seconds,
our result yields an upper limit for the saturation amplitude.

\begin{table}
\caption{\label{tab_onset_nl_m2}
Upper limits on the onset of nonlinearity,
for nonaxisymmetric oscillation modes.
$A_D$ is the initial amplitude $A_I$ at which nonlinear
damping (including mode coupling effects) can be safely
distinguished from the numerical damping.
$\tau$ is an upper limit for the decay timescale of the mode
at amplitude $A_D$.
}
\begin{ruledtabular}
\begin{tabular}{llll}
Model   &Mode     &$A_D / (10^{-3}\usk c)$ & $\tau / \milli\second$\\\hline
MC85   & \Pctr   &10  & 90       \\
MB70    & \Pctr   &35  & 9      \\
MA100    & $\Pctr \equiv \Pcor$   &130  &20      \\
MA65  & \Pctr   &18  & 20      \\
MA65  & \Pcor   &30  & 40
\end{tabular}
\end{ruledtabular}
\end{table}

\subsection{Gravitational luminosity}
\label{sec_gw_lum}
\begin{figure}
  \includegraphics[width=\columnwidth]{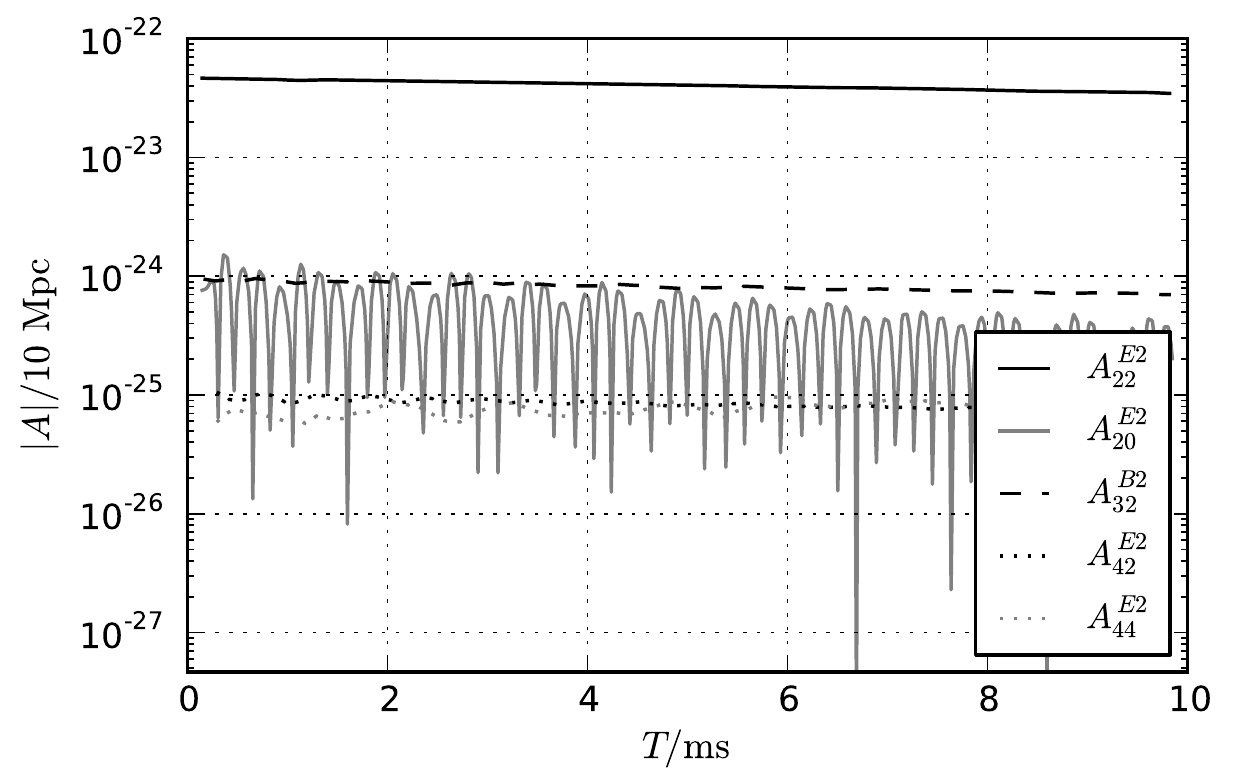}
  \caption{\label{fig_ts_gwstr}
Evolution of GW strain amplitudes for model MA65 perturbed
with the corotating \Pcor~mode at amplitude $A_I=0.01\usk c$.
Shown are only the 5 biggest contributions.
  }
\end{figure}

To estimate the GW production in our simulations,
we use the multipole formalism described in \Sec{sec_gwmultipole}.
For each simulation, we plot the time evolution of the magnitude
of the strain coefficients $\Aelec_{lm}$ and $\Amagn_{lm}$,
as shown in \Figure{fig_ts_gwstr}.

In general,
the only significant contribution is due to the $l=2$ mass
multipoles $q_{22}$ or $q_{20}$.
Unless noted otherwise,
one of the strain amplitudes $\Aelec_{20}$ or $\Aelec_{22}$
is strongly dominant, and will be denoted by $A$ in the following.
For the axisymmetric case, terms with $m \neq 0$ vanish.

For oscillations in the linear regime, the strain is proportional to the
amplitude.
We introduce a normalized mode strain $\mathcal{A}$ and luminosity $\mathcal{L}$ by
\begin{align}
  A &= \frac{\bar{v}}{10^{-3} c} \mathcal{A}, &
  L &= \left(\frac{\bar{v}}{10^{-3} c}\right)^2 \mathcal{L} .
\end{align}
From the simulations with the lowest amplitudes,
which are in the linear regime,
we extracted $\mathcal{A}$ and $\mathcal{L}$ for each mode.
The results are given in \Tab{tab_mode_strain}.

The strain produced by the \Pctr~mode of model MA65 is the lowest one of all modes.
This is mainly due to the low frequency in the inertial frame.
The corotating \Pcor~mode produces a strain larger by a factor
of 143 at the same amplitude; the $\omega^2$ frequency dependency
accounts for a factor of 77.

Since model MA65 is already close to the Kepler limit, the
strain cannot be increased significantly by uniformly spinning up the star.
It might however be possible to construct differentially
rotating models for which the CFS-unstable \Pctr~modes have higher frequencies.

The strain of the \Pctr~mode of model MC85 is surprisingly high given that it's
frequency is even lower than for model MA65.
On the other hand, the radius of model MC85 is $5.1$ times the one of MA65,
which enhances the $l=2$ multipole moments by a factor 25.
We thus consider the high strain amplitude of model MC85 as a consequence of the
unrealistically large radius.

Another noteworthy result is that the strain amplitudes of \Prad~modes of rapidly
rotating stars are comparable to those of the  \Paxi~modes.
As discussed in \Sec{sec_gwmultipole},
cancellations in the integrands of the multipoles prevented
us from computing a meaningful error estimate for the strain produced by the \Prad~modes.
We nevertheless believe that at least for model MA65,
the high strain amplitude of the \Prad~mode
is not just an artifact due to the use of the quadrupole formula,
but a genuine effect of the rapid rotation rate.
However, this can only be verified by means of a fully relativistic study,
using direct wave extraction methods.
The question is relevant for GW astronomy because strong
\Prad-mode oscillations are likely to occur in a number of astrophysical scenarios.

\begin{table}
\caption{\label{tab_mode_strain}
Gravitational luminosity and strain produced by oscillation modes in the linear regime,
normalized to an amplitude $\bar{v}=10^{-3}$.
The values are computed from a fit to the multipole moments.
The error bounds due to the use of the multipole formalism (see \Sec{sec_gwmultipole})
are $\mathcal{A}q^{-1} \dots \mathcal{A}q$ for the strain,
and $\mathcal{L} q^{-2} \dots \mathcal{L} q ^2$ for the luminosity.
For the \Prad~modes, the error is unknown; the values are given only for completeness.
}
\begin{ruledtabular}
\begin{tabular}{lllll}
Model   &Mode     & $\mathcal{A} / 10\usk\mega\parsec$ &  $\mathcal{L} / \watt$ & q\\\hline
MA65    &\Prad    & 2.96\e{-25}     & 1.06\e{40}     & ?  \\
MA65    &\Paxi    & 1.24\e{-24}     & 8.28\e{40}     & 13  \\
MA65    &\Pctr    & 4.09\e{-26}     & 1.37\e{37}     & 13  \\
MA65    &\Pcor    & 5.84\e{-24}     & 2.13\e{43}     & 13  \\
MA100   &\Paxi    & 7.18\e{-24}     & 3.59\e{42}     & 2   \\
MA100   &\Pctr    & 4.96\e{-24}     & 6.85\e{42}     & 2   \\
MB100   &\Paxi    & 7.79\e{-24}     & 1.63\e{42}     & 1.5 \\
MB70    &\Pctr    & 1.10\e{-25}     & 4.92\e{37}     & 4   \\
MC100   &\Paxi    & 5.28\e{-24}     & 3.67\e{40}     & 1.2 \\
MC100   &\Pctr    & 3.74\e{-24}     & 7.39\e{40}     & 1.2 \\
MC95    &\Prad    & 6.11\e{-25}     & 5.38\e{38}     & ? \\
MC95    &\Paxi    & 4.08\e{-24}     & 1.93\e{40}     & 1.5 \\
MC85    &\Prad    & 1.79\e{-24}     & 3.54\e{39}     & ? \\
MC85    &\Paxi    & 3.41\e{-24}     & 9.95\e{39}     & 1.4 \\
MC85    &\Pctr    & 8.65\e{-25}     & 9.44\e{38}     & 1.4 \\
\end{tabular}
\end{ruledtabular}
\end{table}

\begin{figure}
  \includegraphics[width=\columnwidth]{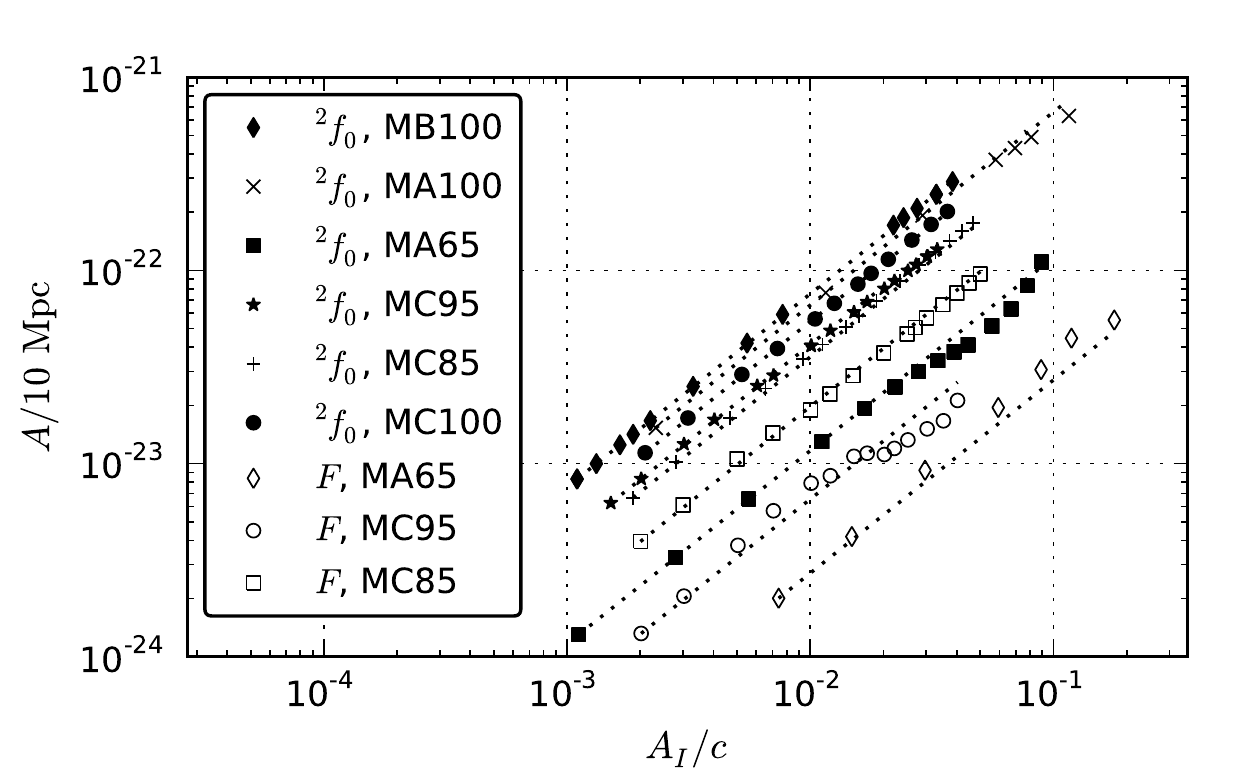}
  \caption{\label{fig_gwstr_axi}
Maximum gravitational wave amplitude $A=|\Aelec_{20}|$ versus initial amplitude $A_I$,
for axisymmetric perturbations.
The contribution of other multipole moments to the strain are negligible in all cases.
The dotted lines are an extrapolation from the lowest amplitude,
assuming linearity.
  }
\end{figure}

An overview over the maximum GW strain observed in the nonlinear
simulations of axisymmetric modes
can be found in \Figure{fig_gwstr_axi}.
As one can see, the maximum strain roughly scales with the amplitude;
by simply extrapolating from the linear regime,
one obtains an estimate  which is correct up to a factor of 2
even for the highest amplitudes.

For each simulation, we computed the Fourier spectra of the strain to determine the
frequency of the dominant contribution.
Usually, the only significant contribution is due to the mode used for
perturbation.
For the \Pctr~mode of MA65 on the other hand,
we find that the GW strain is mainly due to the corotating \Pcor~mode,
as well as the axisymmetric \Prad and \Paxi~modes.
Those modes have a lower amplitude than the \Pctr~mode.
But since the \Pctr~mode of this model is a relatively inefficient GW emitter,
due to the low oscillation frequency,
the other modes dominate the GW strain.

\begin{figure}
  \includegraphics[width=\columnwidth]{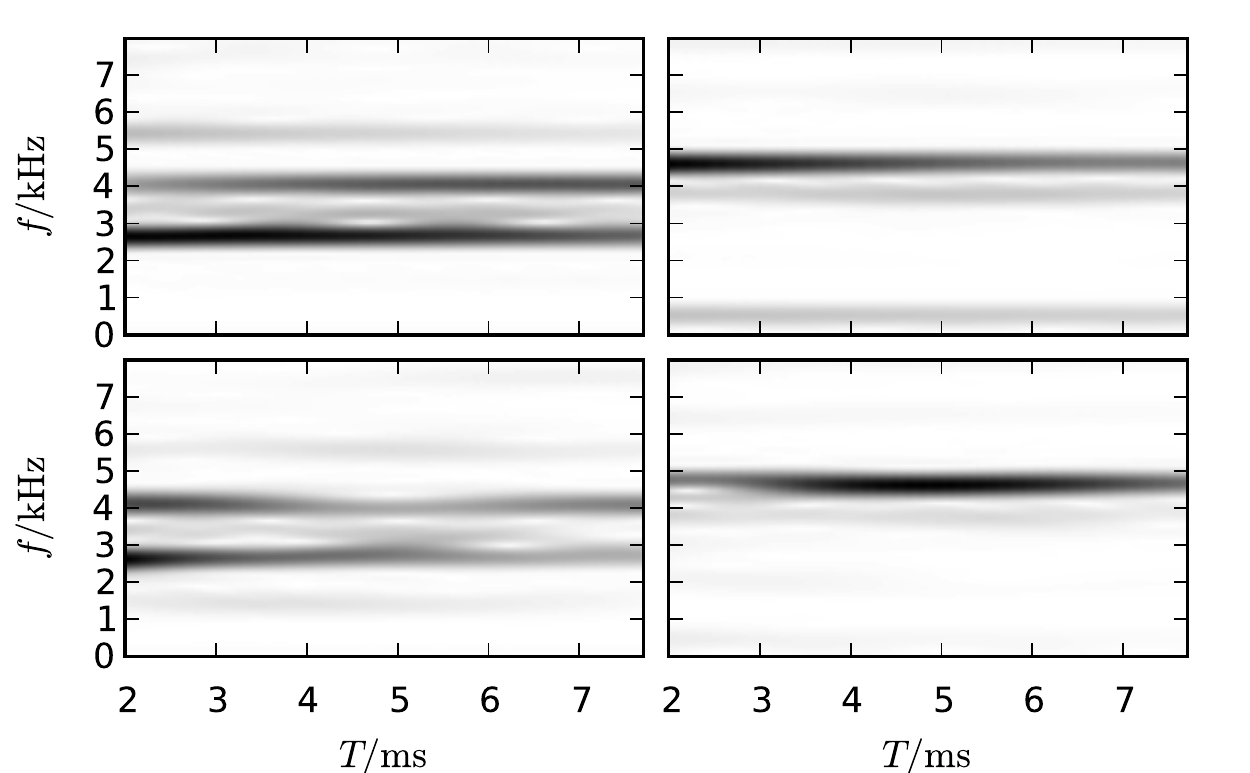}
  \caption{\label{fig_spg3_jega65_fl2m2}
Spectrogram of the gravitational wave amplitude,
for the \Pctr-mode oscillation of model MA65,
at perturbation amplitudes $A_I=0.01$ (TOP) and $A_I=0.04$ (BOTTOM).
Shown is the Fourier spectrum of $\Aelec_{20}$ (LEFT) and $\Re(\Aelec_{22})$ (RIGHT)
inside a running time window of width $4\usk\milli\second$,
versus the time at the center of the window.
  }
\end{figure}

To understand the strain of the \Pctr~mode, we thus need to understand
why the other modes are present.
\Figure{fig_spg3_jega65_fl2m2} shows a spectrogram of the
corresponding strain amplitudes.
By further comparing the absolute amplitudes of the multipole moments $q_{22}$ and $q_{20}$,
we find the following picture:

The \Pcor~mode is excited due to a small contamination of the numerical eigenfunction;
it's initial amplitude is around 4\usk\% of the \Pctr~mode in the linear regime.
For high amplitudes however, it grows up to 20 \% during the evolution.
The reason is unclear, it might be another mode coupling effect.

The \Prad- and \Paxi~modes are present from the beginning as well,
but in the linear regime their amplitude becomes negligible
compared to the amplitude of the \Pctr~mode.
For a high initial amplitude $A_I=0.04\usk c$, the amplitude of $q_{20}$
is still less than 20\usk\% of $q_{22}$.
We explain the presence of the \Prad and \Paxi~modes
by the ad hoc way in which we
scale the eigenfunctions to high amplitudes;
it would be surprising if only one mode was excited.
The relative amplitudes could be totally different for the case of a
\Pctr~mode slowly grown to high amplitudes by the CFS mechanism.

To quantify the strain produced by the \Pctr~mode alone,
we fit a function $q_0 \exp(-t/\tau_m + i \omega t)$ to the $q_{22}$ multipole
and then compute the strain from the multipole moment given by the fit.
This way, we effectively filter out contributions
of frequencies other than the one of the \Pctr~mode.
Note all the strains in the linear regime,
given in \Tab{tab_mode_strain},
have been computed in the same way
(for the axisymmetric modes, we use a damped sinusoidal since $q_{20}$ is real-valued).

\begin{figure}
  \includegraphics[width=\columnwidth]{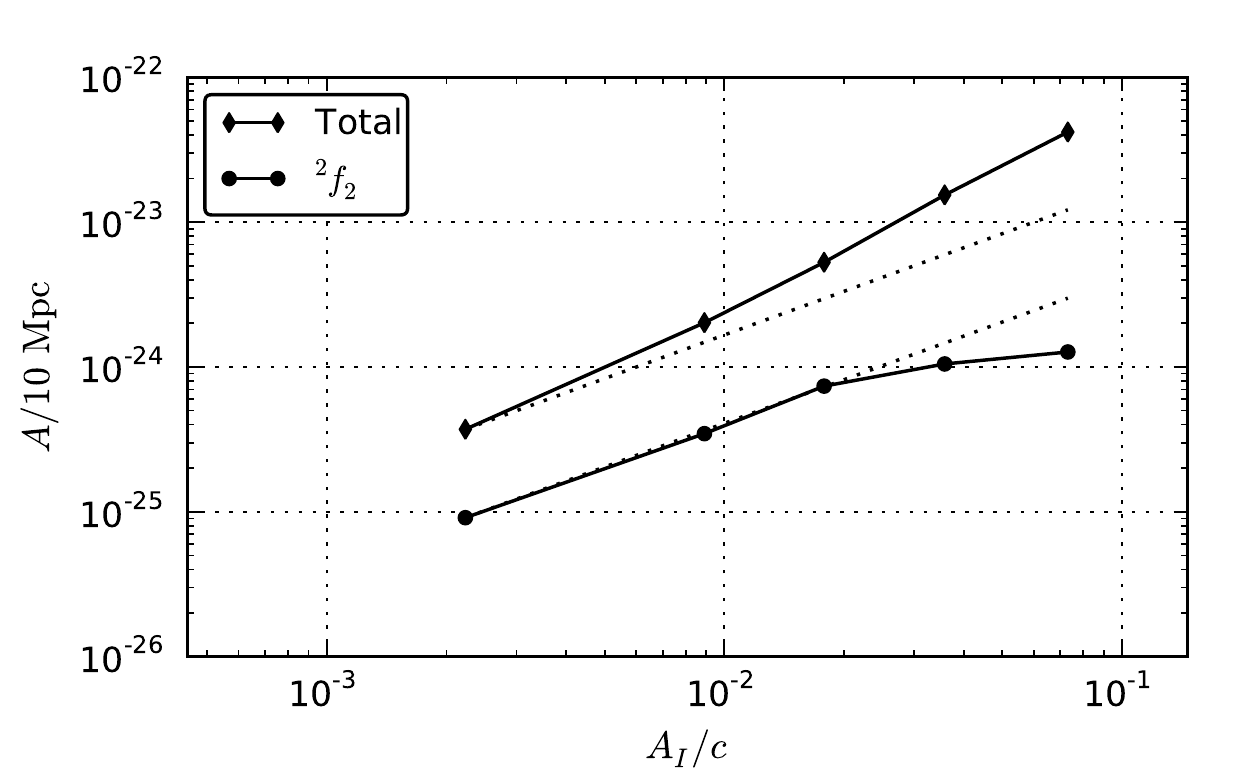}
  \caption{\label{fig_strain_jega65_fl2m2}
Initial GW amplitude $A$ versus initial oscillation amplitude $A_I$,
for model MA65 perturbed with the counter-rotating \Pctr~mode.
Shown is the total strain as well as the strain due to the \Pctr~mode alone (see main text).
The dotted lines are extrapolations from the linear regime.
  }
\end{figure}

The results for different amplitudes are shown in
\Figure{fig_strain_jega65_fl2m2}.
Only  3--24\% of the total strain is due to the main \Pctr~mode.
We stress that the total strain found for our setups
is probably higher than
the strain of a \Pctr-oscillation
grown to high amplitude by the CFS-mechanism.
To accurately compute the latter,
one needs to model carefully the growth of secondary coupled modes,
even if they are dynamically unimportant.

\begin{figure}
  \includegraphics[width=\columnwidth]{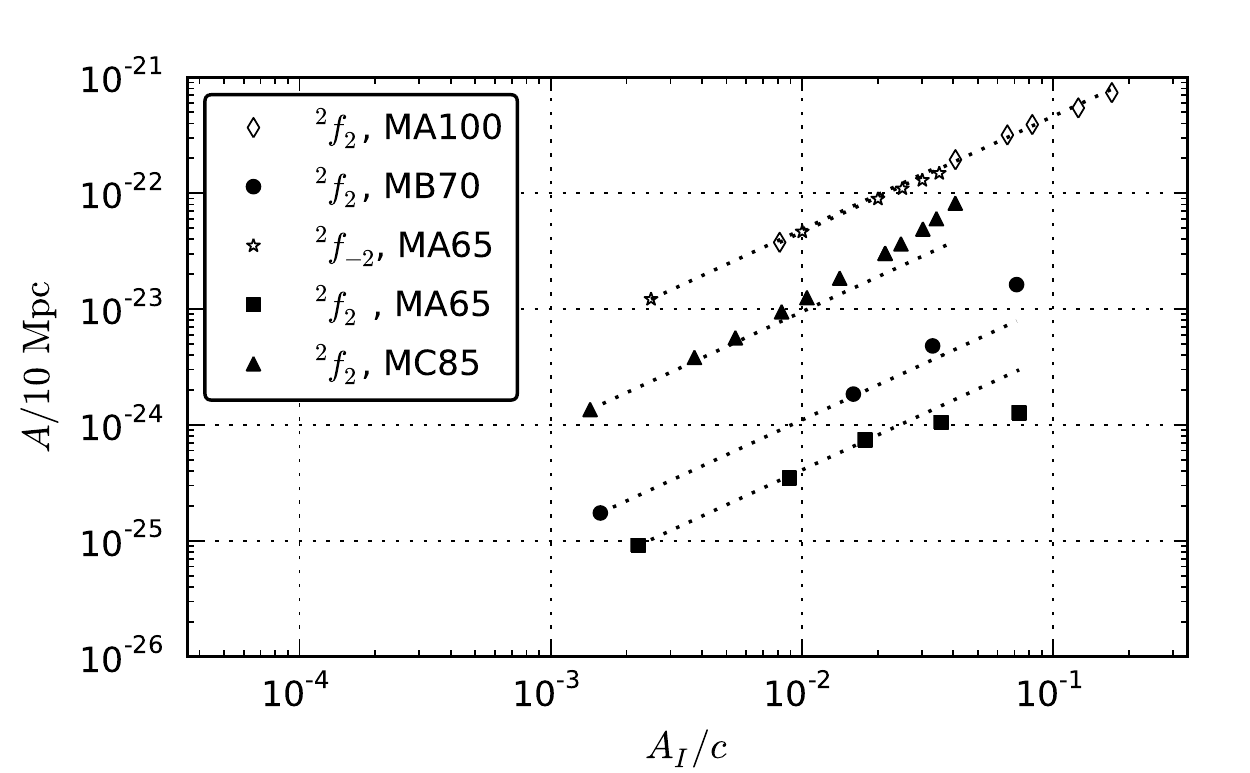}
  \caption{\label{fig_gwstr_m2}
Maximum gravitational wave amplitude $A=|\Aelec_{22}|$ versus initial amplitude $A_I$,
for $m=2$ perturbations.
For the \Pctr~modes of models MB70 and MA65, only the contribution of the \Pctr~mode
itself is plotted (see main text).
The dotted lines are an extrapolation from the lowest amplitude,
assuming linearity.
  }
\end{figure}

For the \Pctr~mode of model MB70, we qualitatively find the  same picture
as for model MA65.
For the \Pctr~mode of models MC85 and MA100, as well as the \Pcor~mode of
MA65, contribution of other modes to the GW strain are small.
The strain at high amplitudes is shown in \Figure{fig_gwstr_m2}.
Again, the extrapolation from the linear regime yields a rough estimate,
correct up to a factor of 3.

\subsection{Gravitational wave detectability}
\label{sec_detect}
In the following, we estimate the prospects of detecting GW
in the case of either a one-time strong excitation or a continuous
excitation at lower amplitudes.

First we investigate the continuous oscillation of one mode at saturation amplitude
of some instability.
Our primary interest here is the CFS instability of the \Pctr~mode of model MA65.
This instability has a growth time of seconds, which is longer than
the timescale of the numerical damping.
Therefore, we can only give upper limits on the saturation amplitude,
namely the amplitude at which we definitely see nonlinear damping.
The actual saturation amplitude may be much smaller.

\begin{figure}
  \includegraphics[width=\columnwidth]{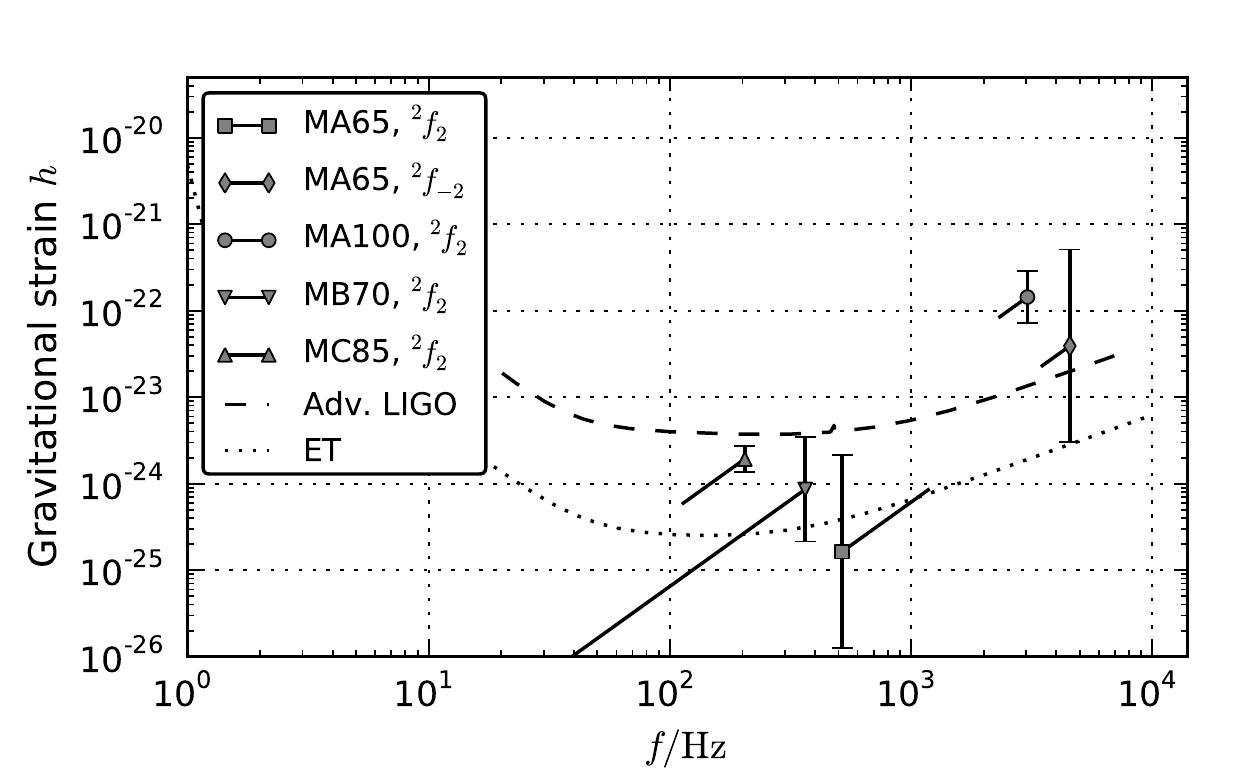}
  \caption{\label{fig_detect_onset}
Gravitational wave strain $h_{\theta\theta}$ and mode frequency,
for \Pctr and \Pcor~modes at amplitudes where
we are confident to observe nonlinear damping.
The strain is given for a source at distance $10\usk\mega\parsec$,
assuming optimal viewing angle.
The lines are estimates for the error due to the Cowling approximation,
assuming that only the frequency changes.
The error bars correspond to the error due to the use of the quadrupole formula.
Other errors are negligible.
For comparison, we plot the sensitivity curves of advanced LIGO and the
proposed Einstein Telescope (ET).
  }
\end{figure}

\Figure{fig_detect_onset} shows the gravitational strain at the observed
onset of nonlinear damping for several oscillation modes.
We plot the $h_{\theta\theta}$ component given by \Eref{eq_gwstrain}, assuming an optimal viewing angle.
The general expression for the spin tensor harmonics can be found in \cite{Mathews62}
(note the standard notation of magnetic and electric type radiation is reversed there).

The detectability of sinusoidal GW signals depends not only on the strain amplitude,
but also on the integration time.
Effectively, the detectable amplitude scales with $\sqrt{N}$,
where $N$ is the number of available wave cycles.
For a continuous signal, this only depends on the detector and the computational resources
for data analysis. 
If we assume at least a few hundred cycles, the \Pctr~mode at the onset
of strong damping is detectable with advanced LIGO at a distance of 
10\usk\mega\parsec.
Under the assumption that the  saturation amplitude of the CFS instability of the \Pctr~mode
is only limited by the strong damping effects investigated in this work,
it would be possible to detect sources as far away as the Virgo galaxy cluster.
To compute the actual saturation amplitude however,
perturbative studies at lower amplitudes are needed.

The numbers above are valid for model MA65, which is an extreme case.
Models rotating slightly slower will produce considerably lower strain.
The reason is that the strain scales with the square of
the mode frequency in the inertial frame,
which for the \Pctr~modes
crosses zero when the rotation rate falls below the neutral point.
Keeping mass, EOS, and mode amplitude $\bar{v}$ fixed,
we can make an expansion of the strain amplitude
$A$ in terms of rotation rate $F_R$ near the neutral point:
\begin{align}
  A &= A_0  \left(\frac{F_R-F_R^N}{F_R^0-F_R^N}\right)^2 ,
\end{align}
where $A_0$, $F_R^0$ are the strain amplitude and rotation rate of MA65,
and $F_R^N$ is the rotation rate at the neutral point.
Further, it is safe to assume that the amplitude marking the onset of nonlinearity
behaves smooth near the neutral point,
since the rotation rate influences the dissipation effects mainly via the 
centrifugal force,
and the inertial modes participating in mode coupling depend on the Coriolis force;
both forces depend on the absolute rotation rate.

For frequencies in the range 0.1--2 times the \Pctr-mode frequency of model MA65,
the sensitivity curve of advanced LIGO is quite flat.
In this range, the \textit{upper limit} we find for the distance at which we can observe the \Pctr~mode
roughly scales with $(F_R-F_R^N)^{2}$.

Since model MA65 is close to the Kepler limit, the strain cannot be increased significantly by
increasing the rotation rate; on the contrary,
we expect that nonlinear effects set in much earlier when approaching
the Kepler limit further, as observed for model MC65.
Further, we doubt that one can construct models where the frequency of the
CFS-unstable \Pctr~modes is more than three times higher
without resorting to extremely unrealistic EOSs.
On the other hand, this statement only applies to uniformly rotating stars;
it might be fruitful to investigate differentially rotating models as well.

Another restriction for detection of continuous signals
is currently given by the limited computing resources for data analysis,
which prevents to search the whole parameter space for signals.
In particular, searches \cite{LIGO09,LIGO09b} for continuous signals in early data from the 5th
science run of LIGO assumed a maximum time derivative of the signal frequency.
Those searches where targeted to signals from spinning nonaxisymmetric neutron stars,
based on estimates on the maximum sustainable ellipticity.
The luminosity produced by such sources is orders of magnitude below the upper limit
we computed for the CFS-unstable \Pctr~mode of model MA65.
Accordingly, our models spin down much faster.
Assuming that the mode frequency $f_c$ in the corotating frame stays constant,
the decrease of the signal frequency, i.e. the mode frequency in the inertial frame,
is $\dot{f}_i = m \dot{F}_R$, provided that $f_c<m F_R$.
The spin-down rate can be computed from the expressions for the
angular momentum lost by gravitational waves given in \cite{Thorne80} and the moment of inertia
computed by the initial data code.
For the \Pctr~mode of model MA65 at the upper limit, we compute a signal frequency decrease
of $\dot{f} = -5 \cdot 10^{-3} \usk\hertz\per\second$, which is orders of magnitude
outside the ranges searched in \cite{LIGO09,LIGO09b}.
For models closer to the neutral point, luminosity and spin-down rate would be smaller.
As a rough estimate, we assume that only the frequencies change,
while maximum amplitude and oscillation pattern stay constant.
The spin-down rate then scales with $f_i^5$. It is still above the maximum
for the whole frequency range $50\dots 1100\usk\hertz$ investigated in \cite{LIGO09,LIGO09b}.
Those searches are clearly only sensitive to neutron star oscillations with amplitudes
well below the onset of strong damping effects investigated here.

\begin{figure}
  \includegraphics[width=\columnwidth]{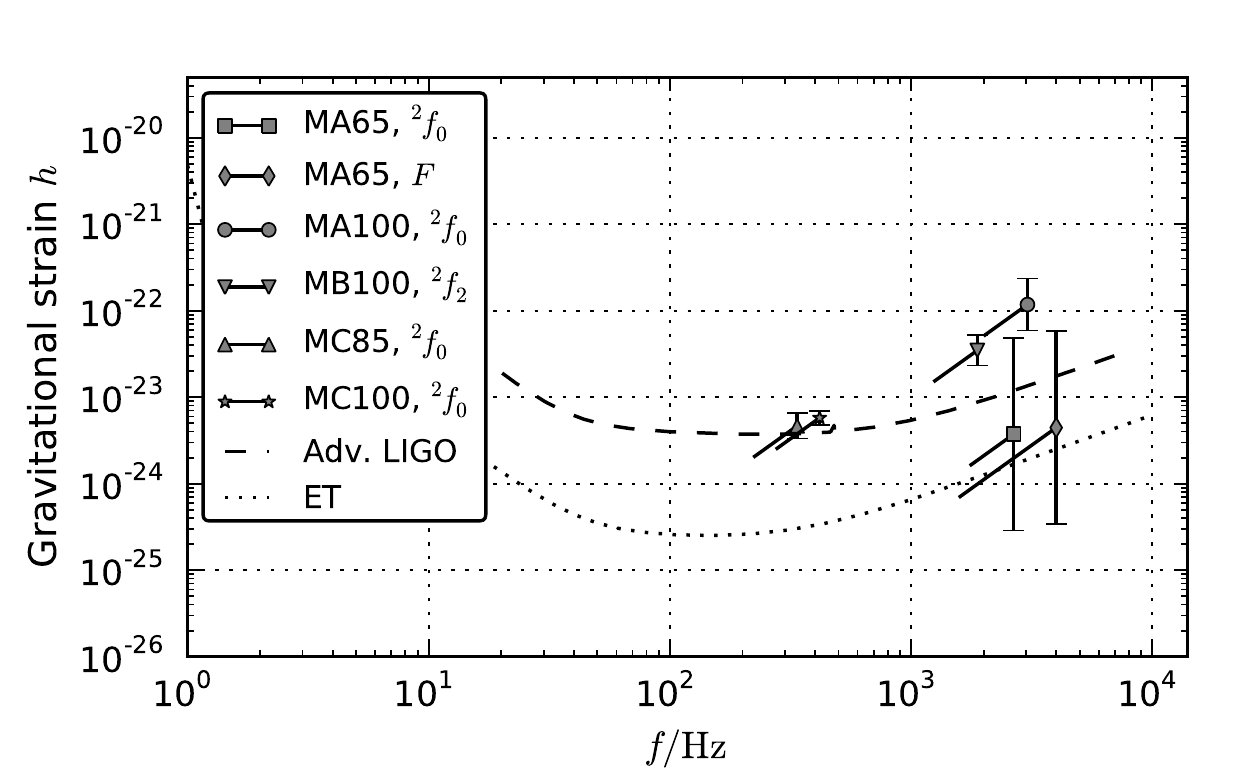}
  \caption{\label{fig_detect_onset_axi}
Like \Figure{fig_detect_onset}, but for axisymmetric modes.
The errorbar of the \Prad~mode is only correct under the assumption that
cancellation effects do not increase the error in comparison to nonradial modes;
see \Sec{sec_gwmultipole}.
  }
\end{figure}

\begin{figure}
  \includegraphics[width=\columnwidth]{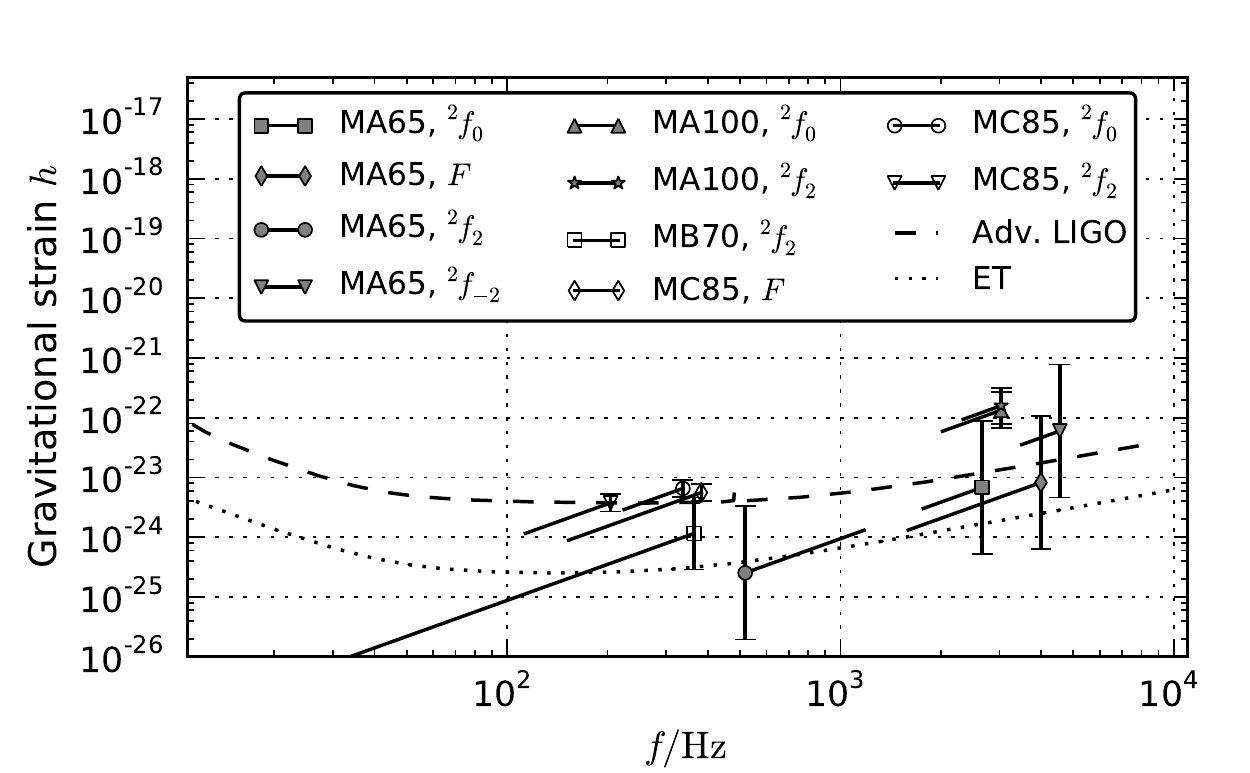}
  \caption{\label{fig_detect_20pf}
Like \Figure{fig_detect_onset}, but for amplitudes such that
the damping timescale satisfies $\tau=20\usk T_F$, where $T_F$ is the oscillation period
of the \Prad~mode of the corresponding model.
  }
\end{figure}

For modes other than the \Pctr~mode of model MA65,
we are not aware of any instability.
Nevertheless, we provide limits for a hypothetical instability.
\Figure{fig_detect_onset} and \ref{fig_detect_onset_axi} shows upper limits
for the strain amplitudes assuming that the instability acts on timescales
slower than the numerical damping.
\Figure{fig_detect_20pf} gives the strain at saturation amplitude
for the optimistic assumption that the hypothetical instability
has growth time $\tau_I=20\usk T_F$, where $T_F$ is the period of the \Prad~mode
of the corresponding stellar model.

Besides slowly growing instabilities,
oscillations can be excited by sudden violent events,
e.g. glitches, magnetar giant flares, phase-change induced collapse (\cite{Ernazar09}),
and mergers forming a metastable rapidly rotating star.
It is beyond the scope of this article to estimate the
amplitude of oscillations excited by such events.
We investigate the detectability for a given amplitude, assuming no further excitation.
In that case, the amplitude starts decaying due to strong nonlinear damping,
until it reaches a threshold where only slow damping occurs.

In order to detect such an event,
one could either search in short segments of the detector data
for the initial strong signal,
or search for the tail by investigating long time intervals.
Since any long tail has an amplitude below the onset of strong damping,
we already know the maximum strain amplitudes, which are given in
\Figure{fig_detect_onset} and \ref{fig_detect_onset_axi}.
The big unknown is the number of available cycles,
which depends on the residual damping timescale
of the tail, and thus cannot be computed by nonlinear evolution.

To estimate the detectability of the initial peak,
we neglect the tail, assume a pure exponential decay,
and limit the integration time to $T_a=1\usk\second$.
We define an effective strain amplitude by
\begin{align}\label{eq_def_eff_str}
  A_\text{eff} &= A \sqrt{\frac{1-e^{-2T_a/\tau}}{1-e^{-2/(f_i\tau)}}} ,
\end{align}
where $f_i$ is the mode frequency in the inertial frame, and $\tau$ is the
initial damping timescale, corrected for the numerical damping.
The enhancement factor in \Eref{eq_def_eff_str} was taken from \cite{Ernazar09}.
For low amplitudes where we cannot disentangle numerical and physical damping,
we set $\tau=1000\usk \second$.

\begin{figure}
  \includegraphics[width=\columnwidth]{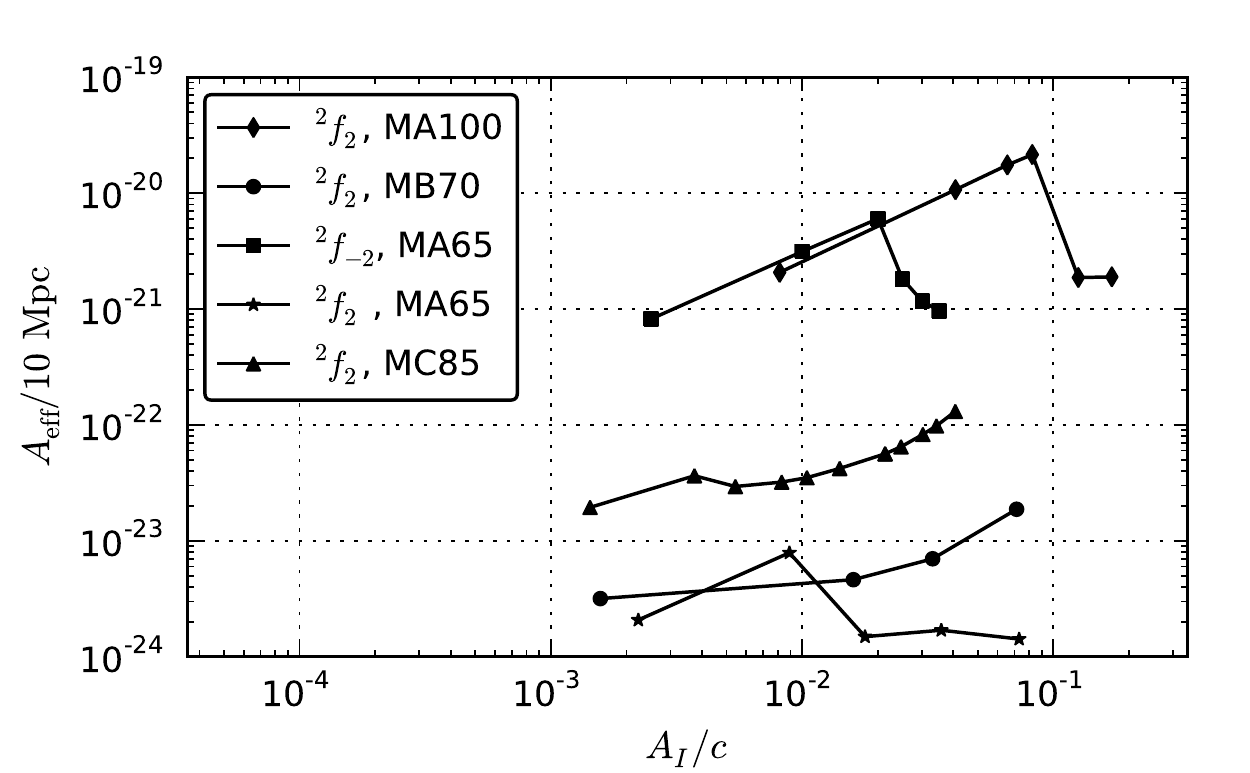}
  \caption{\label{fig_eff_strain}
Effective GW strain amplitude $A_\text{eff}$ defined by \Eref{eq_def_eff_str}
versus perturbation amplitude $A_I$ for the nonaxisymmetric
oscillations.
  }
\end{figure}

The results are plotted in \Figure{fig_eff_strain}.
As one can see, the increase in amplitude after the onset of nonlinearity
competes with the rapid decrease in signal duration,
and in most cases the latter wins.
The effect becomes more pronounced when increasing the integration time.
When considering the tail again, we find that the detectability does not
increase significantly for perturbation amplitudes above the onset of
nonlinearity.

\section{Summary and discussion}
\label{sec_summary}

In our study,
we investigated strong nonlinear damping effects
in high amplitude oscillations of various neutron star models,
and the limits imposed on the gravitational wave signal.
For this, we used a relativistic evolution code,
but kept the spacetime fixed.

Our neutron star models are rigidly rotating ideal fluid configurations
with polytropic EOSs of different stiffness, and typical neutron star masses
in the range 1.4--1.9 solar masses.
The oscillation modes we considered are the
axisymmetric \Prad and \Paxi~modes, the corotating \Pcor~mode, and
the counter-rotating \Pctr~mode.

Our main interest is the \Pctr~mode of one model (MA65),
which can be excited by the CFS instability in full GR,
and which is therefore a candidate for detectable gravitational waves
from rapidly rotating neutron stars.

We found different damping mechanisms for axisymmetric and nonaxisymmetric modes.
Axisymmetric oscillations of high initial amplitude
are rapidly damped by formation of shocks in the outer layers of the star,
until the amplitude falls below a certain threshold.
Below the threshold we observed no further energy dissipation
on the timescales of our evolutions.

Although we are using a one-parametric EOS,
our results accurately determine the onset of shock formation,
and provide a first estimate for the magnitude of energy dissipation by shocks.
We find that the damping timescale decreases with increasing amplitude,
down to a few oscillation cycles.
The damping timescale and the threshold vary by one order of magnitude
for the different EOSs we investigated.
The stiffer the EOS, the higher amplitudes are possible before
shocks form,
and also the damping observed at the highest amplitudes is faster for stiffer EOSs.

With increasing rotation rate,
nonlinear effects start at lower amplitudes.
However, the influence of rotation on the damping of axisymmetric modes
is relatively weak until close to the Kepler limit,
where the damping threshold decreases to zero.
At the Kepler limit, damping by mass shedding occurs.
We explain this behavior by the fact that the material at the equator is
bound less strongly for faster rotation.

For the nonaxisymmetric oscillations,
damping is mainly caused by wave breaking and nonlinear mode coupling effects,
while shock formation only occurs
in low density regions at the stellar surface.
As for axisymmetric modes,
nonlinear damping sets in at lower amplitudes when
increasing the rotation rate or decreasing the stiffness of the EOS.

Since wave breaking is a surface effect,
we feel that an accurate description of the damping at high amplitudes
requires an improved  numerical treatment of the surface
and probably a better physical model as well,
considering e.g. a hot envelope for a proto-neutron star,
a solid crust for a cold star (although in that case, one would also have
to add superfluidity), and the influence of the magnetic field.

Significant mode coupling was only observed for rapidly rotating models,
where the main mechanism seems to be a 1:2 resonance with
high order $m=\pm 2$ inertial modes.
For CFS-unstable models, such a resonance is always possible.
Further, the coupling is most probably active already at lower amplitudes,
but on timescales too long for nonlinear study.
Since the growth time of CFS-unstable modes is longer as well,
it will be necessary to investigate the coupling at lower amplitudes
perturbatively to determine the saturation amplitudes.
Nevertheless, we were able to provide upper limits.

Next, we computed the gravitational wave strain and luminosity of
several modes in the linear regime.
The CFS-unstable \Pctr~mode was found to be a weak emitter;
at the same amplitude, the strain is thirty times lower than for the
axisymmetric \Paxi~mode.
This is mainly due to the low frequency of the \Pctr~mode in the inertial frame.
To construct models for which this mode is more luminous,
one needs to further separate the Kepler limit from the neutral point
(where the \Pctr~mode has zero inertial frame frequency).
This can either be done by increasing the stiffness of the EOS,
or probably by introducing differential rotation.
Since our CFS-unstable model already has a relatively
stiff EOS ($\Gamma=2.46$),
we consider the latter more realistic.

We also found that the quasiradial \Prad~modes of rapidly rotating models
emit gravitational waves almost as efficient as the \Paxi~modes.
For the CFS-unstable model MA65,
the \Prad~mode emits stronger gravitational waves than the \Pctr~mode
at the same amplitude.
The reason is that the eigenfunction of the \Prad~mode
has a considerable quadrupole moment due to the oblateness of the star,
as well as the high oscillation frequency.

Further we investigated the gravitational waves produced by oscillations
in the nonlinear regime.
The maximum strain we observed usually agrees with the values extrapolated from the linear regime
up to a factor of 3 (2 for the axisymmetric modes).
For the CFS-unstable \Pctr~mode however,
we found that the signal is dominated by secondary modes.
We consider those as a feature of
our particular setup;
their amplitudes most likely depends on the exact way
the main mode was excited.
To compute the strain for the case of
a \Pctr~mode grown to high amplitude by the CFS instability,
it will be necessary to model carefully the growth of secondary modes,
especially at higher frequencies.
Secondary modes dominate the strain only for the CFS-unstable \Pctr~mode,
because it is the weakest emitter.
For models where it has a higher frequency, secondary modes
are probably less important.

Finally, we combined the results on strong damping and luminosity
to establish upper limits on the GW detectability of various modes.
There are two astrophysically relevant cases:
a continuous emission by CFS-unstable \Pctr~modes at saturation amplitude,
and stable modes excited by some violent event.

From the upper limit for the saturation amplitude of the CFS instability,
we obtain an upper limit on corresponding gravitational strain.
The detectability of a continuous signal depends on the integration time.
If we consider searches using time windows containing at least a few hundred cycles,
the \Pctr~mode (of model MA65) oscillating at the upper limit
will be detectable at distances around $10\usk\mega\parsec$ with advanced LIGO.
This distance scales approximately quadratically with the mode frequency in the inertial frame.
Current searches for continuous signals in the 5th LIGO science run are not sensitive
to the above source since they exclude the high spindown rates caused
by gravitational waves of such strength.
We stress again that the actual saturation amplitude may be considerably lower than
the upper limits we provided.
We also note that we did not take into account the secondary excited modes;
a careful study of the couplings during the growth phase
might reveal an enhanced detectability.

For stable modes excited by a sudden violent event,
we find that the detectability does not significantly increase
for amplitudes above the onset of strong damping,
since the oscillation is quickly damped below the threshold and stays
there.
The detectability thus depends on the effective number of wave cycles in the tail,
and thus on the residual slow damping at amplitudes near the threshold.
If this information becomes available for a given mode,
e.g. using perturbative techniques,
the detectability can be computed from the luminosities we provided.
However,
we find that the \Paxi~modes oscillating near the strong damping threshold
are detectable with advanced LIGO at least up to distances of $10 \usk\mega\parsec$.
The strong damping effects thus impose only weak limits on the detectability of those;
for most processes,
the important factor will probably be the strength of the excitation.

It is worth mentioning that
although some of the questions we rise can only
be computed with perturbation theory,
only a nonlinear study like ours can provide the necessary
limits for the amplitudes at which perturbation theory
can be safely applied.
Given that the physically most interesting amplitudes are
at the transition from linear to the nonlinear regime,
we feel that it is necessary to combine linear and nonlinear
methods to obtain accurate results on the detectability of oscillation modes.

\begin{acknowledgments}
This work
was supported by the Deutsche Forschungsgemeinschaft DFG (German Research
Foundation) via SFB/TR7, B.W. received funding from Cusanuswerk.
Computations have been carried out on the \textsc{hpc} cluster of the University of
T\"ubingen and the \textsc{blade} cluster at SISSA.
\end{acknowledgments}


\end{document}